\documentclass[preprint2]{aastex}
\usepackage{subfigure}
\usepackage{enumerate}
\usepackage{courier}
\usepackage{epsfig}
\usepackage{lscape}
\usepackage{color,soul}
\usepackage[normalem]{ulem}
\sethlcolor{yellow}
\usepackage{amsmath}
\usepackage{booktabs}
\usepackage{hyperref}

\begin{document}
\shorttitle{RV and Stellar Abundance Analyses of HD 133131A \& B}

\shortauthors{Teske et al.}

\title{The Magellan PFS Planet Search Program: Radial Velocity and Stellar Abundance Analyses of the 360 AU, Metal-Poor
  Binary ``Twins'' HD 133131A \& B$^{\dagger}$}

\altaffiltext{$^{\dagger}$}{This paper includes data gathered with the 6.5 meter Magellan Telescopes located at Las Campanas Observatory, Chile.}

\author{Johanna K. Teske\altaffilmark{1,*}, Stephen
  A. Shectman\altaffilmark{2}, Steve S. Vogt\altaffilmark{3},
  Mat\'{i}as D\'{i}az\altaffilmark{2, 4}, R. Paul
  Butler\altaffilmark{1}, Jeffrey D. Crane\altaffilmark{2}, Ian
  B. Thompson\altaffilmark{2}, Pamela
  Arriagada\altaffilmark{1}}
\altaffiltext{1}{Department of Terrestrial Magnetism, Carnegie
  Institution for Science, Washington, DC 20015}
\altaffiltext{2}{Observatories, Carnegie Institution for Science,, 813 Santa Barbara Street, Pasadena, CA 91101-1292,
  USA}
\altaffiltext{3}{UCO/Lick Observatory, Department of Astronomy and Astrophysics, University of California at Santa Cruz, Santa Cruz, CA 95064}
\altaffiltext{4}{Universidad de Chile, Departmento de Astronom\'{i}a,
  Camino El Observatorio 1515, Las Condes, Santiago, Chile}
\altaffiltext{*}{jteske@carnegiescience.edu; Carnegie Origins
    Fellow, joint appointment between Carnegie DTM and Carnegie Observatories}

\begin{abstract}
We present a new precision radial velocity (RV) dataset that reveals
multiple planets orbiting the stars in the $\sim$360 AU, G2$+$G2
``twin'' binary HD 133131AB. Our 6 years of high-resolution echelle
observations from MIKE and 5 years from PFS on the Magellan telescopes
indicate the presence of two eccentric planets around HD 133131A
with minimum masses of 1.43$\pm$0.03 and 0.63$\pm$0.15
$\mathcal{M}_{\rm J}$ at 1.44$\pm$0.005 and 4.79$\pm$0.92 AU, respectively. Additional PFS observations of HD 133131B spanning 5
years indicate the presence of one eccentric planet of minimum mass 2.50$\pm$0.05
$\mathcal{M}_{\rm J}$ at 6.40$\pm$0.59 AU, making it one of the longest
period planets detected with RV to date. These planets are the first to be reported primarily based on data taken with PFS on Magellan, demonstrating
the instrument's precision and the advantage of long-baseline RV
observations. We perform a differential analysis between
the Sun and each star, and between the stars themselves, to derive
stellar parameters and measure a suite of 21 abundances across a wide
range of condensation temperatures. The host stars are old (likely $\sim$9.5
Gyr) and metal-poor ([Fe/H]$\sim$-0.30), and we detect a $\sim$0.03 dex depletion
in refractory elements in HD 133131A versus B (with standard errors
$\sim$0.017).
 This detection and analysis adds to a
small but growing sample of binary ``twin'' exoplanet host stars with
precise abundances measured, and represents the most metal-poor and
likely oldest in that sample. Overall, the planets around HD 133131A
and B fall in an unexpected regime in planet mass-host star metallicity
space and will serve as an important benchmark for the study of long period
giant planets. 

\end{abstract}


\section{Introduction}

In the study of exoplanets, how unique our Solar System is in
comparison to other planetary systems is one of the most compelling
yet elusive questions. Planet-detecting transit surveys, especially the \textit{Kepler} mission,
as well as radial velocity (RV) surveys have revealed populations of planets
unlike any in our Solar System, of which the most numerous appear to
be ``super-Earths'' that are larger than Earth but smaller
than Uranus and Neptune (Borucki et al. 2011; Batalha et al. 2013), and that may or may not be gas-dominated
(Wolfgang \& Lopez 2015; Rogers 2015; Marcy et al. 2014; Lopez \& Fortney 2014; Ikoma \& Hori 2012;). Both techniques for finding planets are biased toward
short period and large/massive planets, but the long baseline of RV
observations (over 15 years in some cases) makes the RV technique more
suitable for detecting long-period planets, like Jupiter and Saturn, than any current or planned
transit surveys. Current theories of our Solar System formation point to
Jupiter and Saturn as important players in the dynamical
shaping of the Asteroid belt (Walsh et al. 2011; Morbidelli et al. 2010), and to their migratory
dance toward and away from the Sun as the cause of a late delivery
of volatile material to the inner rocky planets (Horner \& Jones 2010; Morbidelli et al. 2000; Owen \& Bar-Nun 1995). The
architecture of our Solar System is likely due in large part to
Jupiter (and to a lesser extent, Saturn); thus, how unique our Solar System is may depend on the
occurrence of long period giant planets.

Recently, the frequency of Jupiter analogs -- defined as $5 < \mathcal{P} < 20$ years, $0.3
< \mathcal{M} < 3~\mathcal{M}_J$, $e < 0.3$ -- was determined from a sample of
over 1100 stars observed with Keck for RV variations due to planets
(Rowan et al. 2016). These authors, correcting for relative
observability around each of the stars, found that $\sim$3\% of stars
host Jupiter analogs, with their 10-90\% confidence intervals
suggesting a frequency of 1-4\%. Wittenmyer et al. (2016) also recently published
a Jupiter analog occurrence rate, corrected for incompleteness, based
on 17 years of AAT observations of 202 solar-type stars, finding a Jupiter analog frequency of 6.2$^{+2.8\%}_{-1.6\%}$. Their ``Jupiter analog'' definition included
  any planets with $\mathcal{M} > 0.3~\mathcal{M}_J$ and $a >3$ AU,
  and their confidence interval is only 65.7\% around the peak of the posterior distribution function. Comparing
           ``apples to apples'', the two studies are consistent within
           errors. 

These studies suggest that Jupiter analogs may indeed be
rare. However, it is also interesting to consider long period giant planets that
fall outside this strict definition, particularly the $e$ limit. Giant
planets with higher eccentricities may have drastically influenced the
formation/survival of any interior rocky planets (Matsumura et
al. 2013; Raymond 2006; Th\'ebault et al. 2002), or may
be relics of a multi-planet system where one or more
planets were ejected, potentially shedding light on the existence of 
free-floating planets (e.g., Rasio \& Ford 1996; Weidenschilling \&
Francesco 1996). In fact, the working theory of
giant planet evolution in the Solar System proposes that an ice giant scattered off of Jupiter to cause it to
``jump'' over a 2:1 mean motion resonance with Saturn, that this
planet-planet scattering excited Jupiter's eccentricity (e.g., Batygin et al. 2012; Morbidelli et al. 2010; Nesvorn\'y 2011; Brasser et al. 2009;
Tsiganis et al. 2005), and that early-formed short period planets
would be scattered into the Sun (Batygin \& Laughlin 2015). If long period giant planets are
more frequently eccentric, this would be an important clue to understanding
our Solar System formation and planet habitability in general (e.g., Kaib \& Chambers 2016). 

Here we report the detection of three giant planets on
long, eccentric orbits from data collected primarily with the Planet Finder Spectrograph on
Clay/Magellan II at Las Campanas Observatory. In addition, we perform a precision stellar
parameter and abundance analysis on the two host stars, which form a
$\sim$360 AU binary and are ``twins'' of identical spectral
type. Interestingly, we find small but significant differences in the
refractory element (with condensation temperatures $> 1000$ K)
abundances between the two host stars. We discuss how these
differences may be related to differences in planet
formation/composition, and the nature of the host star binary.

\section{Prior Characterization of HD 133131A \& B}
HD133131 (HIP 73674), composed of a pair of bright (V = 8.40 and 8.42)
G2V stars, was first recorded as a 7$\arcsec$ binary by Stock \&  Wroblewski (1972) on
objective prism plates taken with the 24'' Curtis Schmidt Telescope
at CTIO. The two stars are included in the Geneva-Copenhagen Survey
(GCS), a comprehensive catalog providing kinematics and Galactic
orbits, companion detections, and distances of nearly 17000 late-type stars in the solar neighborhood
(Nordstr\"om et al. 2004; Holmberg et al. 2007; Holmberg et
al. 2009). The latest large update to the catalog was by Casagrande et
al. (2011), who used the Infrared Flux Method (IRFM; Casagrande et
al. 2010) to derive new temperatures, metallicities (which are
calibrated to high-resoultion spectroscopy), and ages. They report
updated effective temperatures ($T_{eff}$) of 5791 K and 5768 K and updated [Fe/H]\footnote{[X/H]=log(N$_{\rm{X}}$/N$_{\rm{H}}$) - log
  (N$_{\rm{X}}$/N$_{\rm{H}}$)$_{\rm{solar}}$} values of -0.40 and
-0.42 for HD 133131A and B, respectively, and report an age between
5.52 and 5.96 Gyr based on a Bayesian approach using BASTI
(Pietrinferni et al. 2004, 2006, 2009) and Padova (Bertelli et
al. 2008, 2009) isochrones and log($T_{eff}$), absolute Johnson $V$
magnitude, and metallicity. However, their age distributions are
wide, and extend to $\sim$12.8 Gyr. The average distance reported in
Nordstr\"om et al. (2004) and Holmberg et al. (2009) is 47 pc, adopted from the trigonometic distance from Hipparcos (original reduction
  from ESA 1997 and new reduction from van Leeuwen
  2007). Tokovinin (2014a) included the HD
133131 system in his imaging survey of wide
binaries, and found no evidence of a
companion around HD 133131B between 0.042$\arcsec$ and
1.5$\arcsec$. 

Desidera et al.\,(2006a,b) also included HD 133131A and B in their
study of abundance differences between components of wide binaries,
measured with the FEROS spectrograph at ESO La Silla. The
  separation between the two stars is 7.4$\arcsec$ (from
  Hipparcos; ESA 1997), translating to a physical separation of
360 AU. We calculate a period of $\sim$4240 years for HD
133131AB from the median separation (Washington Double Star Catalog\footnote{\url{http://ad.usno.navy.mil/ad/wds/wds.html}}),
parallax (van Leeuwen 2007), and assumed solar mass of both
stars. Tokovinin (2014b), which compiled results from several of his own
works, also reported a period of $\sim$4240 years for the HD 133131AB
pair. This makes HD 133131A and B the smallest-separation ``twin''
binary system analyzed for high precision abundance differences (the
next closest is 16 Cyg AB, at $\sim$860 AU; Eggenberger et
al. 2003). The Desidera stellar properties of HD 133131A and B, listed in
Table \ref{params}, were determined with a methodology very similar to
that used here (described below), so represent the best comparison. Desidera
et al. used their spectroscopic $T_{eff}$ and abundances with
isochrones of Girardi et al. (2002) and the bolometric corrections of
Kurucz (1995) to iteratively derive masses of 0.95 $\mathcal{M}_{\odot}$ and 0.93
$\mathcal{M}_{\odot}$ for the A and B components, respectively, which are the
stellar masses we assume in our radial velocity search for planets
below. They derive a much older age for the HD 133131 system, of
$\sim$9.86 Gyr, from chromospheric activity measured in the Ca H \& K
lines (log$R'$(HK)).  

\section{Radial Velocity Observations}

The radial velocity (RV) observations of HD 133131A and B are part of the
large Magellan Planet Search Program, which began in 2002 and is
surveying a sample of $\sim$500 of the nearest stars ($<$100 pc). The
survey was started with observations from the MIKE echelle
spectrograph (Bernstein et al. 2003), mounted for a limited time on the
Magellan I (Baade), but mostly on Magellan II (Clay), 6.5 m
telescopes at Las Campanas Observatory. In 2010, the survey switched
to using the Carnegie Planet Finder Spectrograph (PFS) (Crane et al. 2006,
2008, 2010), a temperature-controlled high resolution echelle
spectrograph built for precision radial velocity observations, on
Magellan II. Both instruments contain an iodine absorption cell (Marcy \& Butler 1992)
that imprints the reference iodine spectrum on the incoming starlight,
providing a measurement of the instrument point spread function and a
precise wavelength solution. Doppler shifts from
the spectra are obtained using the technique described in Butler et
al. (1996). In brief, the iodine region of the stellar spectrum
(between $\sim$5000-6200 {\AA}) is divided into 2 {\AA} chunks, on which
a forward modeling procedure is performed, providing an independent
measurement of the instrument PSF, wavelength and Doppler shift. The
measured velocity for each spectrum is calculated from the weighted mean of the
independent chunk velocities; the reported internal uncertainty is the standard
deviation of all of the chunk velocities measured from one
spectrum. The weighted mean Doppler shift and the internal uncertainty
for each observation are reported in Tables \ref{tab:rvdata_MIKE},
\ref{tab:rvdataA_PFS}, \ref{tab:rvdataB_PFS}.

\subsection{Magellan/MIKE RV Observations}

Only HD 133131A observations from MIKE are included here. Using a
0.35x5$\arcsec$ slit, MIKE provides spectra with $R\sim$70,000 in the
blue and $\sim$50,000 in the red and covers 3900-6200\,{\AA}. Only the red MIKE orders are used
for radial velocity determination, while the blue orders provide
coverage of the Ca II H and K lines for monitoring stellar activity. 

The MIKE observations of HD 133131A span June 2003 to July 2009, with
total exposure times ranging from 150 s to 600 s, depending on
observing conditions. Calibrations, taken at the beginning and end of
each night, consist of 20, 21, or 30 flat-field images taken when the
slit is illuminated by an incandescent lamp, two exposures of the
incandescent lamp passing through the iodine cell, two exposures of
rapidly rotating B stars taken through the iodine cell, and two ThAr
exposures, which are not used in the data reduction of the RV determination described above. 

Reduction of the raw CCD images and spectral extraction were carried
out using a custom IDL-based pipeline that performs flat fielding,
removes cosmic rays, and measures and subtracts scattered light. No
sky subtraction is done, as our targets are all relatively bright.

\begin{deluxetable}{cccc}
\tablecaption{MIKE radial velocities and $S$-index values for HD 133131A \label{tab:rvdata_MIKE}}
\tablecolumns{4}
\tablewidth{0pc}
\tablehead{\colhead{JD} &\colhead{RV
    [m\,s$^{-1}$]}&\colhead{Uncertainty [m\,s$^{-1}$]} & \colhead{$S$-index\tablenotemark{a}}}
\startdata
2452808.68 & -34.92 & 9.67 & \nodata\\ 
2453041.86 & 39.84 & 9.64 & \nodata\\ 
2453128.70 & -45.78 & 5.06 & 0.1507\\ 
2453215.59 & -24.23 & 9.97 & \nodata\\ 
2453872.71 & -22.16 & 4.94& 0.1380\\ 
2454190.80 & 40.61 & 5.22 & 0.1343\\ 
2454277.62 & 30.32 & 7.70 & 0.1336\\ 
2454299.52 & 28.35 & 6.24& 0.1412\\ 
2454300.56 & 30.84 & 5.24 & \nodata\\ 
2454501.86 & -3.36 & 5.41& 0.1310\\ 
2454522.87 & -24.05 & 5.65 & 0.1353\\ 
2454650.64 & -17.76 & 8.85& 0.1403\\ 
2454925.83 & 38.12 & 4.79& \nodata\\ 
2454963.74 & 44.61 & 5.06 & 0.1621\\ 
2454993.67 & 21.47 & 4.61& 0.1496\\ 
2455001.67 & 0.00 & 4.62& 0.1511\\ 
2455018.62 & -6.87 & 4.61& \nodata\\ 
\enddata
\tablenotetext{a}{Missing $S$-index values are due to the absence of
  a thorium argon (ThAr) calibration frame, which is necessary to
  derive a wavelength solution and an accurate $S$-index value.}
\end{deluxetable}

\subsection{Magellan/PFS Observations}

Both HD 133131A and B were observing with PFS, the former observations
ranging from February 2010 to September 2015, and the latter from
August 2010 to September 2015. PFS has a more limited wavelength range
than MIKE (3880-6680\,{\AA}), but still covers the entire iodine
wavelength region, Ca II H and K, and H$\alpha$. We use a 0.5x2.5$\arcsec$
slit for target observations, providing $R\sim$80,000 in the iodine
region. The total exposure times for the A component range from 285 to
720 s, and for the B component range from 282 to 800 s. 

Similar to MIKE, PFS calibrations taken at the beginning of each night
include 20-30 flat-field images, two iodine exposures, two rapidly
rotating B star exposures, and one or two ThAr exposures. A
modification of the MIKE pipeline is used for PFS raw frame reduction and spectral extraction.

\begin{deluxetable}{cccc}
\tablecaption{PFS radial velocities and $S$-index values for HD 133131A \label{tab:rvdataA_PFS}}
\tablecolumns{4}
\tablewidth{0pc}
\tablehead{\colhead{JD}&\colhead{RV
    [m\,s$^{-1}$]}&\colhead{Uncertainty [m\,s$^{-1}$]} & \colhead{$S$-index}}
\startdata
2455254.88451  &   -3.54 &  1.24 & 0.1494  \\
2455339.75139   &   6.65  & 1.06 &  0.1502 \\
2455428.50899   &  23.76  & 1.41 &  0.1516 \\
2455671.72991   &  12.44 &  1.45 &  0.1523 \\
2456087.61006   &  16.21 &  1.63 &  0.1553 \\
2456137.54073  &   31.91 &  1.40 &  0.3851 \\
2456343.84765   &   0.00 &  1.38 &  0.1563 \\
2456355.85373 &    -3.28 &  1.39 &   0.1505\\
2456357.86100  &   -2.51 &  1.44 &   0.1561\\
2456428.73879 &   -10.79 &  1.72 &  0.1562 \\
2456434.70128 &   -13.01 &  1.43 &  0.1549 \\
2456508.53596 &   -18.00 &  1.43 & 0.1527\\
2456701.87954 &     7.10 &  1.37 &   0.1527\\
2456702.84446 &    10.45  & 1.51 &   0.1579\\
2456731.80915 &    20.76 &  1.64 &   0.1712\\
2456735.82306 &    16.46 &  1.57 &   0.1549\\
2456816.66188 &    52.13 &  1.80 &   0.1639\\
2456867.49782 &    58.84 &  1.43 &   0.1559\\
2456877.50060 &    57.33 &  1.39 &   0.1554\\
2457059.87029 &    -7.71 &  1.43 &   0.1545\\
2457117.77736 &   -12.15 &  1.53 &   0.1543\\
2457122.79501 &    -8.63 &  1.47 &   0.1532\\
2457199.61947 &    -8.18 &  1.57 &   0.3073\\
2457202.54461  &  -12.58 &  1.60 &   0.1544\\
2457260.51644  &   -1.19  & 1.37 &   0.1561\\
2457268.49079  &   -6.22  & 1.44 &   0.1544\\
\enddata
\end{deluxetable}

\begin{deluxetable}{cccc}
\tablecaption{PFS radial velocities and $S$-index values for HD 133131B \label{tab:rvdataB_PFS}}
\tablecolumns{4}
\tablewidth{0pc}
\tablehead{\colhead{JD}&\colhead{RV [m\,s$^{-1}$]}
  &\colhead{Uncertainty [m\,s$^{-1}$]} & \colhead{$S$-index}}
\startdata
2455428.51343 &    58.42&   1.27 &  0.1512   \\
2455671.73461  &   63.18 &  1.37 &   0.1515 \\ 
2456087.62415 &    59.12 &  1.87 &   0.1611\\  
2456137.55012 &    59.12 &  1.27 &  0.1580 \\
2456343.85348 &    32.00 &  1.35 &   0.1595 \\
2456355.85926 &    27.74 &  1.47 &    0.1540 \\
2456357.86754 &    30.47 &  1.54 &    0.1531 \\
2456358.84334 &    31.97 &  1.30 &     0.1486\\
2456428.74626 &    15.37 &  1.70 &   0.1544\\
2456434.70757 &    15.75 &  1.40 &   0.1546\\  
2456508.54383 &    -0.09 &  1.48  &    0.1521\\
2456701.88398 &   -11.38&   1.37  &   0.1515\\ 
2456702.84978 &   -11.59 &  1.40 &   0.1488\\ 
2456731.81891 &   -10.39  & 1.68 &    0.1665\\ 
2456735.82875 &    -7.91  & 1.62  &   0.1545 \\
2456816.67181 &   -11.78 &  1.82  &   0.1538 \\
2456867.50604  &  -12.35 &  1.49 &   0.1551\\  
2456877.50858 &    -6.64 &  1.43 &   0.1567\\ 
2457060.87758  &   -3.55 &  1.23 &   0.1470\\ 
2457117.78230 &    0.00  & 1.52 &    0.1491\\ 
2457122.80056 &    -1.58 &  1.40  &   0.1523\\ 
2457199.62601 &    -0.90  & 1.63  &   0.1565\\ 
2457202.55159 &    -2.30 &  1.60 &    0.1513\\ 
2457260.52516 &     1.87 &  1.39 &   0.1586\\
2457268.49663 &     0.86 &  1.46 &   0.1589 \\
\enddata
\end{deluxetable}

\section{Best-fitting Keplerian Solutions}
We use the SYSTEMIC console for Keplerian fitting of the radial
velocity data (Meschiari et al. 2009).  First, we computed the
error-weighted ($w_j$=1/$\sigma^2_j$),
normalized Lomb-Scargle periodogram (Zechmeister \& K\"{u}rster 2009;
Gilliland \& Baliunas 1987)
of each star's observations (MIKE$+$PFS for HD 133131A, PFS for
HD 133131B), shown in Figure \ref{fig:periodogram_AB}. The False Alarm Probabilities were computed by
scrambling the datasets 10,000 times and sampling the periodogram at
80,000 frequencies, to calculate the probability that the power at
each frequency could be exceeded by random chance. Overplotted in red
on the periodograms is the spectral window, or periodogram of the
sampling, which shows the expected peaks from the observing cadence, the
sidereal and solar days, and the solar year (Dawson \& Fabrycky
2010). 

\begin{figure}
\centering
\includegraphics[width=0.5\textwidth]{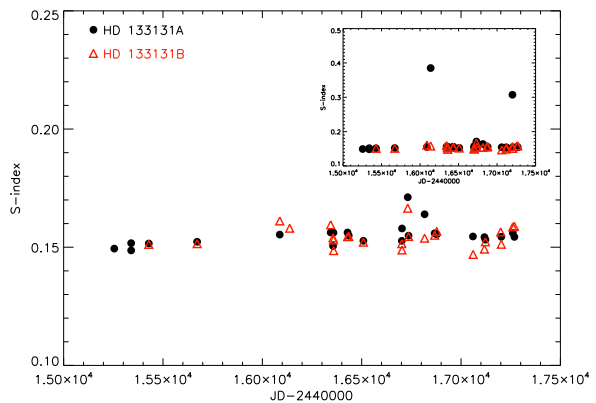}
\caption{The change in $S$-index measurements in HD 133131A (black
  dots) and HD 133131B (red triangles) with time. The inset plot is a
  zoomed-out version, showing two instances where the $S$-index for HD
133131A was significantly larger, caused by
small errors in the wavelength calibration, which is not
well-constrained outside of the iodine region. The final median
S-index value for HD 133131A does not change ($\Delta$ median $\sim$3$\times 10^{-5}$) when we discount the two large
outliers. The plotted values are tabulated in Tables \ref{tab:rvdata_MIKE},
\ref{tab:rvdataA_PFS}, and \ref{tab:rvdataB_PFS}.}
\label{fig:svar}
\end{figure}

To evaluate the stellar activity levels of HD 133131A and B, we measure
Mt. Wilson $S$-index values in every radial velocity observation
spectrum; these values are listed in Tables \ref{tab:rvdata_MIKE},
\ref{tab:rvdataA_PFS}, and \ref{tab:rvdataB_PFS}. The $S$-index
compares the flux in triangle-weighted bins with full width at half
maximums of 1.09\,{\AA} centered on the Ca II H\&K lines (at 3968.47\,{\AA} and
3933.66\,{\AA}) to the flux in two rectangular 20\,{\AA}-wide continuum regions
centered on 3901 and 4001\,{\AA} (Duncan et al. 1991). This
index is known to be correlated with spot activity on the stellar
surface, and serves as a proxy for chromospheric activity that could
cause radial velocity shifts that mimic those induced by
planets. Previous measurements of $S$-indices for stars in the Magellan Planet
Search Program monitored with the MIKE spectrograph were detailed in
Arriagada (2011). In Figure \ref{fig:svar} we show how these values change with time in HD 133131A and B,
specifically focusing on the PFS data, since we do not have MIKE data
for HD 133131B. In this work we only measure the Ca II H line and surrounding flux
for PFS $S$-index values, effectively taking the ratio of the flux in
the H line to the flux in the \textit{R} continuum region, centered at
3996.5\,{\AA} as in Santos et al. (2000), referred to as $S_{COR}$ in
that work. The two large outliers in the $S$-index values for
HD133131A, shown in the inset of Figure \ref{fig:svar}, are caused by
small errors in the wavelength calibration, which is not
well-constrained outside of the iodine region. The $S$-index is
thus measured on the upward slope of the Ca II H line instead of centered
in the middle, causing an increase in its value. We include these
outliers to avoid any ``special'' treatment of specific spectra, and
the effect is small -- the final median $S$-index value for HD 133131A
changes by $\sim$3$\times 10^{-5}$ when we discount the two large
outliers. We show in Figure \ref{fig:periodogram_ABsval} that the peaks in the L-S periodograms for the Mt. Wilson $S$-index
measurements, which are a proxy for stellar magnetic activity modulation, are not
as significant as and do not correspond to the peaks in the RV
periodograms. 

\begin{figure}
\includegraphics[width=0.5\textwidth]{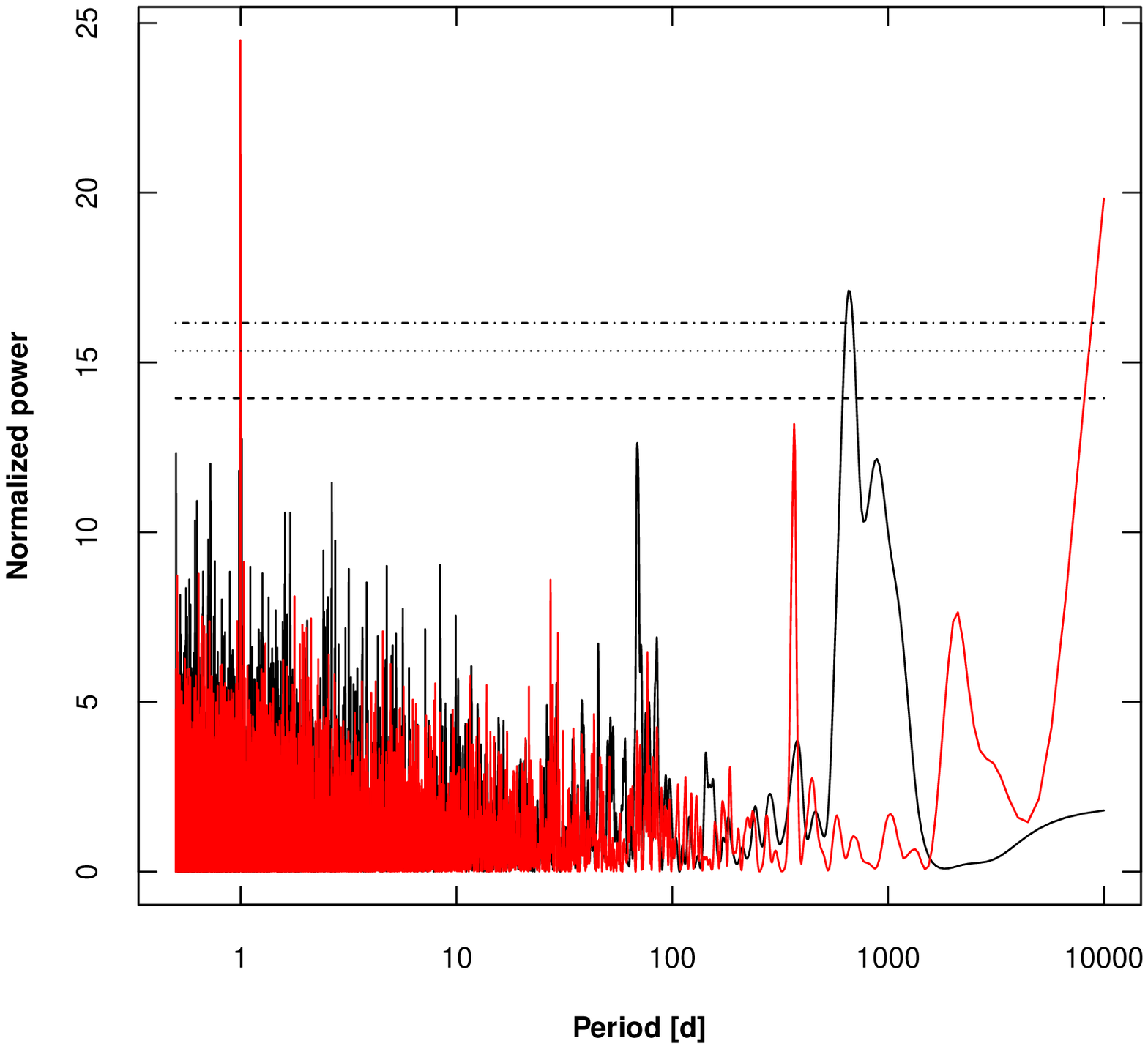}
\includegraphics[width=0.5\textwidth]{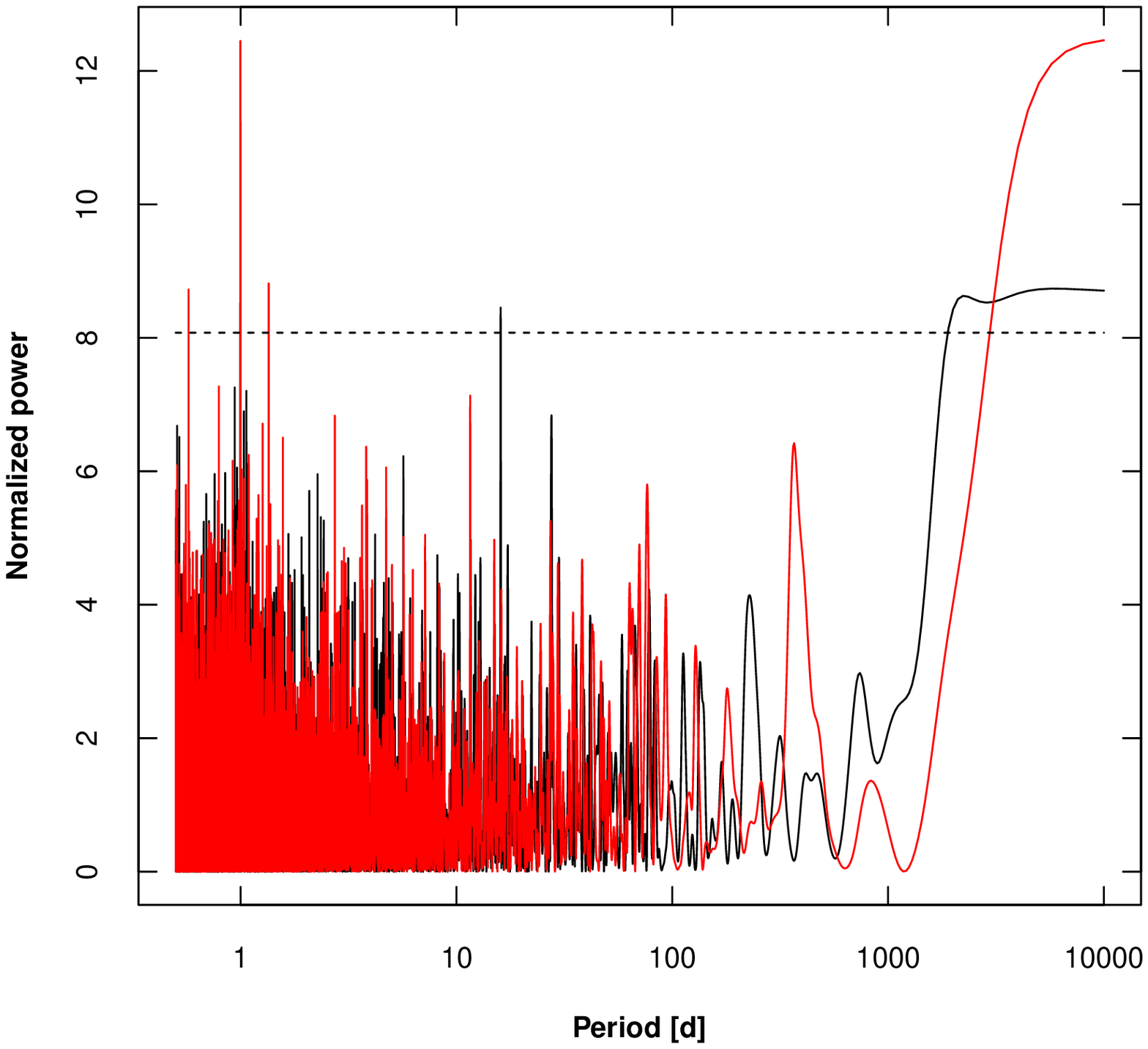}
\caption{Periodogram of the RVs of HD 133131A (left), from both Magellan/MIKE
  and Magellan/PFS, and HD 133131B (right), from Magellan/PFS, in black, with
  periodogram of sampling overlaid in red. False-alarm probability
levels are shown at the 10\%, 1\%, and 0.1\% levels, from bottom to
top, respectively; the 10\% false-alarm
  probability level for HD 133131B is shown as a dashed line. }\label{fig:periodogram_AB}
\end{figure}

\begin{figure*}
\includegraphics[width=0.5\textwidth]{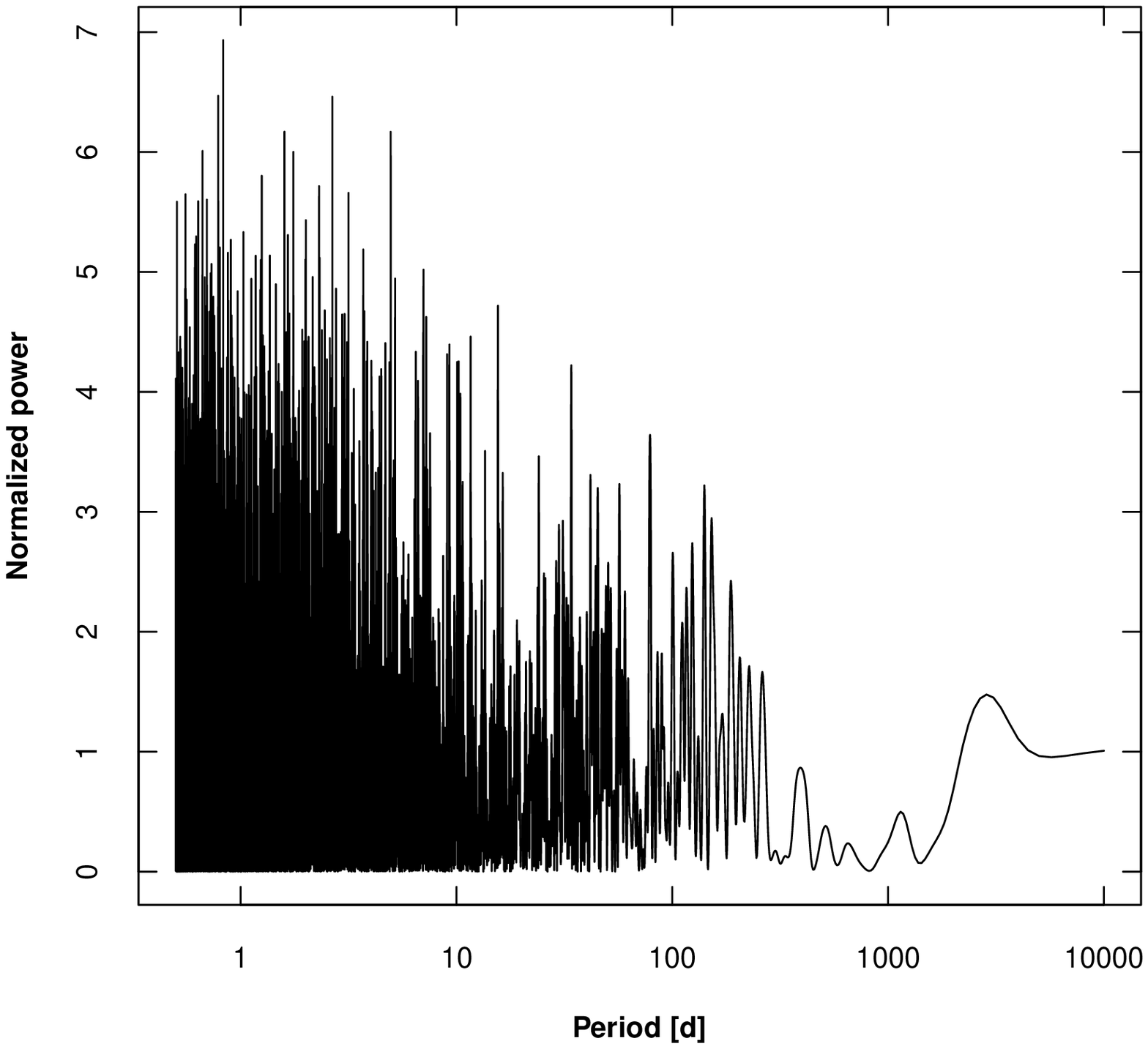}
\includegraphics[width=0.5\textwidth]{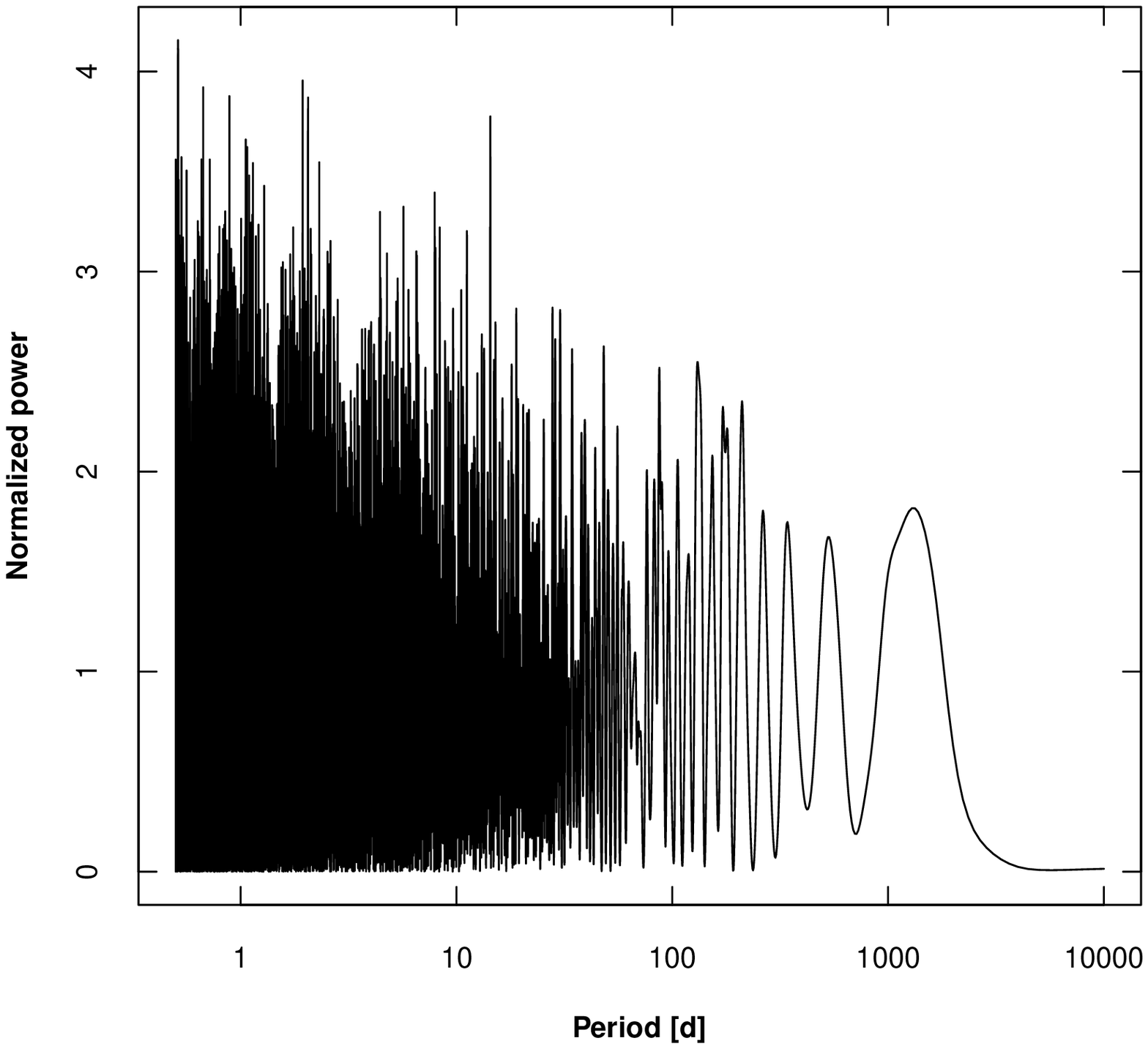}
\includegraphics[width=0.5\textwidth]{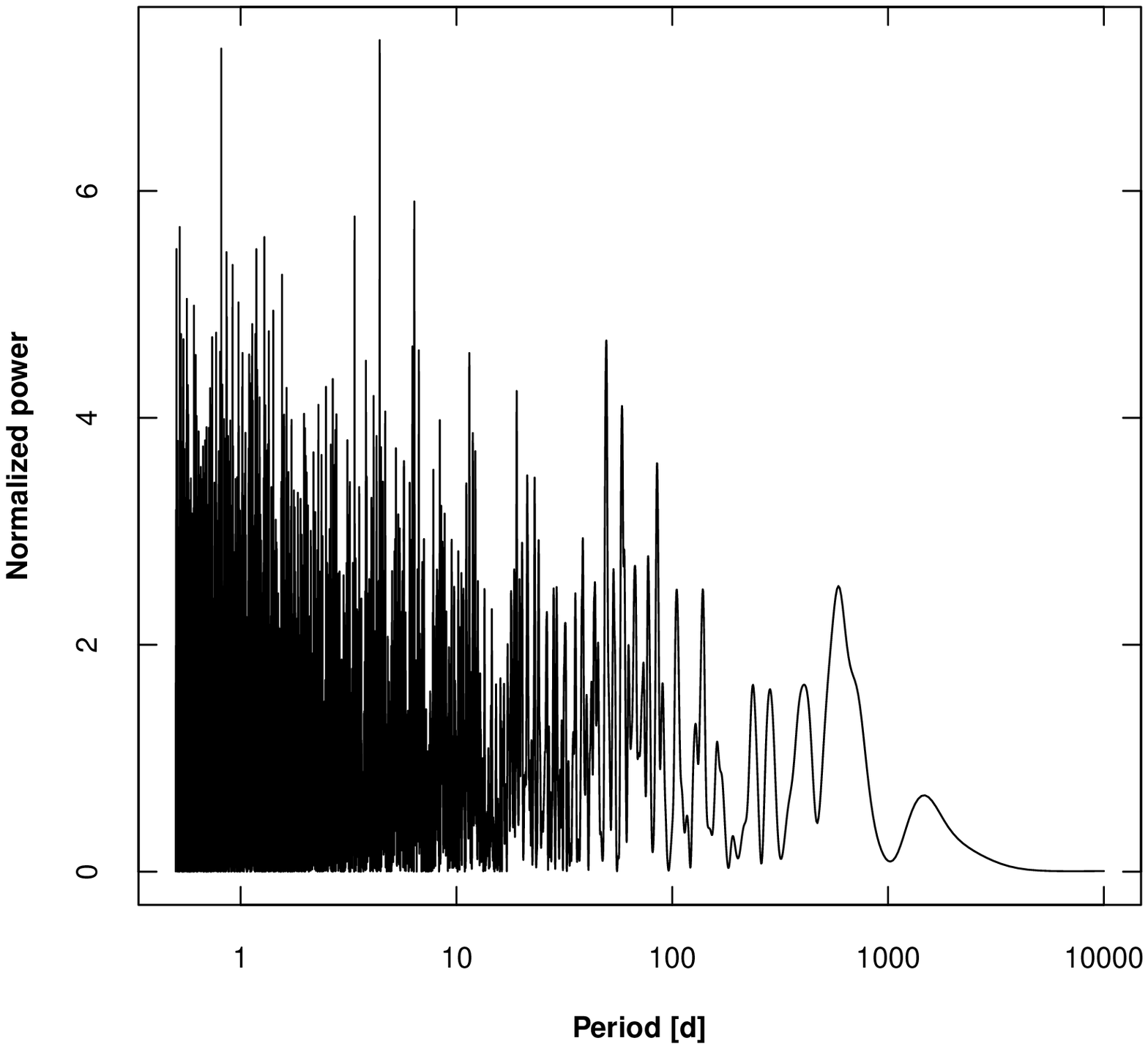}
\caption{Periodogram of the $S$-values of HD 133131A, from both
  Magellan/MIKE (top left)
  and Magellan/PFS (top right), and HD 133131B, from Magellan/PFS (bottom). The peaks here are not
as significant as and do not correspond to the peaks in the RV
periodograms in Figure \ref{fig:periodogram_AB}.}
\label{fig:periodogram_ABsval}
\end{figure*}

The highest peaks in the Lomb-Scargle periodogram provide a first
guess of the periods of planets around the two
stars. For HD 133131A, the highest periodogram peak (that is not at an
expected period due to time sampling) is at $\sim$660 days with a FAP of
1.28$\times10^{-8}$. For HD 133131B, the highest peak is at $\sim$6100
days with a FAP of 2.03$\times10^{-2}$. The procedure that SYSTEMIC
uses to model the radial
velocity signals is described in detail in Meschiari et al. (2009) and
Vogt et al. (2015), but briefly, we begin by fitting a 1-planet
Keplerian model with six free parameters (period, mass, mean anomaly,
eccentricity, longitude of pericenter, and a vertical offset to account
for differences in the velocity zero point between datasets). Each radial
velocity measurement is represented by the predicted velocity, the
formal (observing error) uncertainty, and an additional error term
accounting for scatter about the fit (e.g., from underestimated
measurement errors, stellar jitter, other astrophysical sources of
variation) that is the same for each observation in the data set. The
best-fit parameters are then derived by optimizing the log-likelihood
of the model:

\begin{equation}
\rm{log}~\mathcal{L} = -\frac{1}{2} \bigg[\chi^{2} +\sum_{i=1}^{N_o} \rm{log}~(e^{2}_{i} +s^{2}_{i})+N_{o} \rm{log}~(2\pi) \bigg]
\end{equation}

where

\begin{equation}
\chi^2 = \sum_{i=1}^{N_o} (V_i-v_i)^2/(e^2_i+s^2_i),
\end{equation}

\noindent and $V_i$ is the predicted velocity, $v_i$ is the observed velocity,
$e_i$ is the formal error, and $s_i$ is the additional error term. 

In a semi-automatic way, SYSTEMIC chooses the best parameters for a
planet fit using a downhill simplex algorithm (AMOEBA; Press et
al. 1992; Nelder \& Mead 1965).\footnote{In our analysis, we assume 0.95 $\mathcal{M}_{\odot}$ and
  0.93 $\mathcal{M}_{\odot}$ for HD 133131A and B, respectively, although our
  results do not depend strongly on small changes in these values.} The first planet fit is almost always the lowest FAP peak
in the RV periodogram; this is the case for HD 133131A and B. After the
first planet fit, the periodogram is re-calculated and the FAPs of the
peaks reassessed. In the case of the B component, there are no
remaining significant peaks (see Figure
\ref{fig:periodogram_resid_B}); stopping with a single
planet fit, displayed in Figure \ref{fig:B_rvfits_1p}, results in a final RMS of
1.59 m~s$^{-1}$. The detailed parameters of HD 133131Bb are given in
Table \ref{tab:Bfit}, with the errors explained below. 

The next highest peak in the periodogram of HD 133131B after
removal of planet b is at 5.88 days with a FAP of
4.63$\times10^{-1}$, as shown in Figure
\ref{fig:periodogram_resid_B}. This is not high enough to merit a
detection designation, and when including a planet at 5.88 days in our
fit to the HD 133131B RV data, the reduced $\chi^2$ is significantly
less than one (0.45), suggestive of over-fitting. However, with a
best-fit $\mathcal{M}~\sin{i}$ of 0.018 [$\mathcal{M}_{J}$], this
potential planet would be sub-Neptune in mass, likely falling into the
Super-Earth planet regime, perhaps at the boundary between planets
that are dominated by a volatile envelope and those that are not
(e.g., Rogers et al. 2011; Lopez et al. 2012; Zeng \& Sasselov 2013;
Wolfgang \& Lopez 2015). Only $\sim$15 other planets have both masses
and periods less than or equal to the values for the potential HD
133131Bc planet. Adding a low-mass short period planet to the system does not
disrupt the stability -- a 100,000 year Bulirsch–Stoer integration of
the two planets' orbits shows no orbital or eccentricity
overlap, and no significant changes in semi-major axis. Further implications of this potential planet are explored in
\S5; in any case HD 133131B warrants continued monitoring. 

In the case of HD 133131A, after removing a one planet fit, there is
still significant power, with a FAP of 1.37$\times10^{-4}$, at $\sim$3000 days as seen in Figure
\ref{fig:periodogram_resid_A}, left. We fit this as a second planet
around the star, resulting in a periodogram with no remaining
significant power (Figure
\ref{fig:periodogram_resid_A}, right). Thus we stop with a two planet
fit, displayed in Figure \ref{fig:A_rvfits_2p} (top), with a final RMS of
9.38 m~s$^{-1}$. The detailed parameters of HD 133131Ab and Ac are given in
Table \ref{tab:Afit}, with the errors explained below. We note that
only including the 26 Magellan/PFS data points in the HD 133131A fits, and
not the 17 Magellan/MIKE data points, decreases the RMS to 1.82 m~s$^{-1}$, and does not
alter the fitted planet parameters outside their errors (see below),
with the exception of the mean anomalies for both planets (which are
basically reversed in the sans-MIKE data fits, and are not well
constrained in either case).

\begin{figure}
\includegraphics[width=0.5\textwidth]{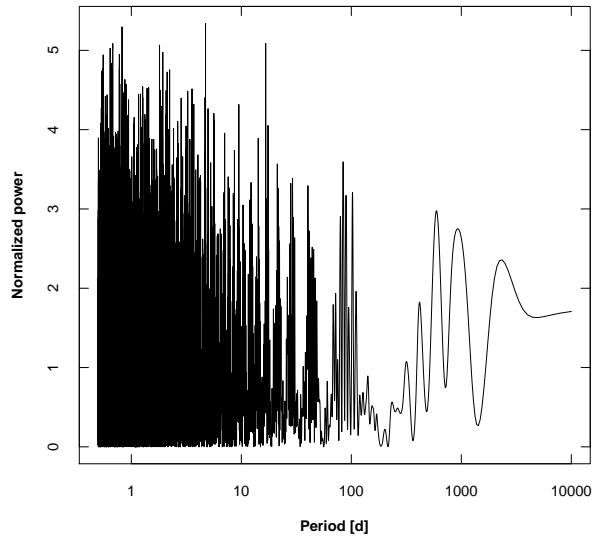}
\caption{Periodogram of $S$-index values measured from PFS spectra, after
  removal of two large outliers show in inset of Figure \ref{fig:svar}. There
  are no significant peaks, lending evidence to the planetary nature
  of the residual signal present in the HD 133131A RV data after
  fitting for a single planet.} \label{fig:Sval_periodogram_cut}
\end{figure}

\begin{figure}
\includegraphics[width=0.5\textwidth]{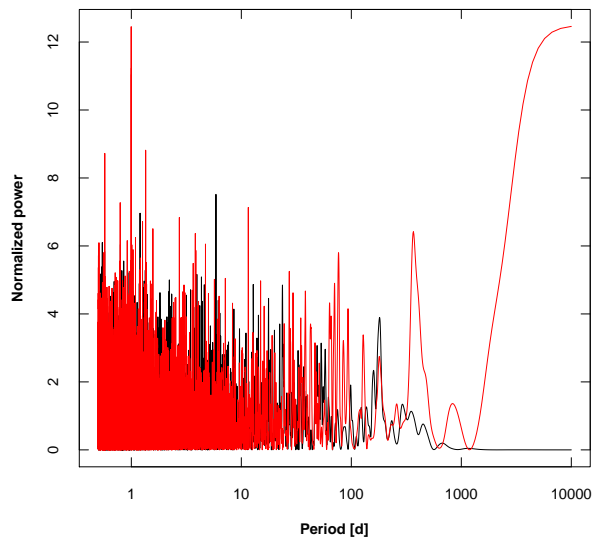}
\caption{Periodogram of residuals after fitting planet b around
  HD 133131B. Overplotted in red is the sampling periodogram.} \label{fig:periodogram_resid_B}
\end{figure}

\begin{figure}
\includegraphics[width=0.5\textwidth]{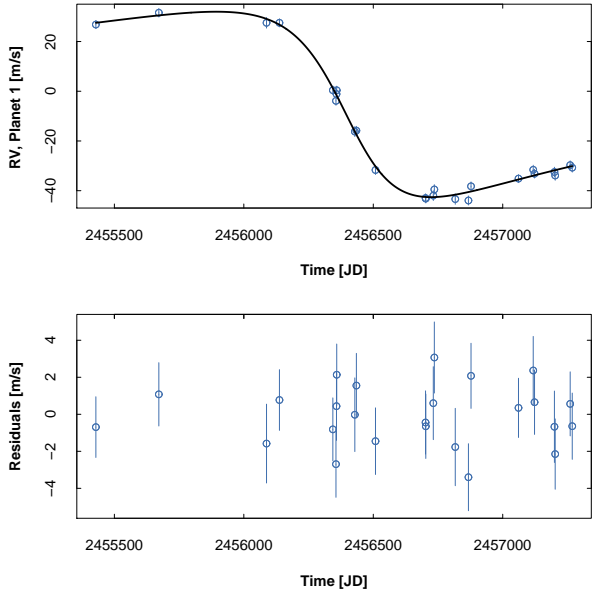}
\caption{Best-fit 1 planet solution to the RV data presented in this paper for
  HD 133131B. The parameters of the fit are detailed in Table \ref{tab:Bfit}.}
\label{fig:B_rvfits_1p}
\end{figure}

\begin{figure}
\includegraphics[width=0.5\textwidth]{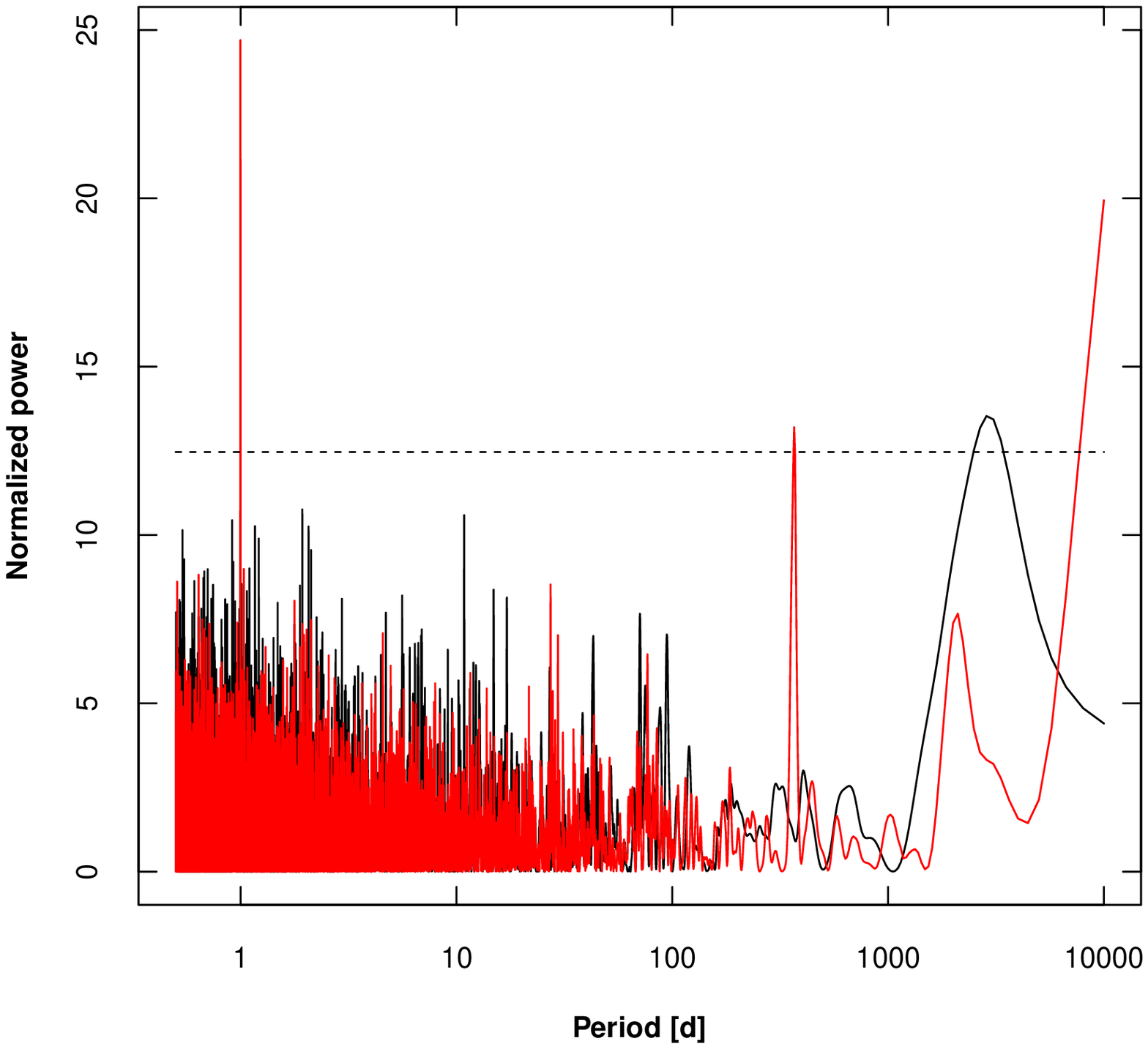}
\includegraphics[width=0.5\textwidth]{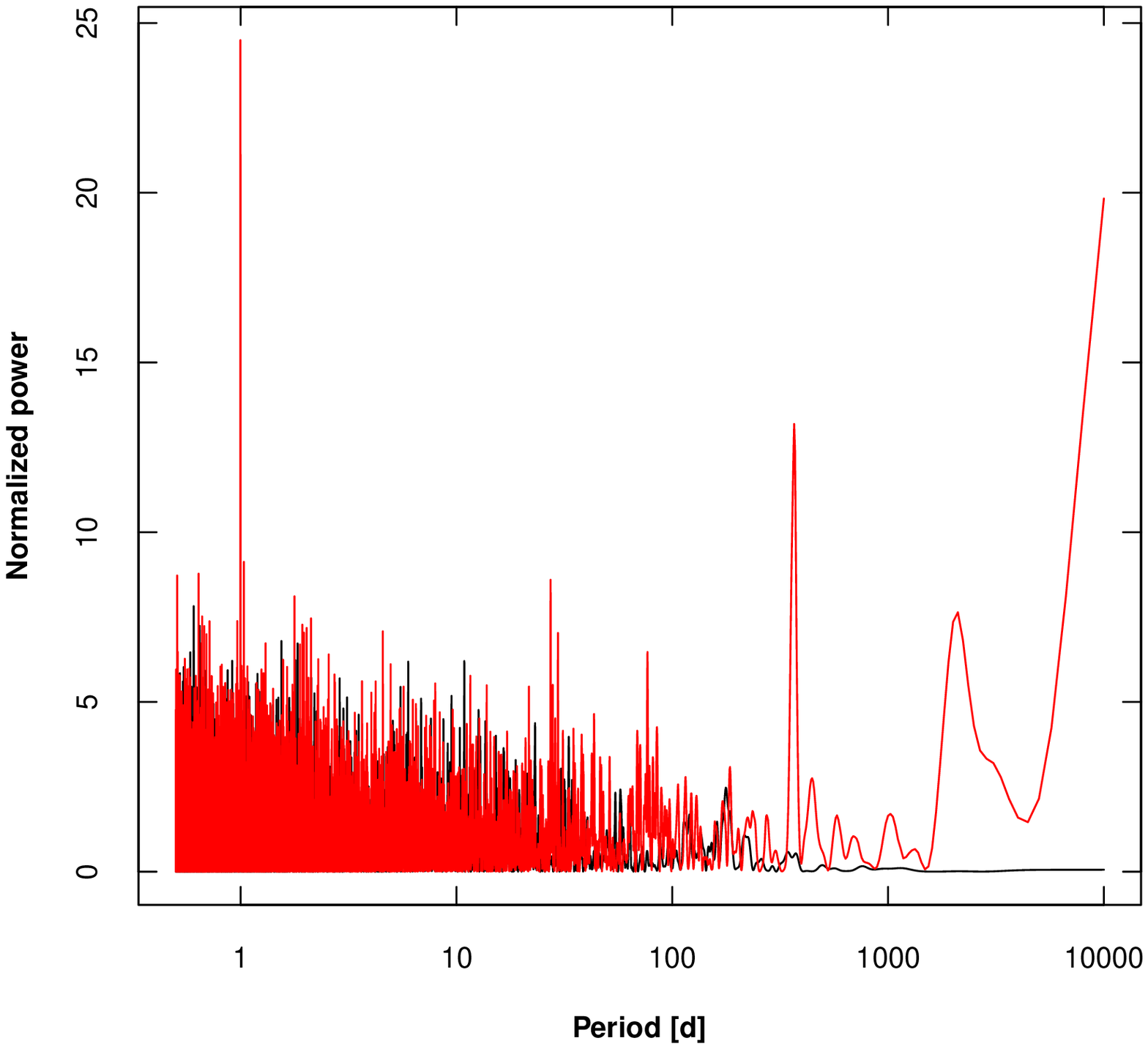}
\caption{Periodogram of residuals after fitting planet b (left) and
  planet c (right) around HD 133131A. Overplotted in red is the sampling periodogram.}\label{fig:periodogram_resid_A}
\end{figure}

\begin{figure}
\includegraphics[width=0.5\textwidth]{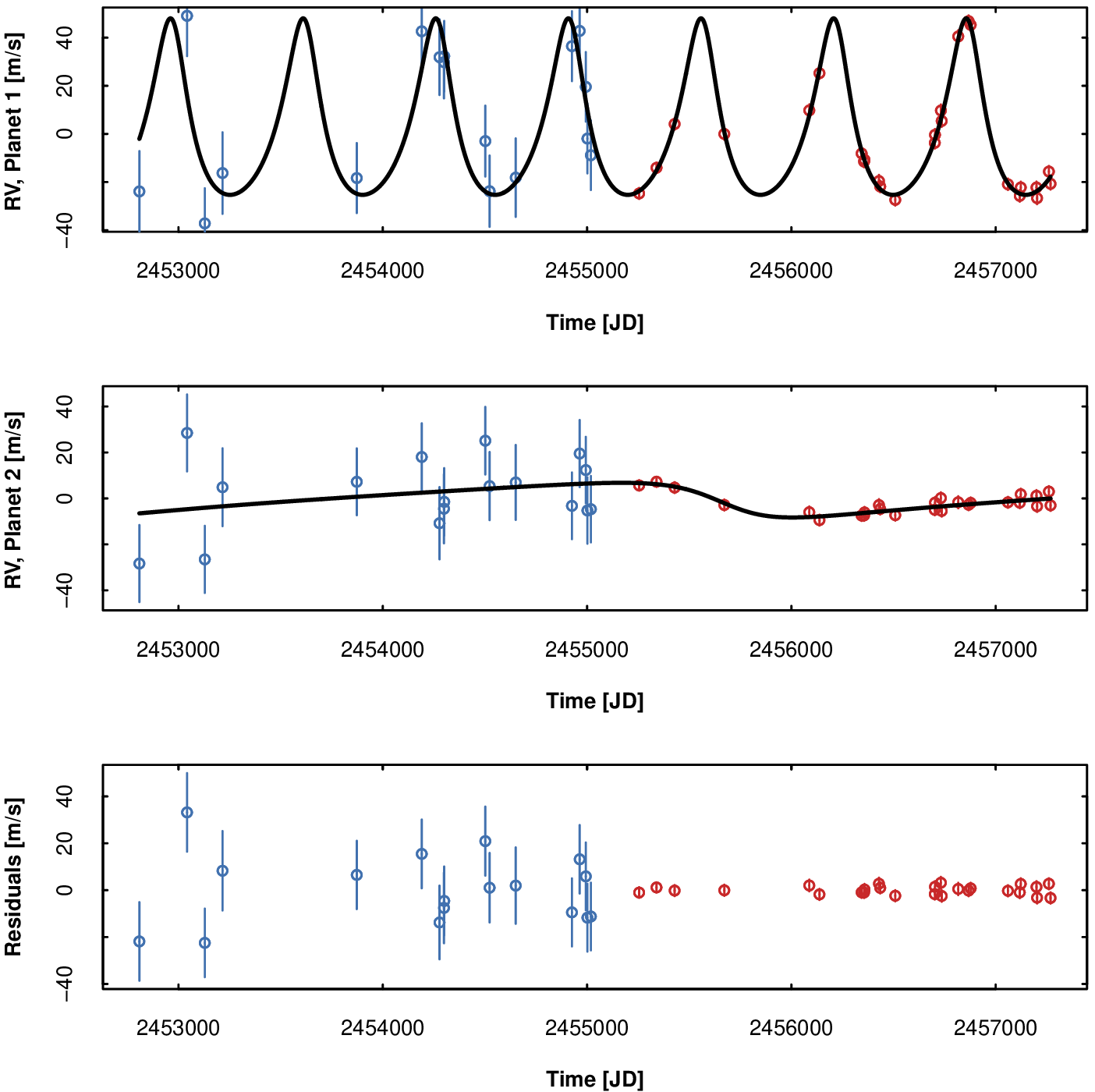}
\includegraphics[width=0.5\textwidth]{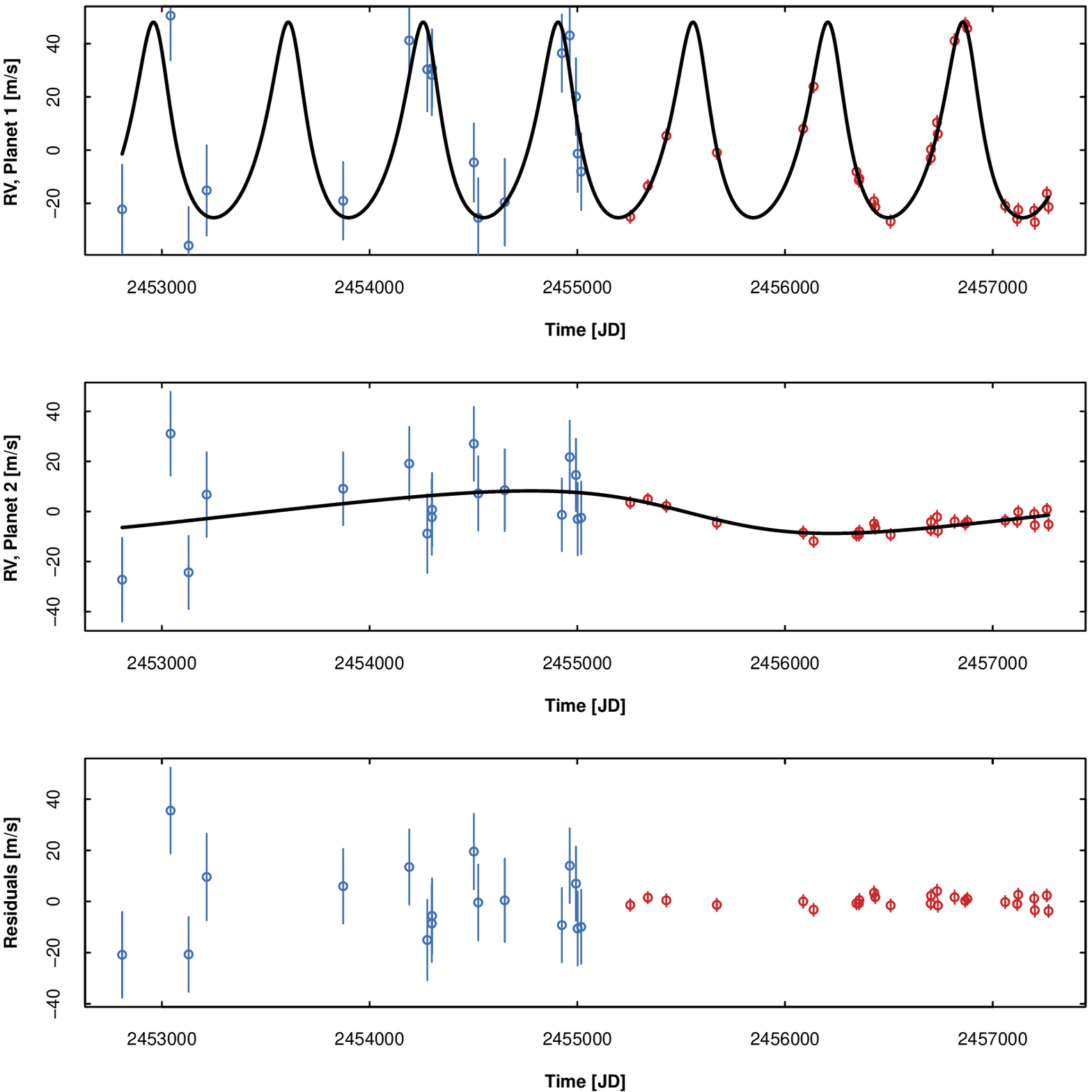}
\caption{Best-fit solutions to the RV data presented in this paper for
  HD 133131A. In blue are Magellan/MIKE data, and in red are
  Magellan/PFS data. The left panel shows our high eccentricity
  fit for planet c (RMS 9.38 m~s$^{-1}$), and the right panel shows our low eccentricity
  fit for planet c (RMS 9.39 m~s$^{-1}$.} \label{fig:A_rvfits_2p}
\end{figure}

Considering the low amplitude and incomplete phase coverage of this
second planet around HD 133131A, we performed some additional tests as
suggested by the referee to rule out stellar activity as the cause of
the RV variation. First, we checked that excluding the two outliers in
the $S$-index values measured from PFS (Figure \ref{fig:svar}) did not significantly alter the
$S$-index periodogram. The heights of all of the peaks are slightly
increased in the outlier-free periodogram (below, Figure
\ref{fig:Sval_periodogram_cut}), but none of the peaks are significant
(no FAPs below 1). The peaks around 1000 days in both Figure \ref{fig:periodogram_ABsval} and
Figure \ref{fig:Sval_periodogram_cut} are at shorter periods ($\sim 1280$ days in Figure
\ref{fig:periodogram_ABsval} and $\sim 940$ days and $\sim 2350$ days
in Figure \ref{fig:Sval_periodogram_cut}) than the $\sim 3500$ day peak in the residuals of HD 133131A after one
planet is removed (Figure \ref{fig:periodogram_resid_A}).

Second, we checked the correlation between the $S$-index and
  the residuals in the RV data after fitting for a single planet
  (shown in Figure \ref{fig:periodogram_resid_A}). We find no
  significant correlations -- Spearman's correlation coefficient is
  -0.097, and the Pearson correlation coefficient is -0.15; this
  remains true after the removal of the same two $S$-index outliers as
above (resulting in correlation coefficients of -0.045 and 0.11,
respectively). These tests lend confidence to the planetary nature of
the residual signal in the RV measurements of HD 133131A after fitting
for one planet (b).

\begin{figure}
\includegraphics[width=0.5\textwidth]{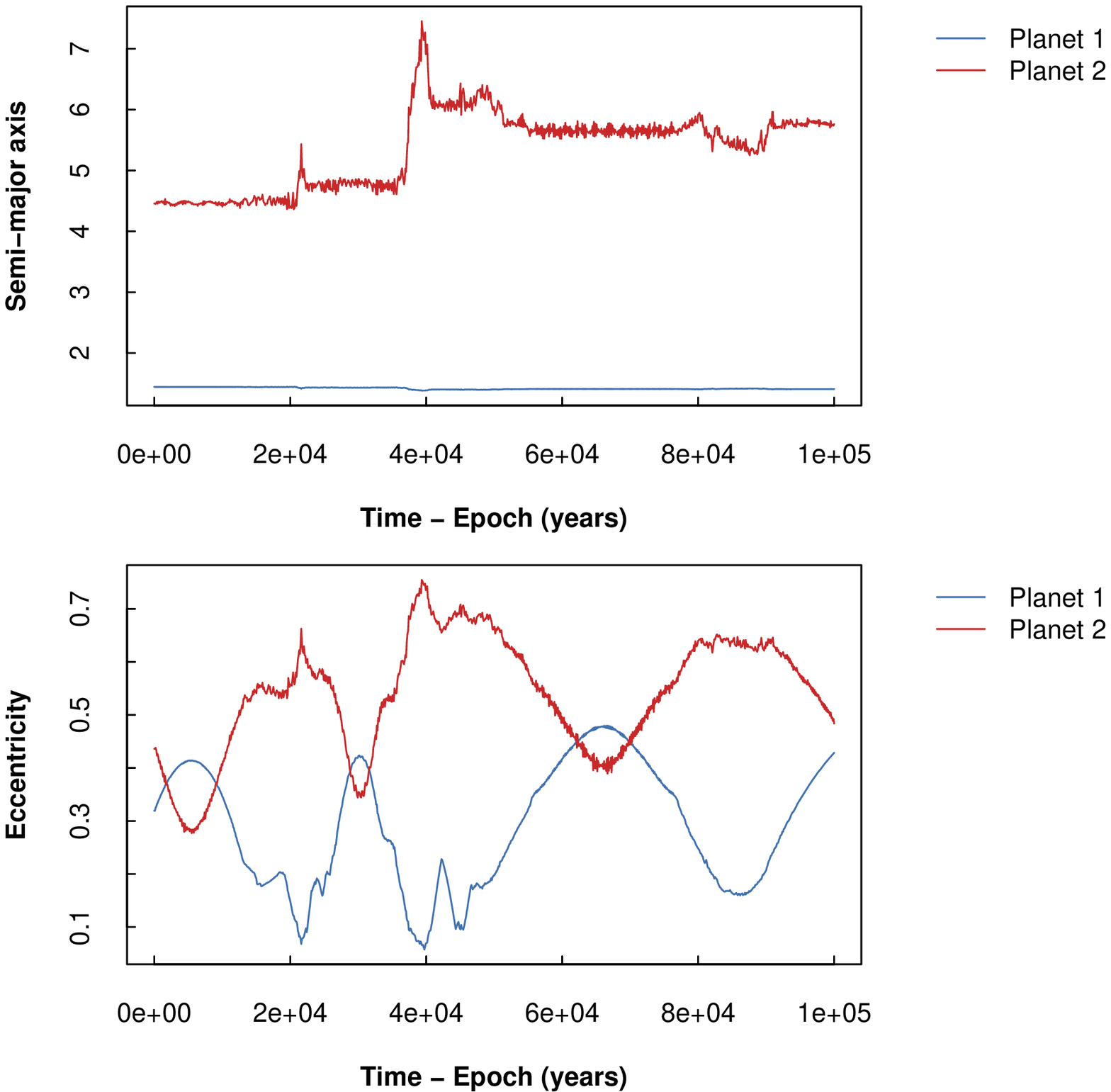}
\includegraphics[width=0.5\textwidth]{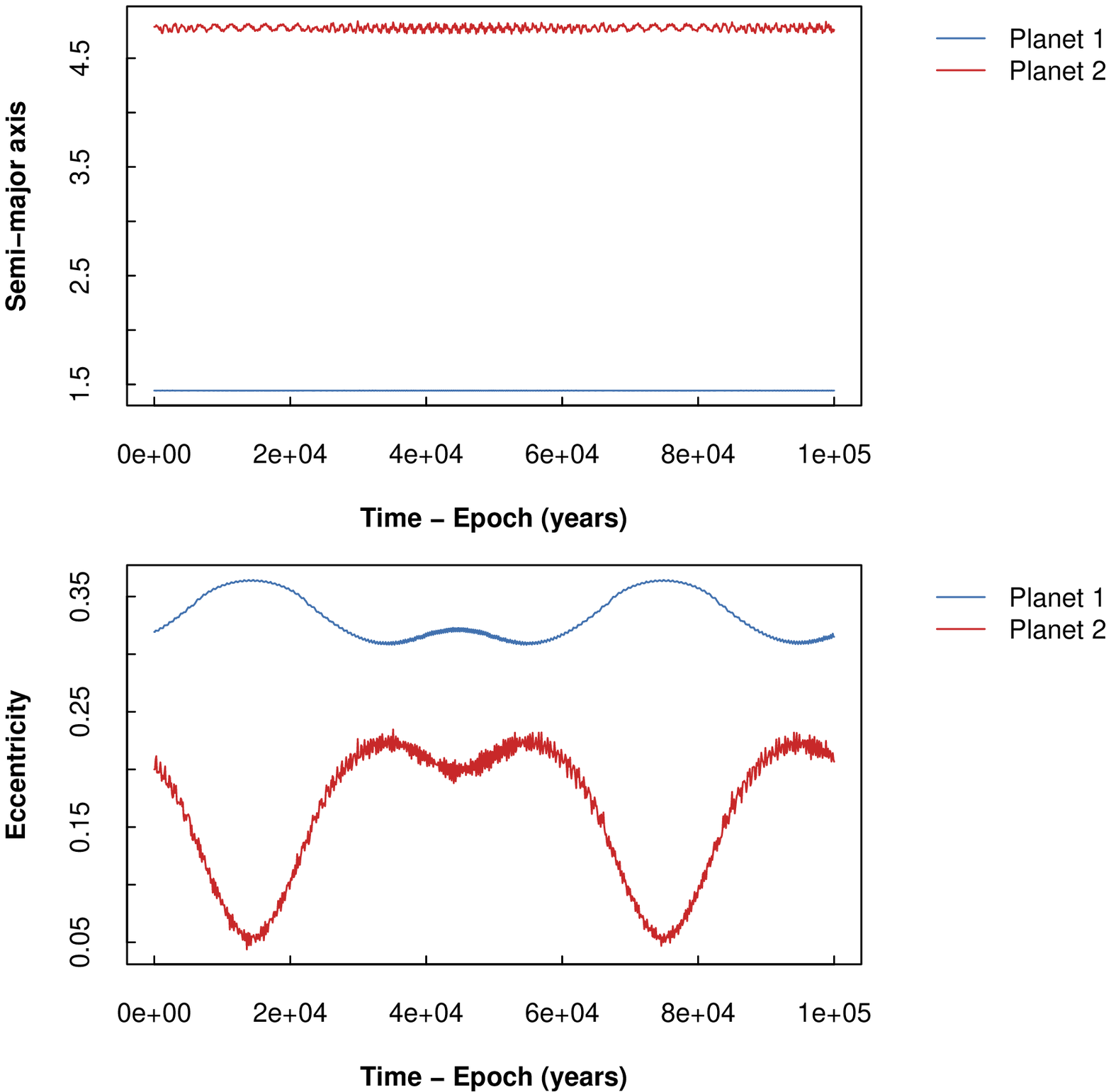}
\caption{Stability of the semi-major axis and eccentricities over
  100,000 years for the planets around HD 133131A using a higher
  eccentricity (left) and a lower eccentricity (right) for planet
  c. See Table \ref{tab:Afit} for details.}\label{fig:A_stability}
\end{figure}

As with HD 133131B, we integrated the orbits of the b and c
planets around HD 133131A over 100,000 years with the Bulirsch–Stoer
extrapolation algorithm in SYSTEMIC. In this case, the fits decribed
in columns 5-7 of Table \ref{tab:Afit} result in an
unstable configuration, defined as a $>$10\% change in semi-major axis in
planet c (see
Figure \ref{fig:A_stability}, top). However, if we manually reduce and fix
the eccentricity of HD 133131Ac at 0.20, instead of the median value
of $\sim$0.5, a similar orbit integration results in a more stable
configuration that does not include eccentricity overlap between the
two planets (Figure \ref{fig:A_stability}, bottom). We report the low
eccentricity fit parameters for planet c in the last three columns of Table
\ref{tab:Afit}; the parameters for planet b are unchanged. This fit is
shown in the bottom panel of Figure \ref{fig:A_rvfits_2p}. Though
there is almost total agreement between the parameters for planet HD
133131Ac in the high and low eccentricity fits when errors are
included, and there is no change within the errors to the
  parameters of planet HD133131Ab, we favor the low eccentricity fit
for planet c based on its greater stability.

\begin{deluxetable}{cccccccccc}
\rotate
\tablecaption{Best-fit solution for planets b and c around HD 133131A \label{tab:Afit}}
\tablecolumns{10}
\tablewidth{0pc}
\tablehead{ \colhead{Parameter} & \multicolumn{3}{c}{b} &\multicolumn{3}{c}{c, high $e$} &\multicolumn{3}{c}{c, low $e$}\\
\cmidrule(lr){2-4} \cmidrule(lr){5-7} \cmidrule(lr){8-10}
\colhead{} & \colhead{Best-fit} & \colhead{Median} & \colhead{MAD} & \colhead{Best-fit} & \colhead{Median} & \colhead{MAD} & \colhead{Best-fit} & \colhead{Median} & \colhead{MAD} }
\startdata
Period [days] & 649 & 649 &  3 & 3407 & 3686  &970 & 3925 & 3568 & 1084\\ 
  $\mathcal{M}~\sin{i}$ [$\mathcal{M}_{J}$] & 1.43 & 1.42 & 0.04 &0.48 & 0.44  &0.14 & 0.63 & 0.42 & 0.15 \\ 
  Mean anomaly [deg] & 268 & 267 & 11 & 56 & 79  & 82 & 106 & 101 & 87\\ 
  Eccentricity  & 0.32 & 0.33 & 0.03 & 0.47 & 0.50  & 0.22 & 0.20[fixed] & 0.49 & 0.22 \\ 
  $\varpi$ [deg] & 14 & 16 & 4.7 & 104 & 104  &31 & 98 & 100 & 37\\ 
  $K$ [m\,s$^{-1}$] & 36.68 & 36.52  & 0.93 & 7.57 & 7.15 & 2.10 & 8.56& 6.89 & 2.20\\ 
  Semi-major axis [AU] & 1.44 & 1.44  & 0.005 & 4.36 & 4.59 & 0.82&4.79 & 4.49 & 0.92\\ 
  Periastron passage time [JD] & 2452326 & 2452327 & 23 & 2452277 &2451701 & 1171 & 24523231 & 2452327 & 21\\ 
 \hline
  Stellar mass [$\mathcal{M}_{\odot}$] & 0.95 &  &  & & & & & &  \\ 
 Reduced $\chi^2$ & 0.980 &  &  & & & &0.965 & & \\
  RMS [m\,s$^{-1}$] & 9.38 &  &  & & & &9.39 & & \\ 
  Stellar jitter [m\,s$^{-1}$] & 13.7 (MIKE) & &  & & & & & &  \\ 
                                            & 1.92 (PFS) & &  & & & & & & \\ 
  Instrument RMS [m\,s$^{-1}$] & 6.60 (MIKE)& &  & & &
  & & &  \\ 
Instrument RMS [m\,s$^{-1}$] & 1.47 (PFS) & &  & & & & & &  \\
  Data points  & 43 &  &  & & & & & & \\ 
  Span of observations [JD] & 2452808.68 $-$ &  &  & & & & & &\\ 
                                          &2457268.49 &  &  & & & & & & \\ 

\enddata
\tablecomments{All elements are defined at epoch JD = 2452808.68. Uncertainties are reported in brackets.}
\end{deluxetable}

\begin{deluxetable}{cccc}
\tablecaption{Best-fit 1 planet solution for HD 133131B \label{tab:Bfit}}
\tablecolumns{4}
\tablewidth{0pc}
\tablehead{\colhead{Parameter} & \multicolumn{3}{c}{b} \\
\cmidrule(lr){2-4}
\colhead{} & \colhead{Best-fit} & \colhead{Median} & \colhead{MAD}}
\startdata
Period [days] & 6119 & 5769 & 831 \\ 
  $\mathcal{M}~sin {i}$ [$\mathcal{M}_{J}$] & 2.50 & 2.50 & 0.05 \\ 
  Mean anomaly [deg] & 302 & 299 & 6 \\ 
  Eccentricity  & 0.62 & 0.61& 0.04\\ 
  $\varpi$ [deg]  & 103 & 103 & 3\\ 
  $K$ [m\,s$^{-1}$] & 37.29 & 37.41 & 0.65\\ 
  Semi-major axis [AU] & 6.40 & 6.15 & 0.59 \\ 
  Periastron passage time [JD] & 2450298 & 2450644& 828\\ 
\hline
  Stellar mass [$\mathcal{M}_{\odot}$] & 0.93 & &  \\ 
 Reduced $\chi^2$ & 0.998 & & \\
  RMS [m\,s$^{-1}$] & 1.59 & &  \\ 
 Stellar jitter [m~s$^{-1}$] & 1.03 & &  \\ 
 Instirument RMS [m~s$^{-1}$] & 1.49 & &  \\ 
  Data points  & 25 \\ 
  Span of observations [JD] & 2455428.51$-$ 2457268.50 & & \\ 
\enddata
\tablecomments{All elements are defined at epoch JD = 2455428.51. Uncertainties are reported in brackets.}
\end{deluxetable}

\subsection{Error Analysis}
To characterize the marginal distribution of the parameters of the
model above, we used the functionality of SYSTEMIC to implement Markov-Chain Monte Carlo
algorithm (MCMC; e.g., Ford 2005, 2006; Gregory 2011) fitting paired with flat
priors on log $P$, log $\mathcal{M}$, and the other parameters. In
previous works using SYSTEMIC (e.g., Rowan et al. 2015; Vogt et
al. 2015), the noise parameter $s_j$ was fit with a modified Jeffrey
function as a prior (see Vogt et al. 2015, \S4); the MCMC routine then
also returns a best-fit $s_j$ ``jitter'' term and its marginal
distribution. 

Instead of relying solely on the MCMC resulting best-fit $s_j$ values,
here we outline an analytic approach that can be applied to any RV
data set to estimate the stellar jitter term without including them
as free parameters. In the
case of HD 133131A, the Magellan/MIKE and Magellan/PFS data have
significantly different errors (with MIKE errors being on average
$\sim$4$\times$ as large as PFS errors), meaning that the $s_j$
stellar jitter values for the two instruments will, in reality, be
different. We estimated
the $s_j$ jitter term separately for each data set with the
following procedure, and added it in quadrature to the formal error
of each RV data point to get new errors, which we used as the initial input into
SYSTEMIC. 

The purpose of the $s_j$ is to compensate for any noise, beyond the
formal errors, that separates the data from the final best fit, as
defined by a reduced $\chi^2=1$. The
contribution to the $\chi^2$ of each data point is
$\sigma^2_{fit}/\sigma^2_{inst}$, where $\sigma_{fit}$ is the residual
on the fit, and $\sigma_{inst}$ is the instrumental ($+$jitter)
error. To find the appropriate $s_j$, we wanted the sum of these
values, divided by the number of data points minus the number fitted
parameters, to equal one:

\begin{equation}
\bigg[\sum_{i}^{N_o} \frac{\sigma^2_{fit,\ i}}{(\sigma^2_{inst,\ i}+s_j^2)} \bigg] \times \frac{1}{N_o-N_{param}}=1
\end{equation}

We separately fit the 26 PFS and 17 MIKE
RV data points to find the residuals, for the $\sigma_{fit,\ i}$
values. To decide how $N_{param}=12$ should be divided between the
PFS and MIKE data points, we estimated the effective weight for each
data set as the square of the ratio of the
RMS values of each data set, (1.47 m~s$^{-1}$/6.60 m~s$^{-1}$)$^{2}
=$0.05, so that that number of free parameters allocated to the MIKE
data are 0.05$\times$12 $\sim$1, and PFS is allocated 11 free
parameters. The final $s_j$ value that solved the equation above for
the PFS data is 1.92 m~s$^{-1}$, and for the MIKE data is 13.70
m~s$^{-1}$. These values are reported in Table \ref{tab:Afit}. 

Even though the HD 133131B data are only from PFS, we can
apply the same procedure to keep the analysis consistent, this time
allocating all six parameters to the PFS data. We derive a $s_j$ value
of 1.03 m~s$^{-1}$ for HD 133131B, as reported in Table
\ref{tab:Bfit}. 

\begin{figure*}
\centering
\includegraphics[width=0.4\textwidth]{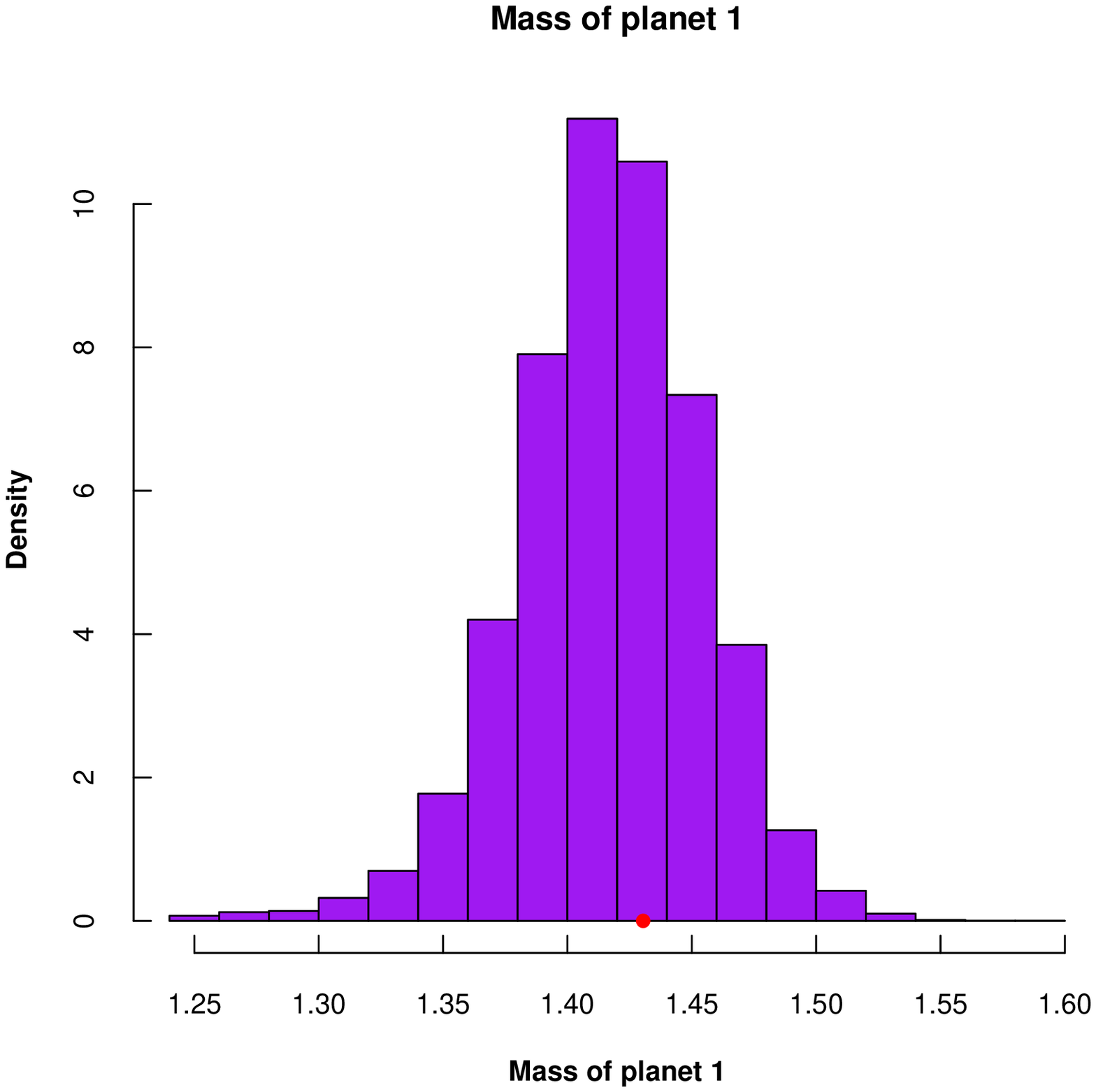}
\includegraphics[width=0.4\textwidth]{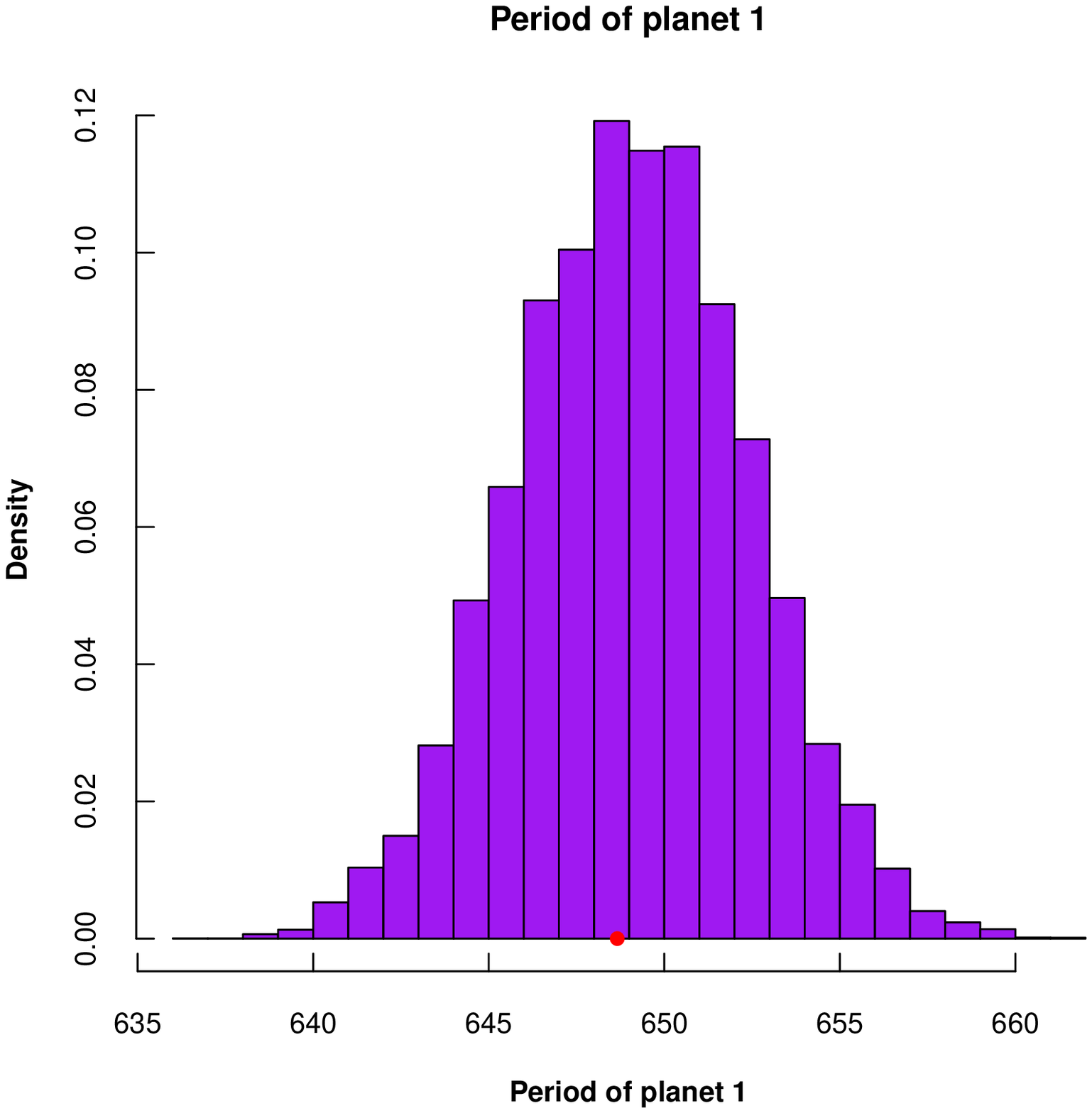}
\includegraphics[width=0.4\textwidth]{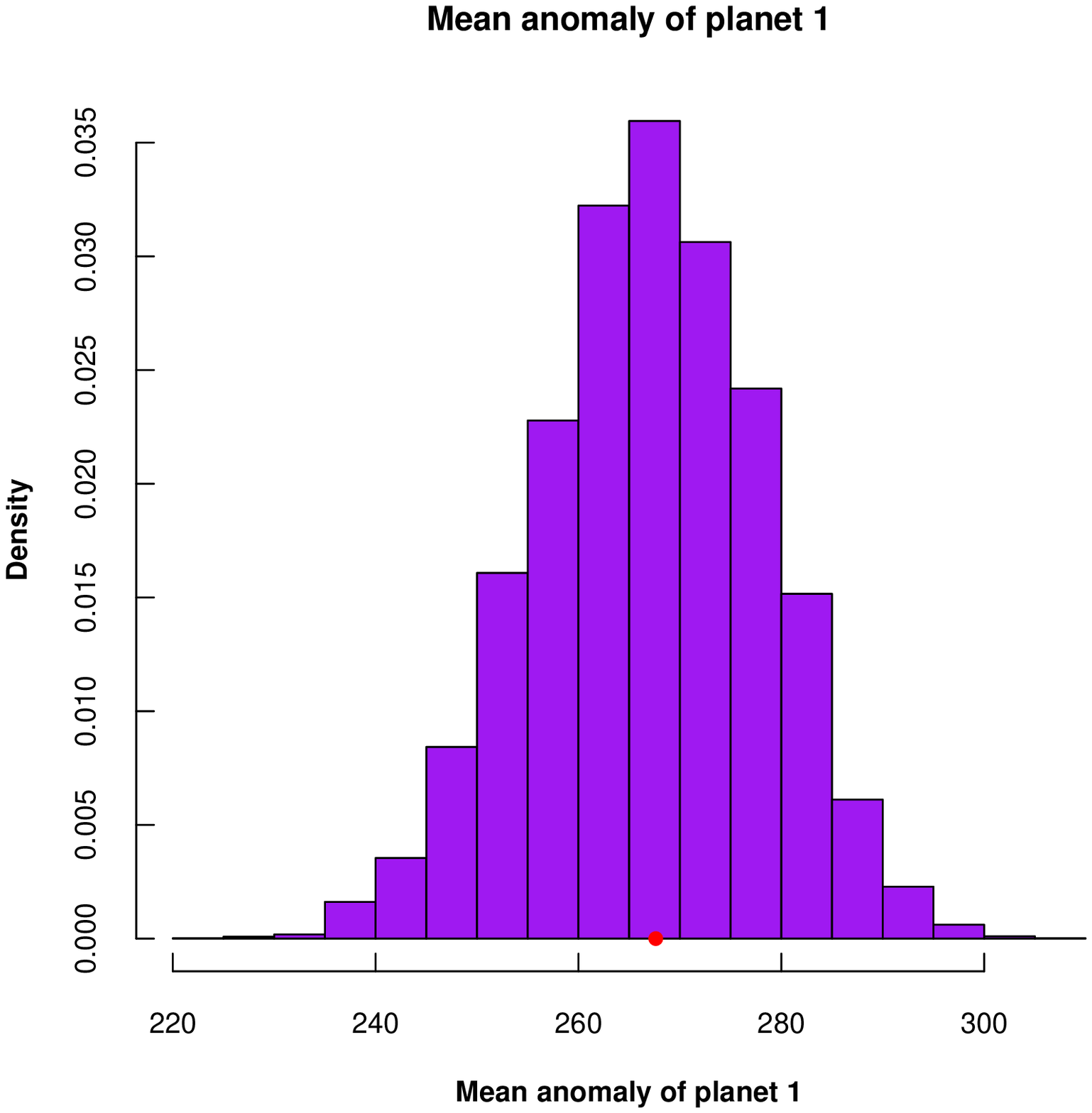}
\includegraphics[width=0.4\textwidth]{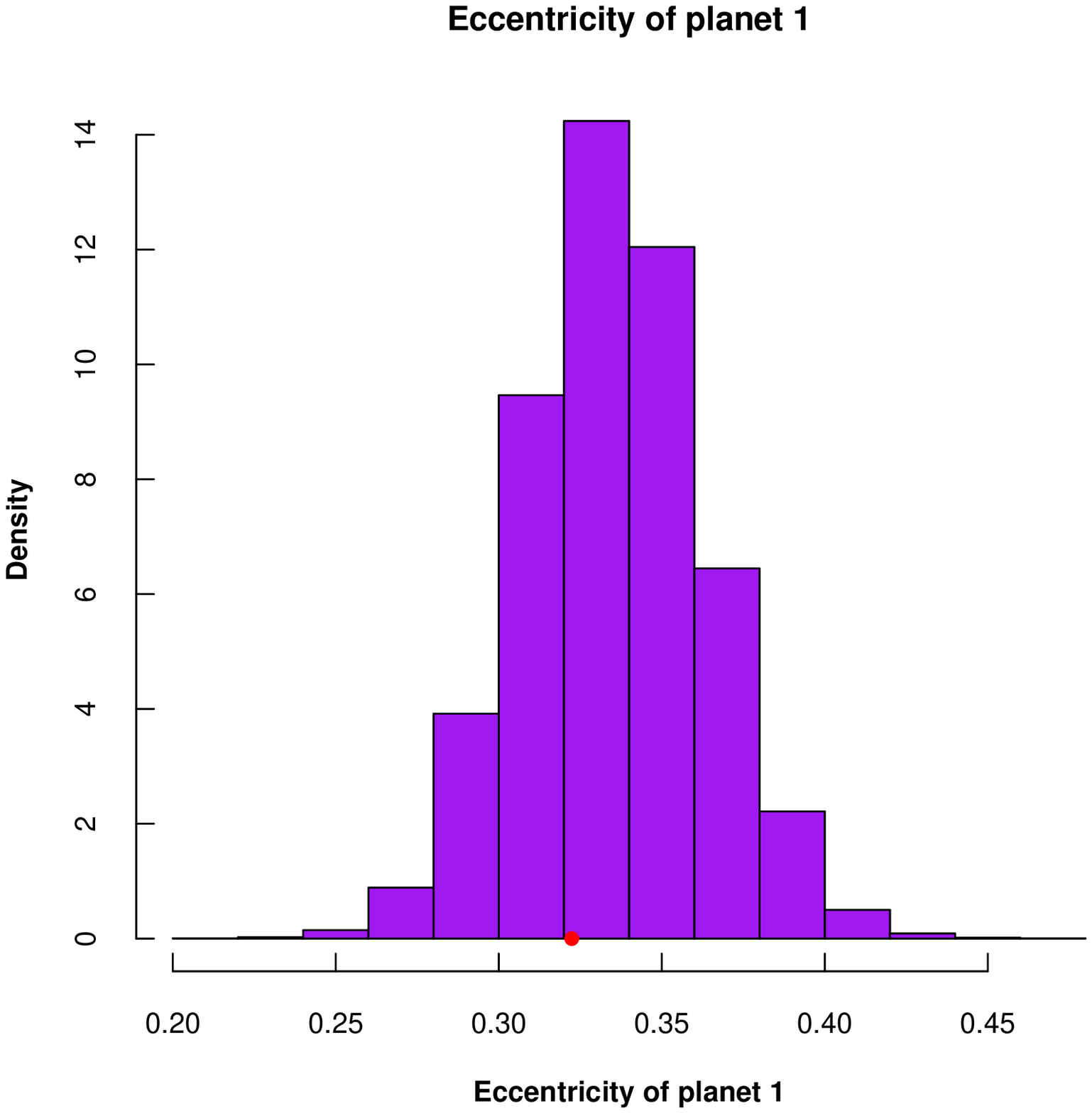}
\includegraphics[width=0.4\textwidth]{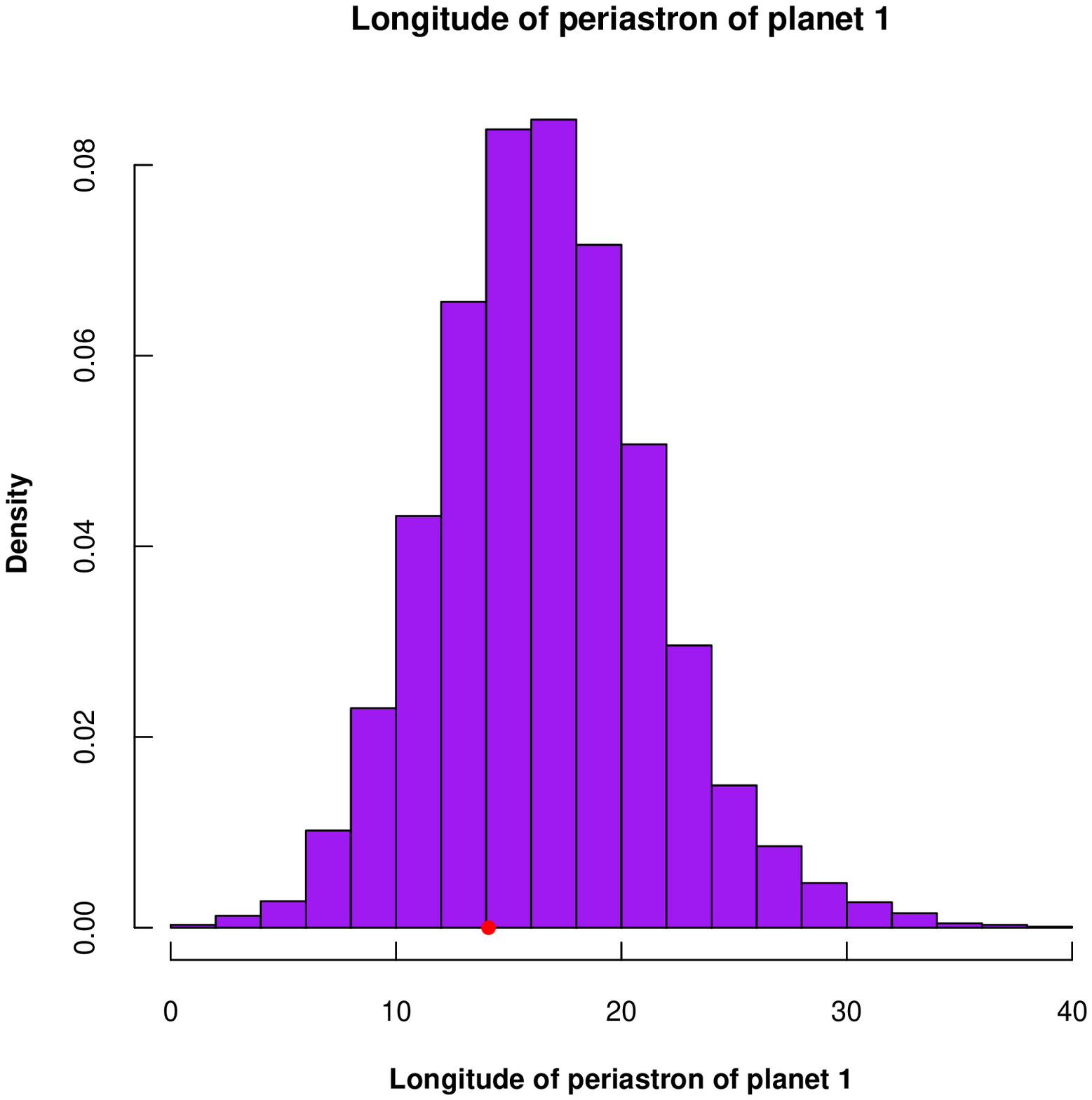}
\includegraphics[width=0.4\textwidth]{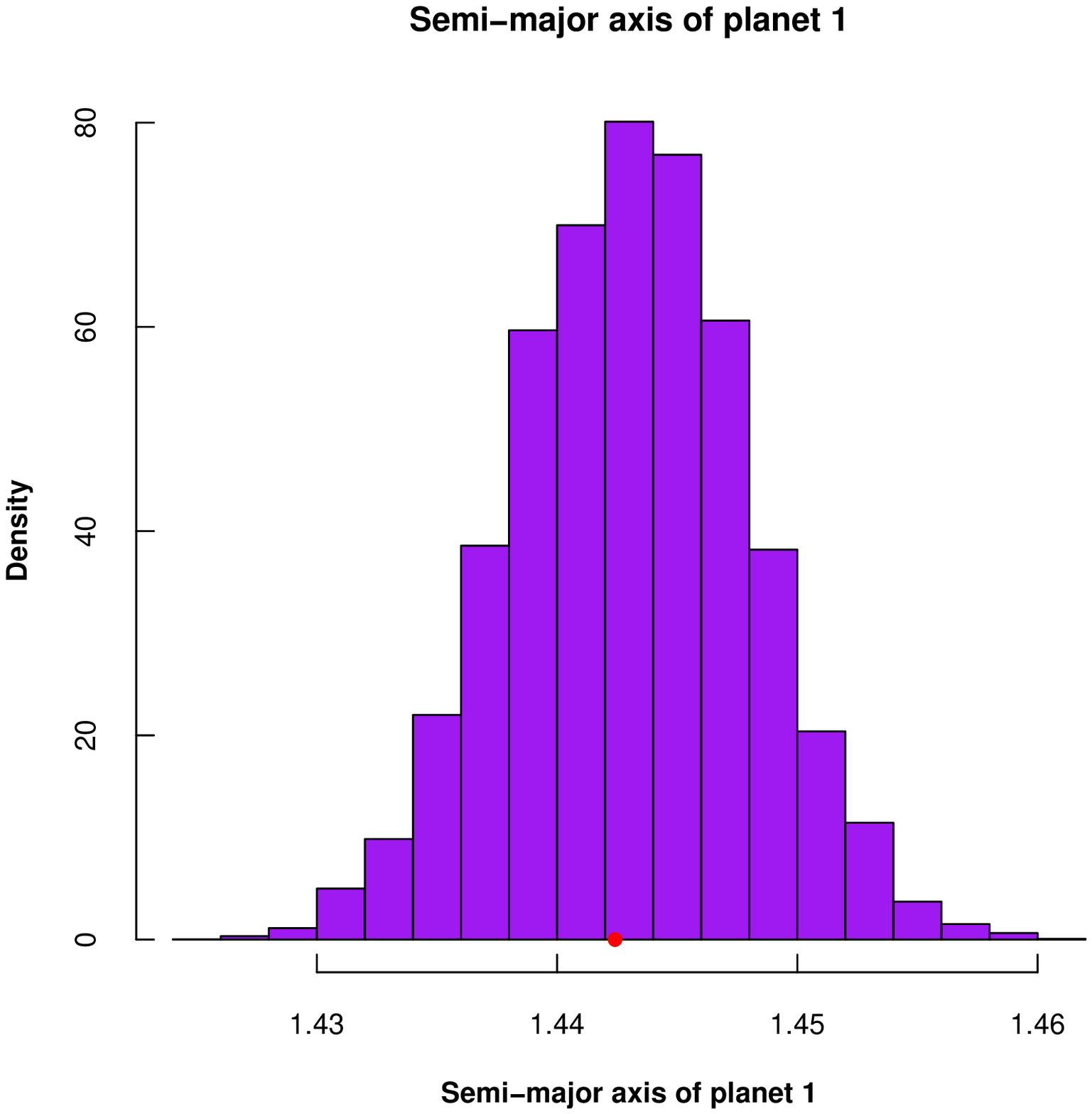}
\caption{Marginal distributions of the orbital elements for planet 1
  of the two planet fit resulting from
our MCMC analysis of HD 133131A RV data. The best-fit values from
Table \ref{tab:Afit} are marked with red dots.} \label{fig:errorsA1}
\end{figure*}

\begin{figure*}
\centering
\includegraphics[width=0.4\textwidth]{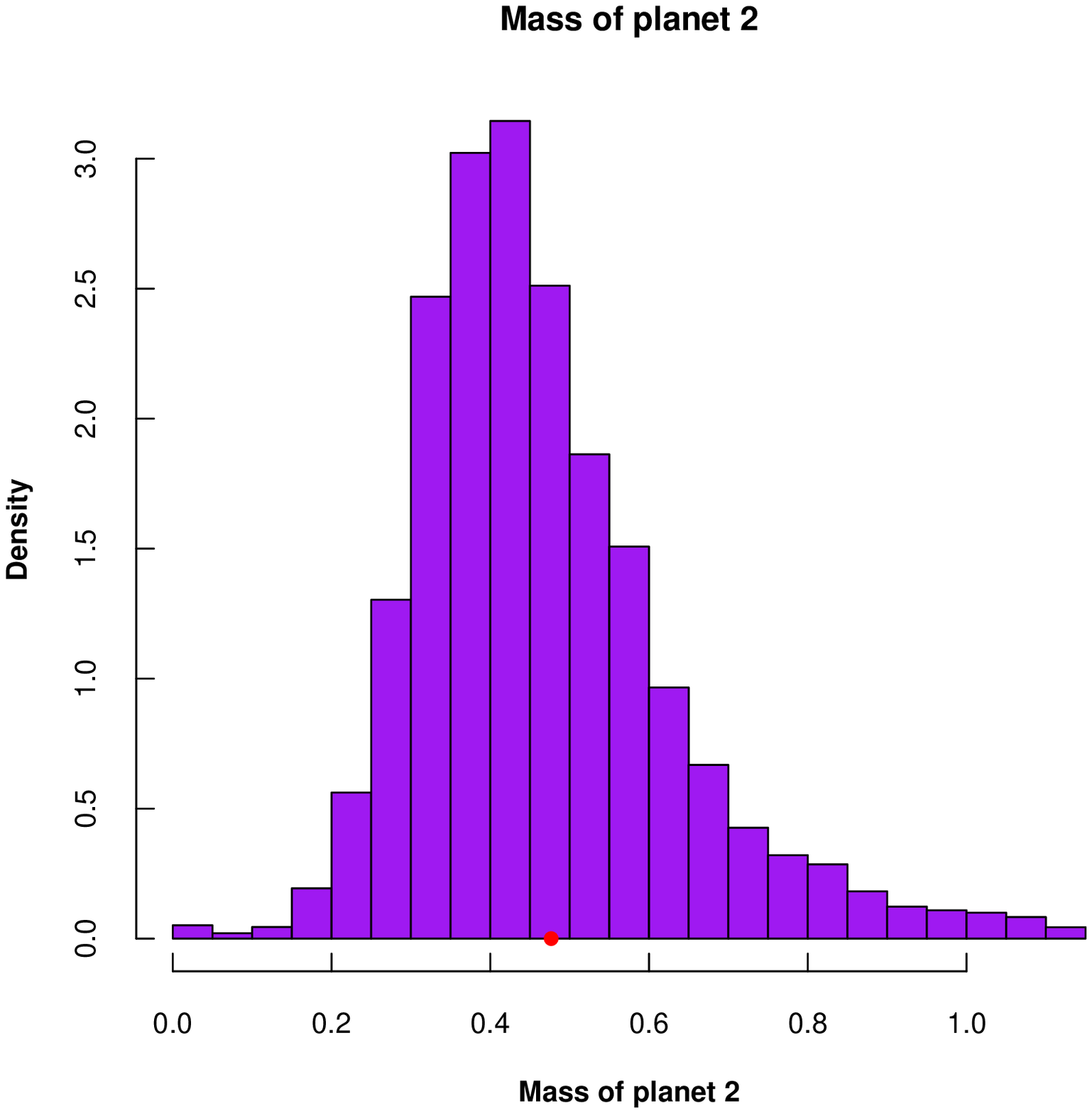}
\includegraphics[width=0.4\textwidth]{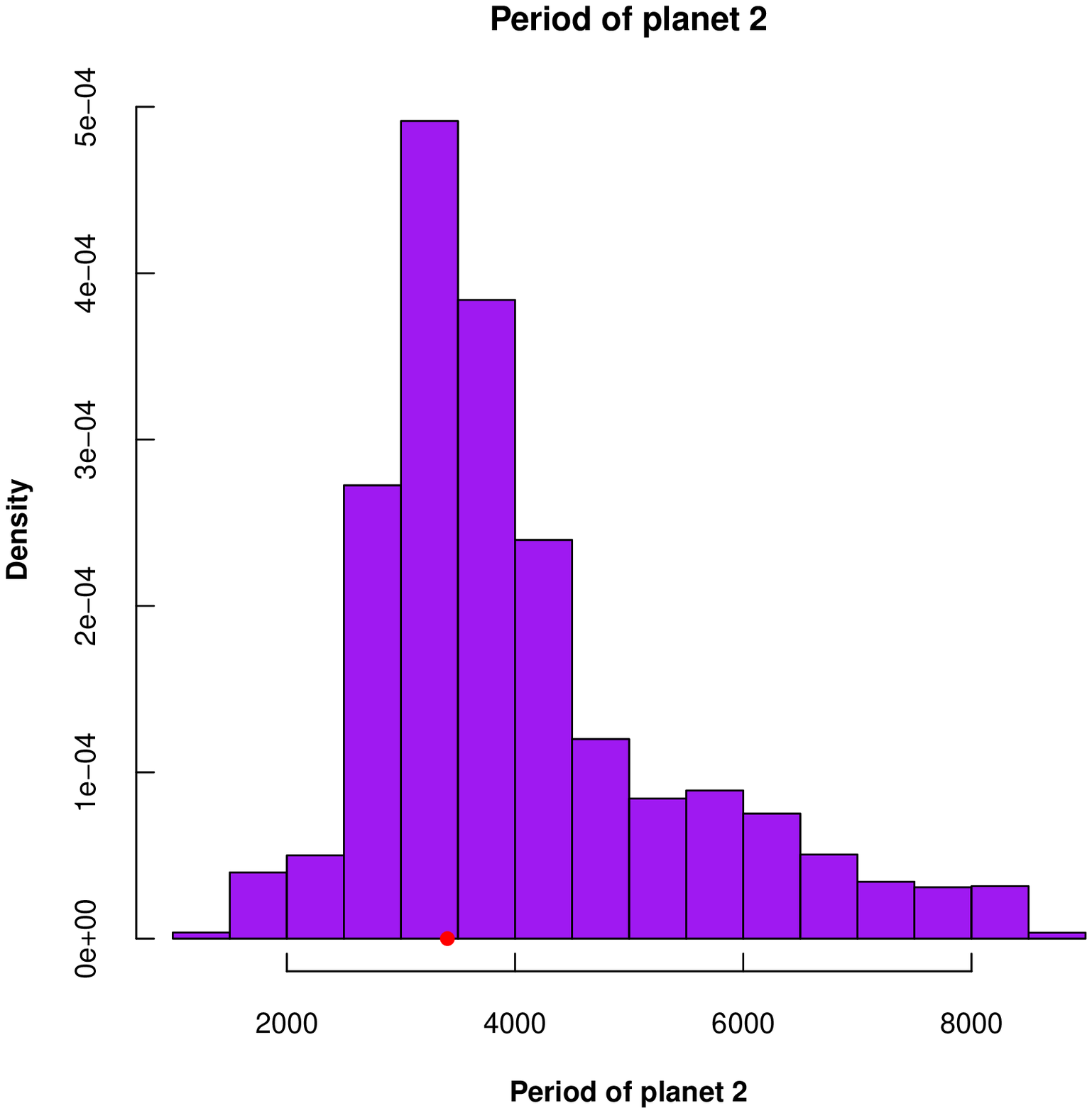}
\includegraphics[width=0.4\textwidth]{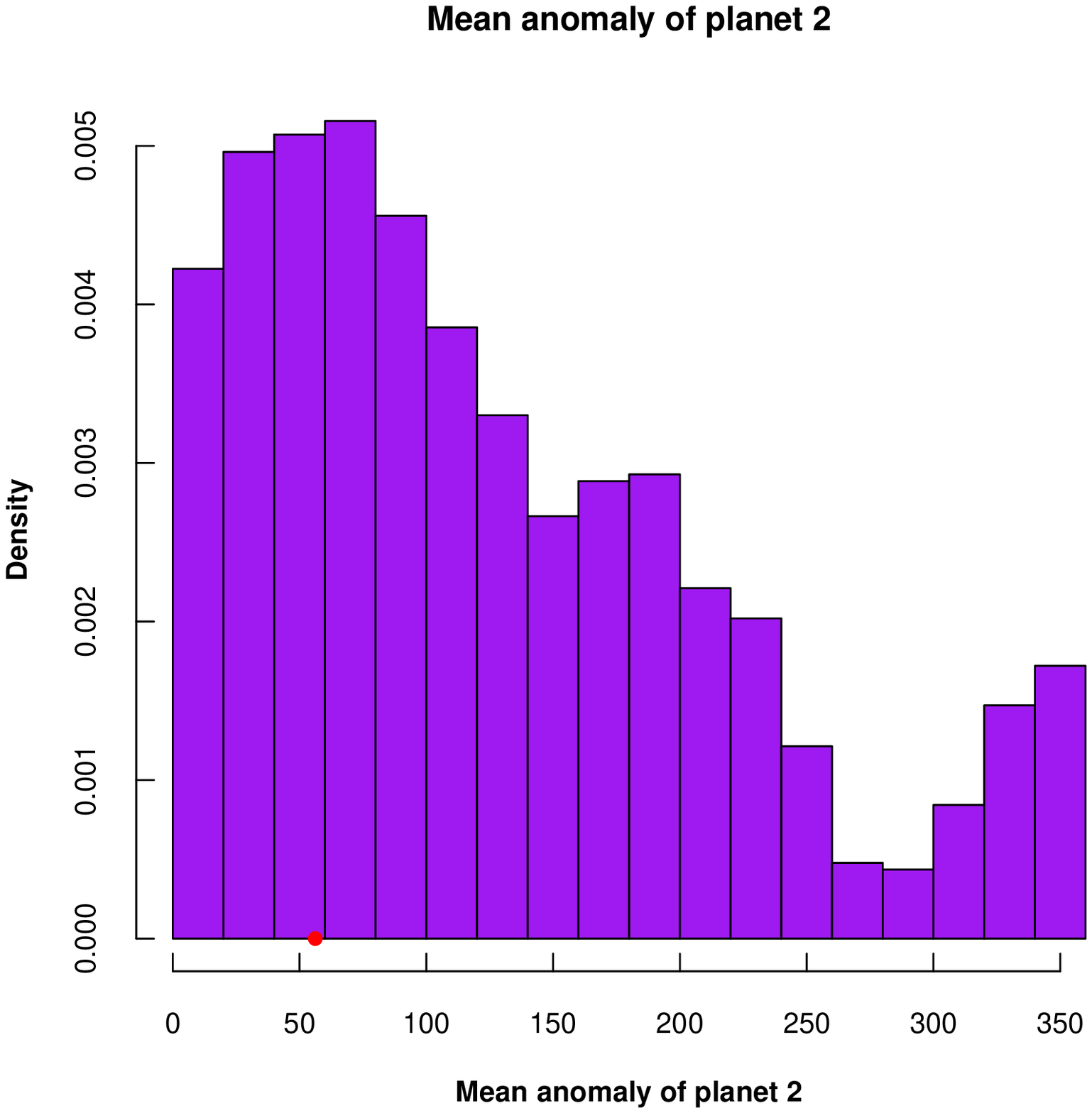}
\includegraphics[width=0.4\textwidth]{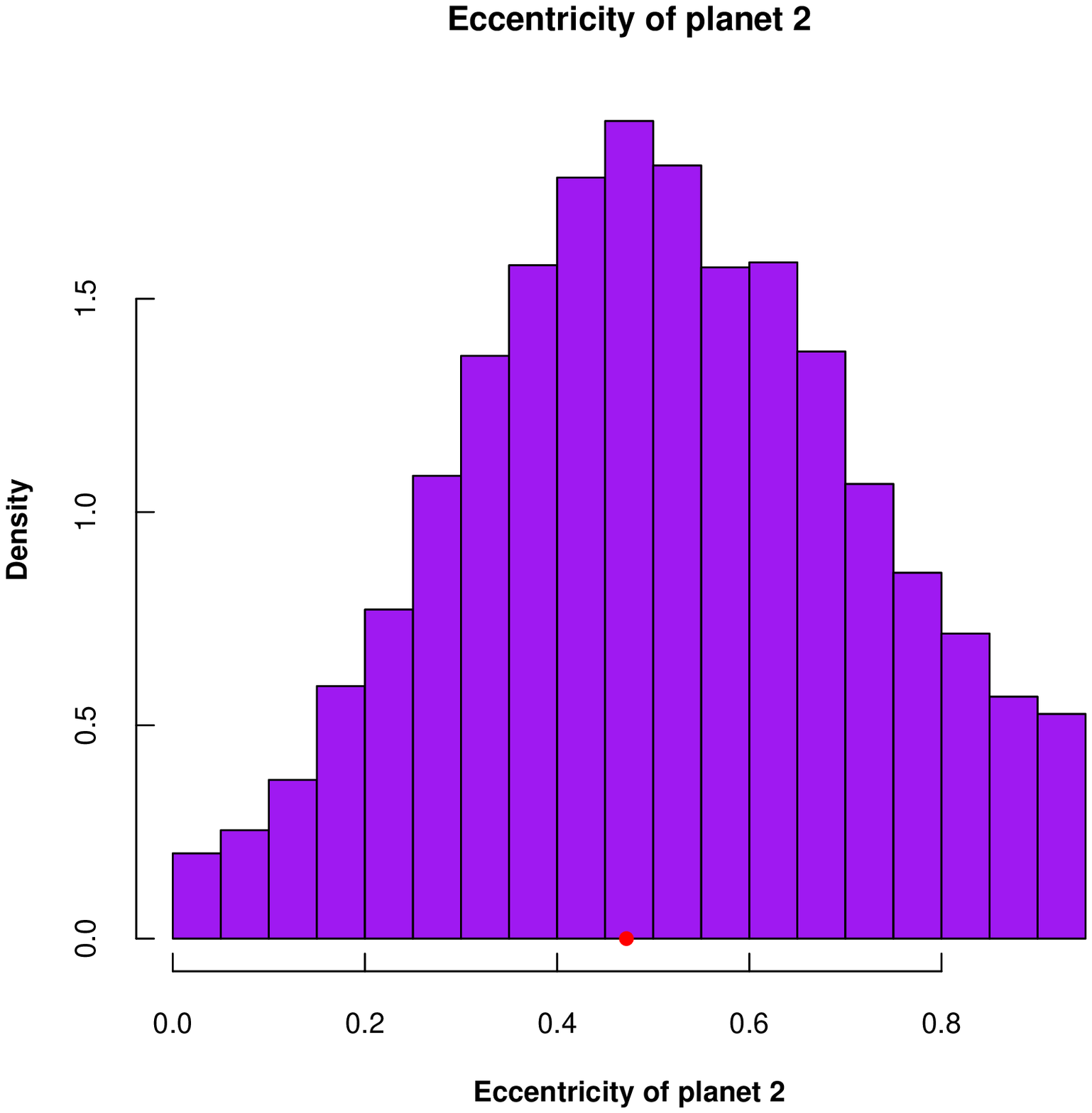}
\includegraphics[width=0.4\textwidth]{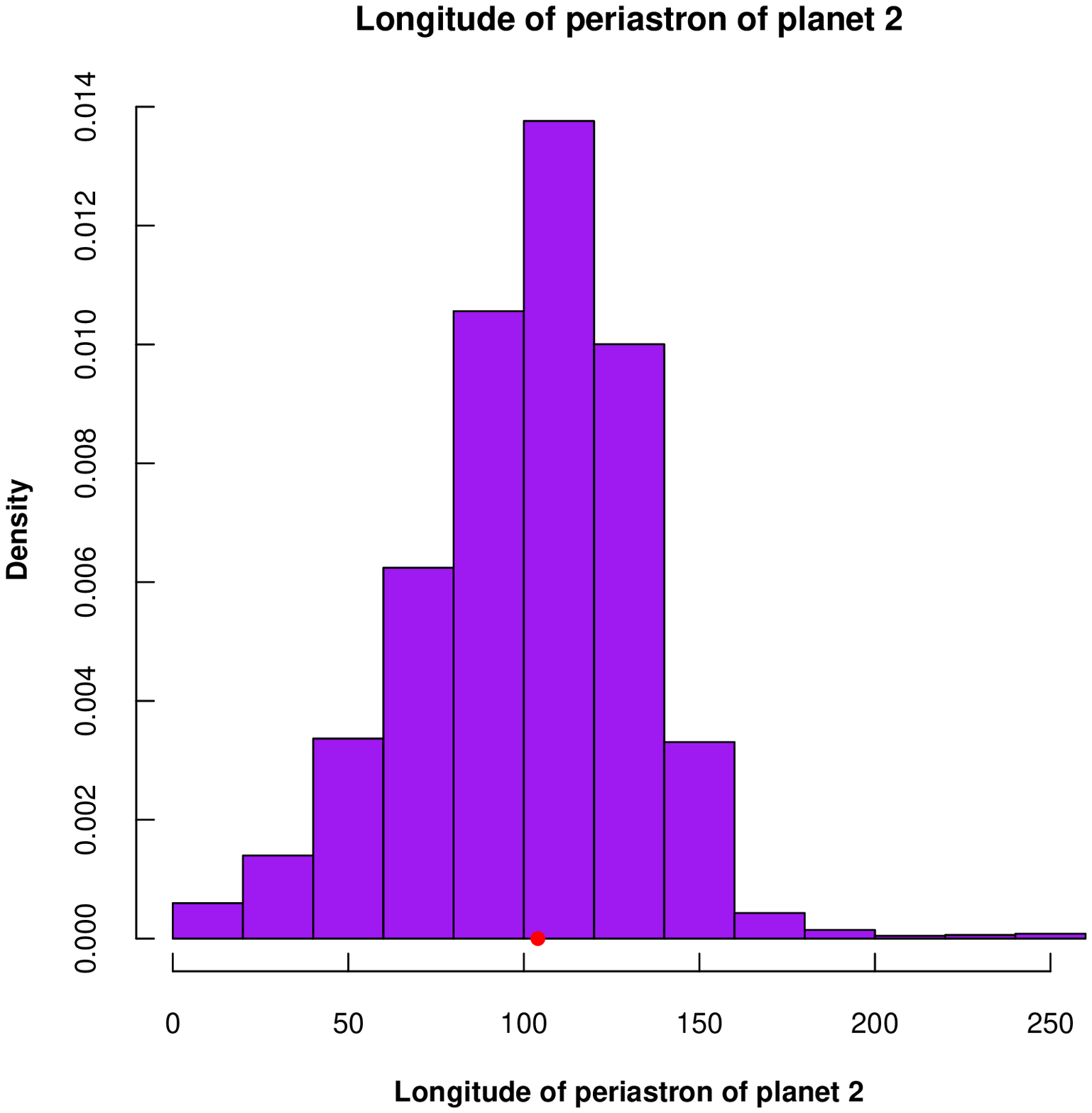}
\includegraphics[width=0.4\textwidth]{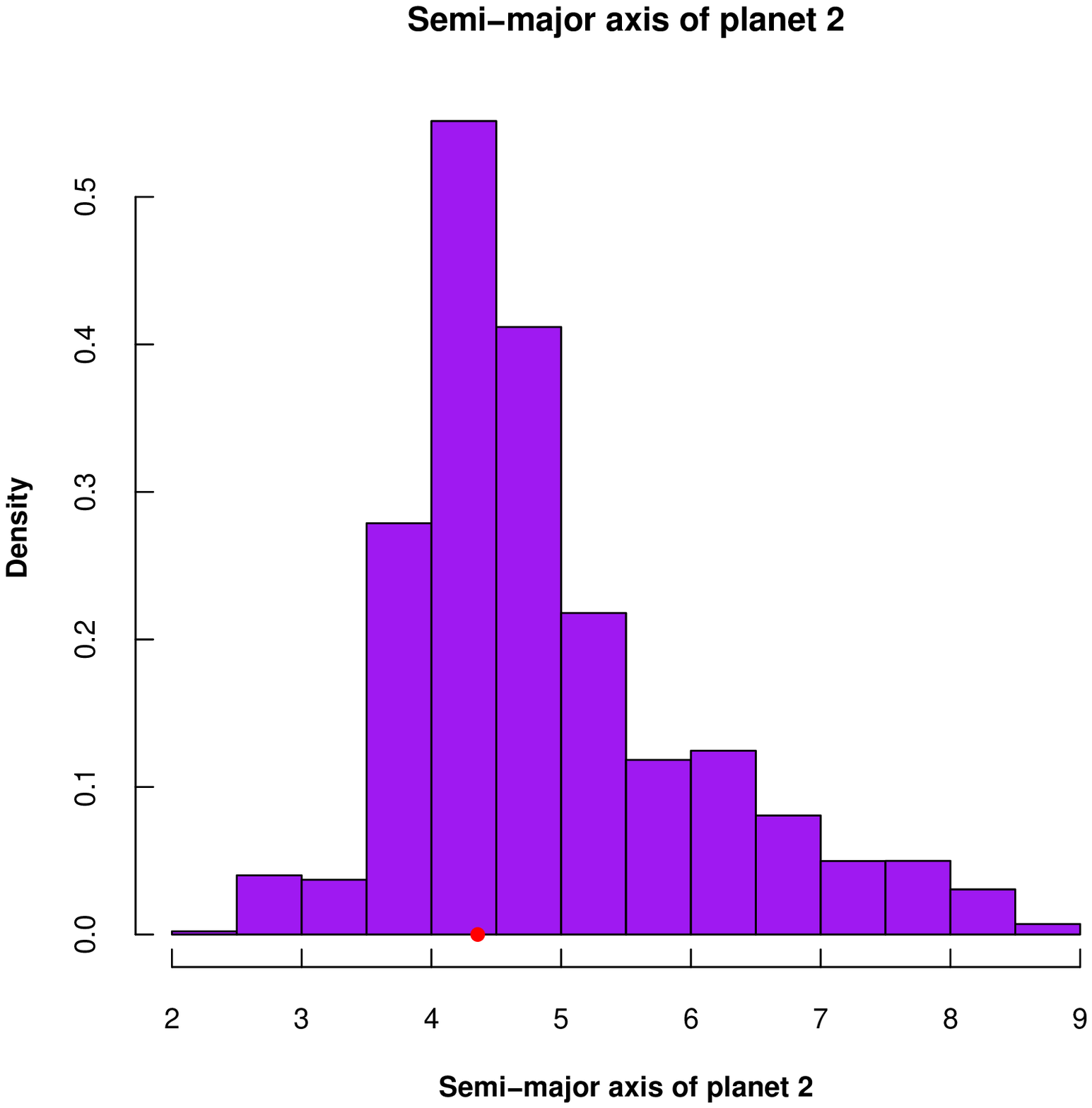}
\caption{Marginal distributions of the orbital elements for planet 2, of the two planet fit resulting from
our MCMC analysis of HD 133131A RV data; these plots show the high
eccentricity fit distributions. The best-fit values from
Table \ref{tab:Afit} are marked with red dots.}\label{fig:errorsA2_highecc}
\end{figure*}

\begin{figure*}
\centering
\includegraphics[width=0.4\textwidth]{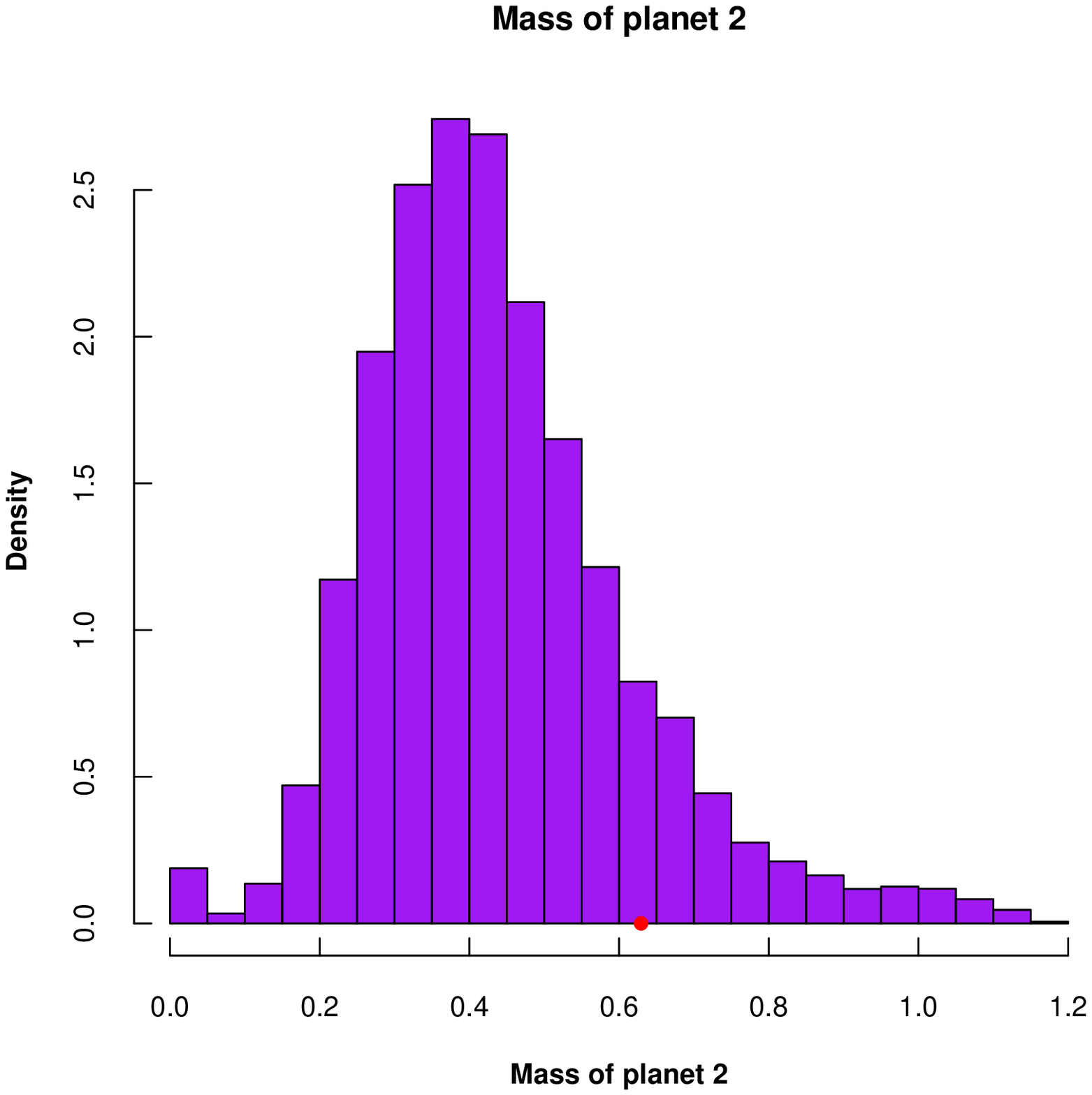}
\includegraphics[width=0.4\textwidth]{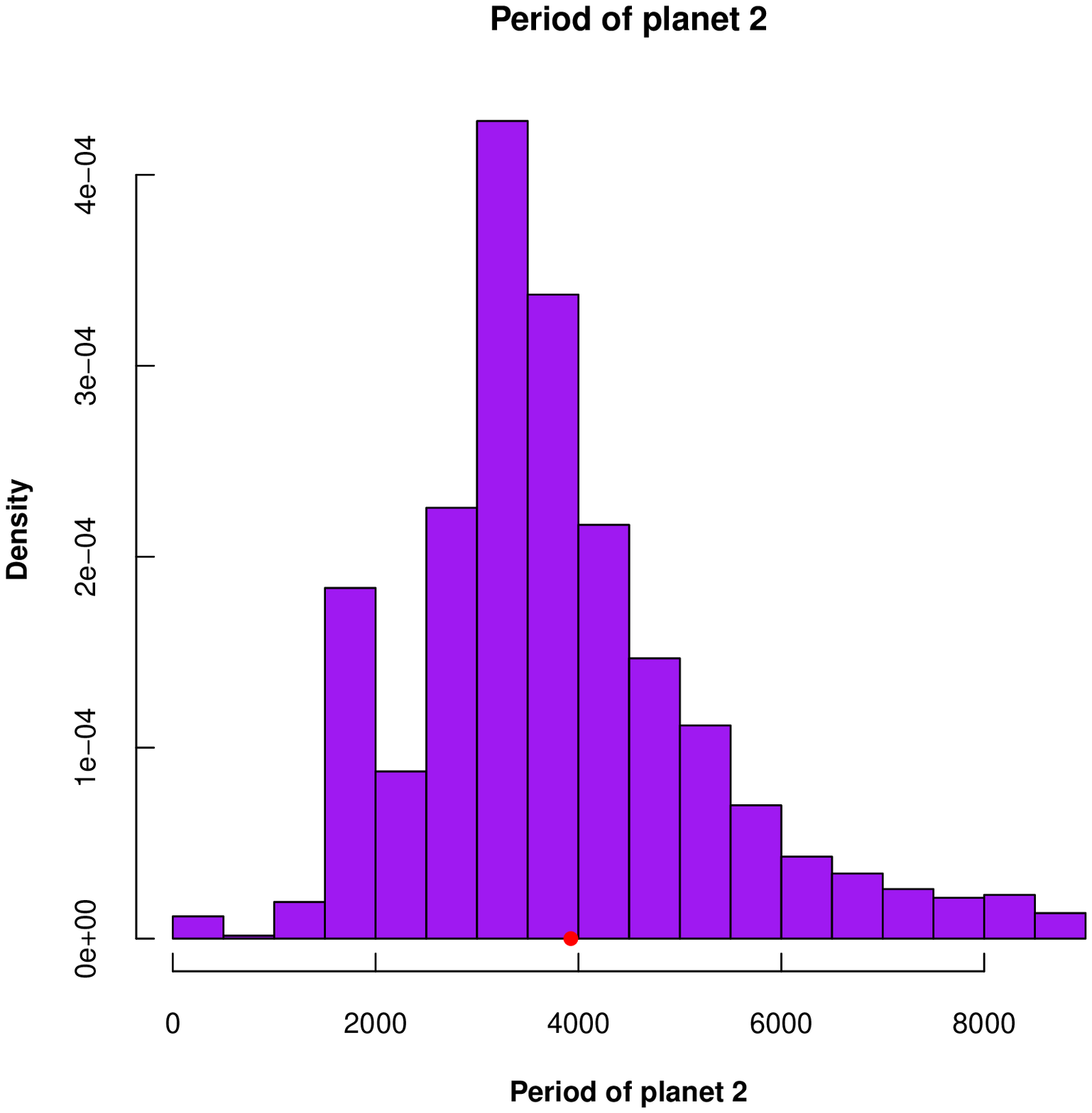}
\includegraphics[width=0.4\textwidth]{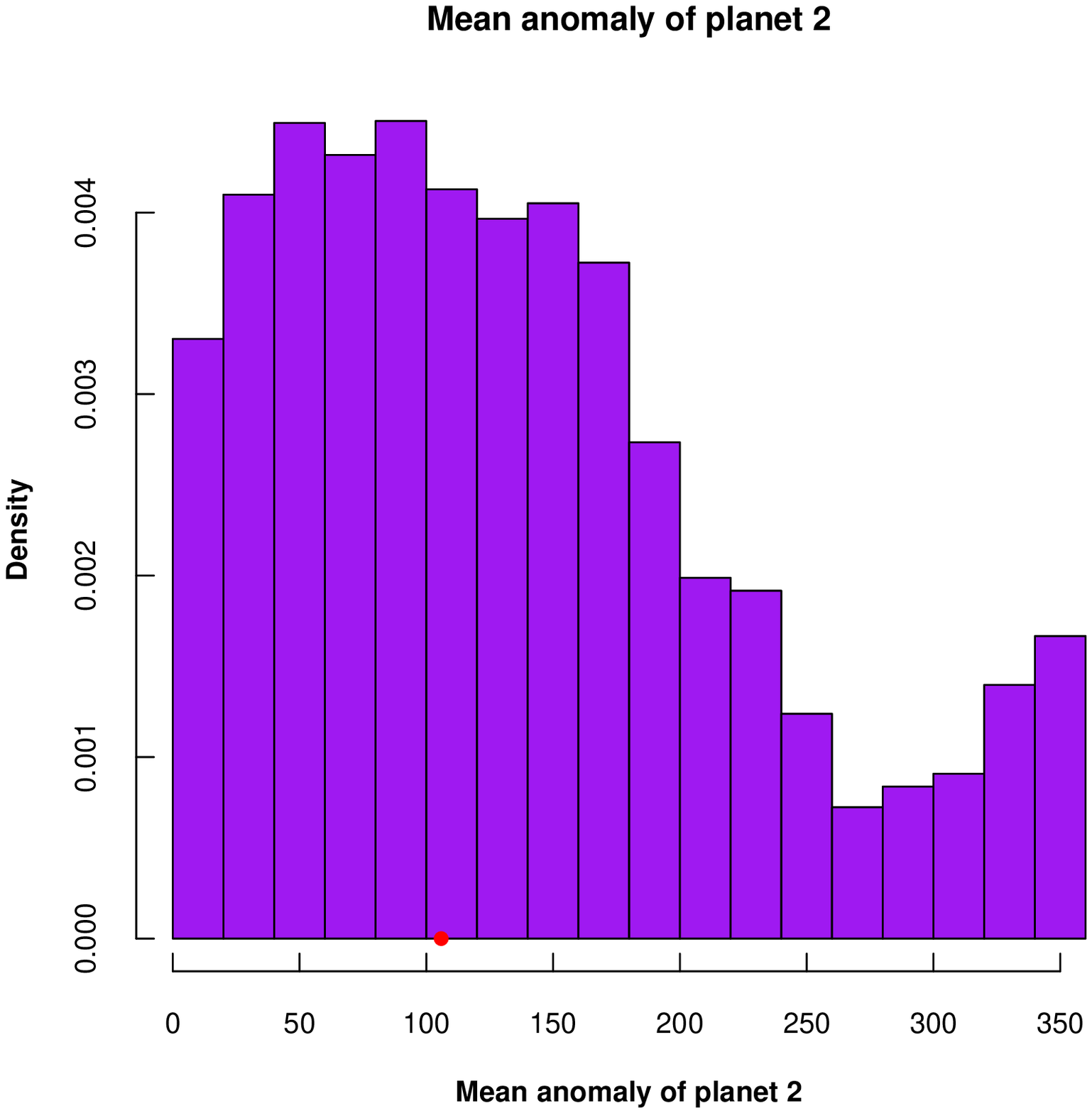}
\includegraphics[width=0.4\textwidth]{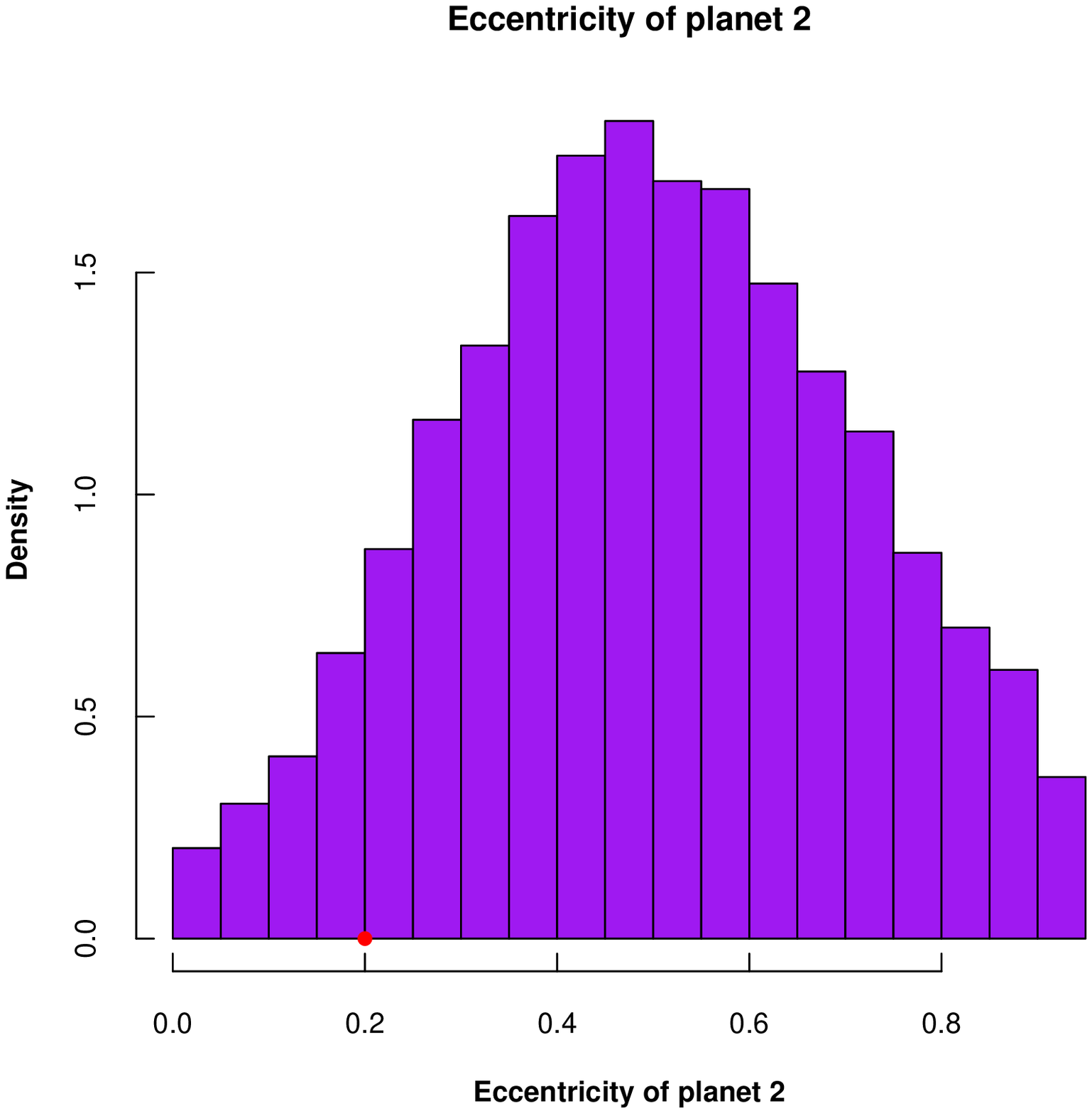}
\includegraphics[width=0.4\textwidth]{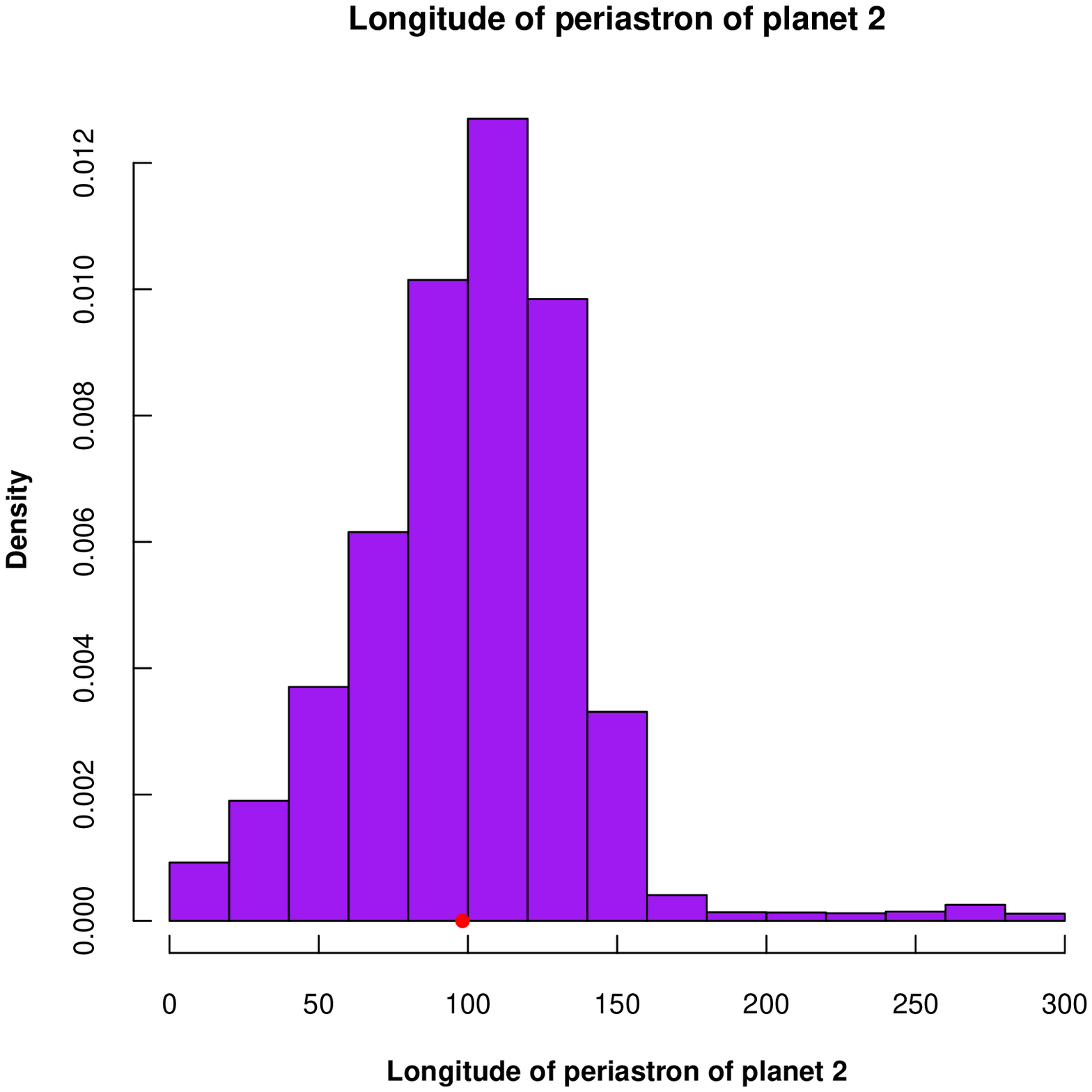}
\includegraphics[width=0.4\textwidth]{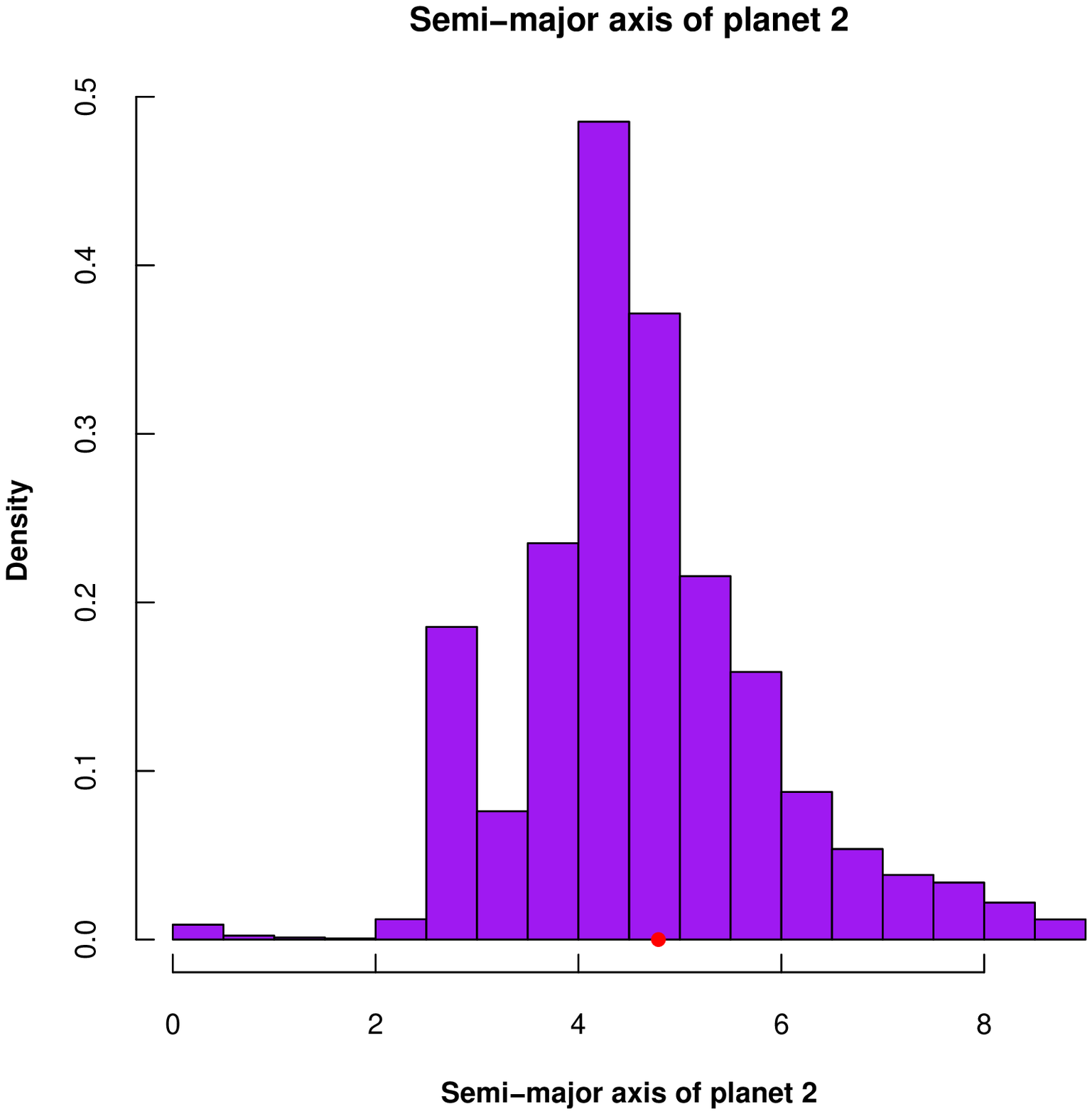}
\caption{Marginal distributions of the orbital elements for planet 2, of the two planet fit resulting from
our MCMC analysis of HD 133131A RV data; these plots show the low
eccentricity fit distributions. The best-fit values from
Table \ref{tab:Afit} are marked with red dots.}\label{fig:errorsA2_lowecc}
\end{figure*}

\begin{figure*}
\centering
\includegraphics[width=0.4\textwidth]{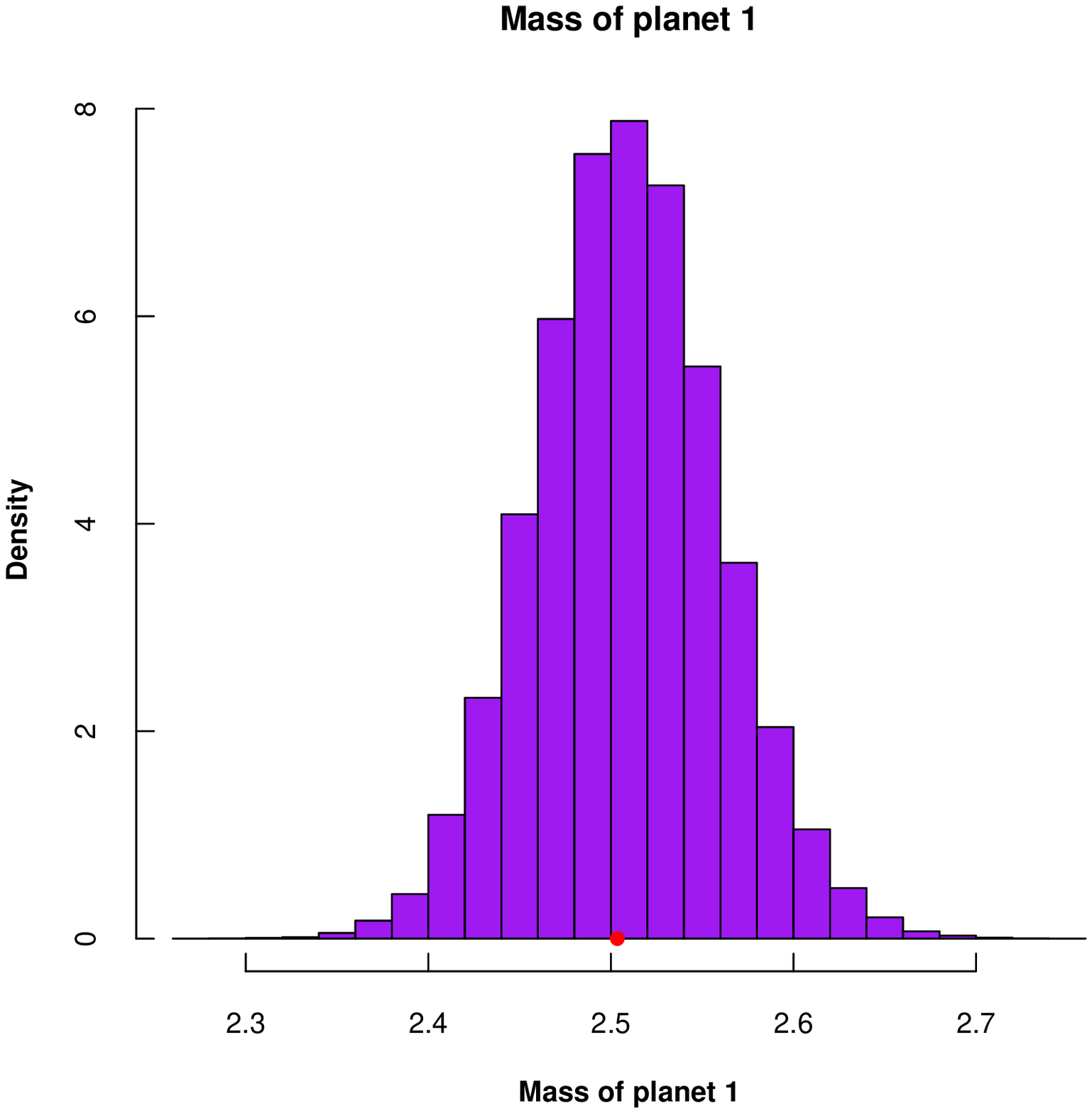}
\includegraphics[width=0.4\textwidth]{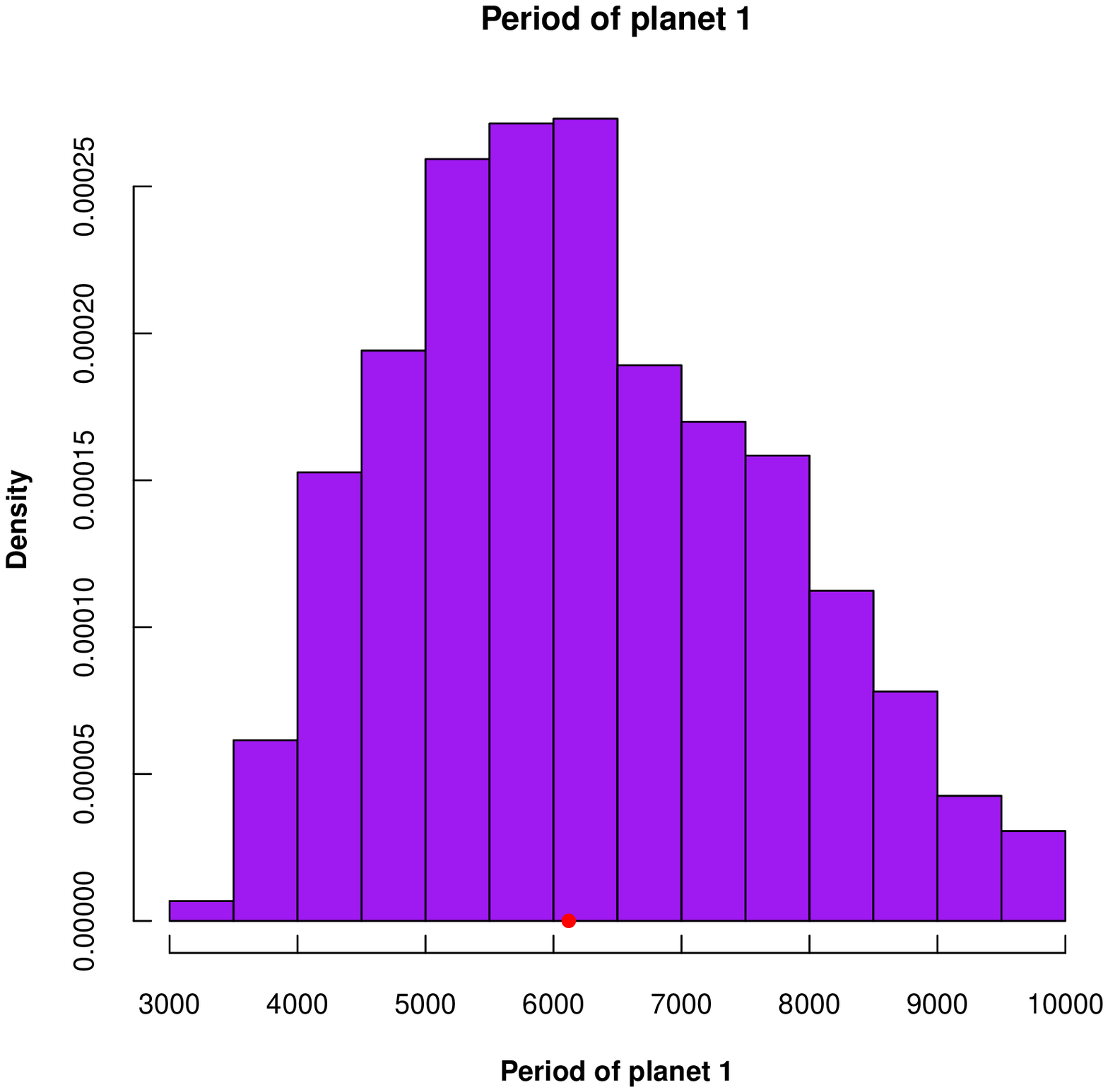}
\includegraphics[width=0.4\textwidth]{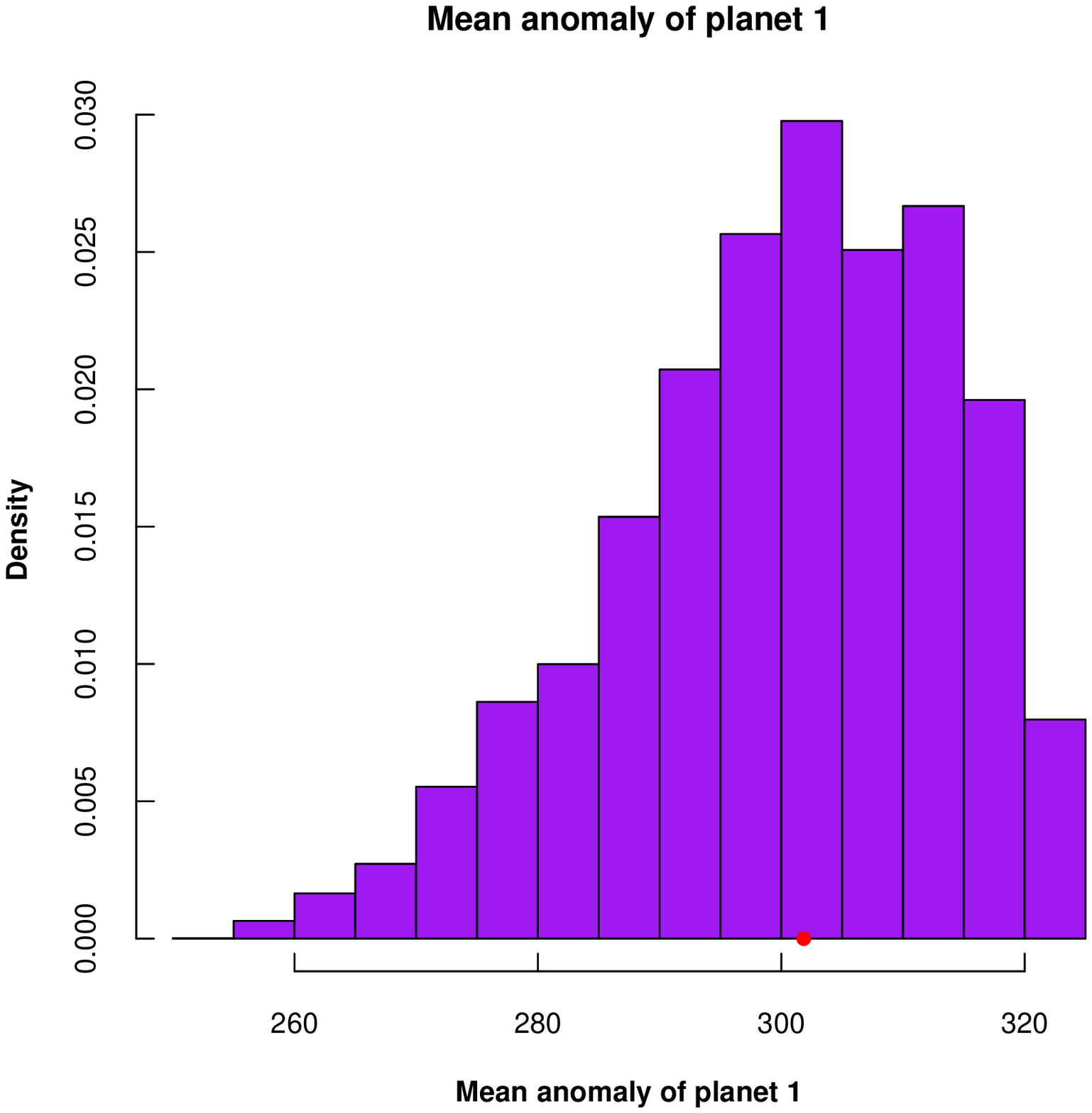}
\includegraphics[width=0.4\textwidth]{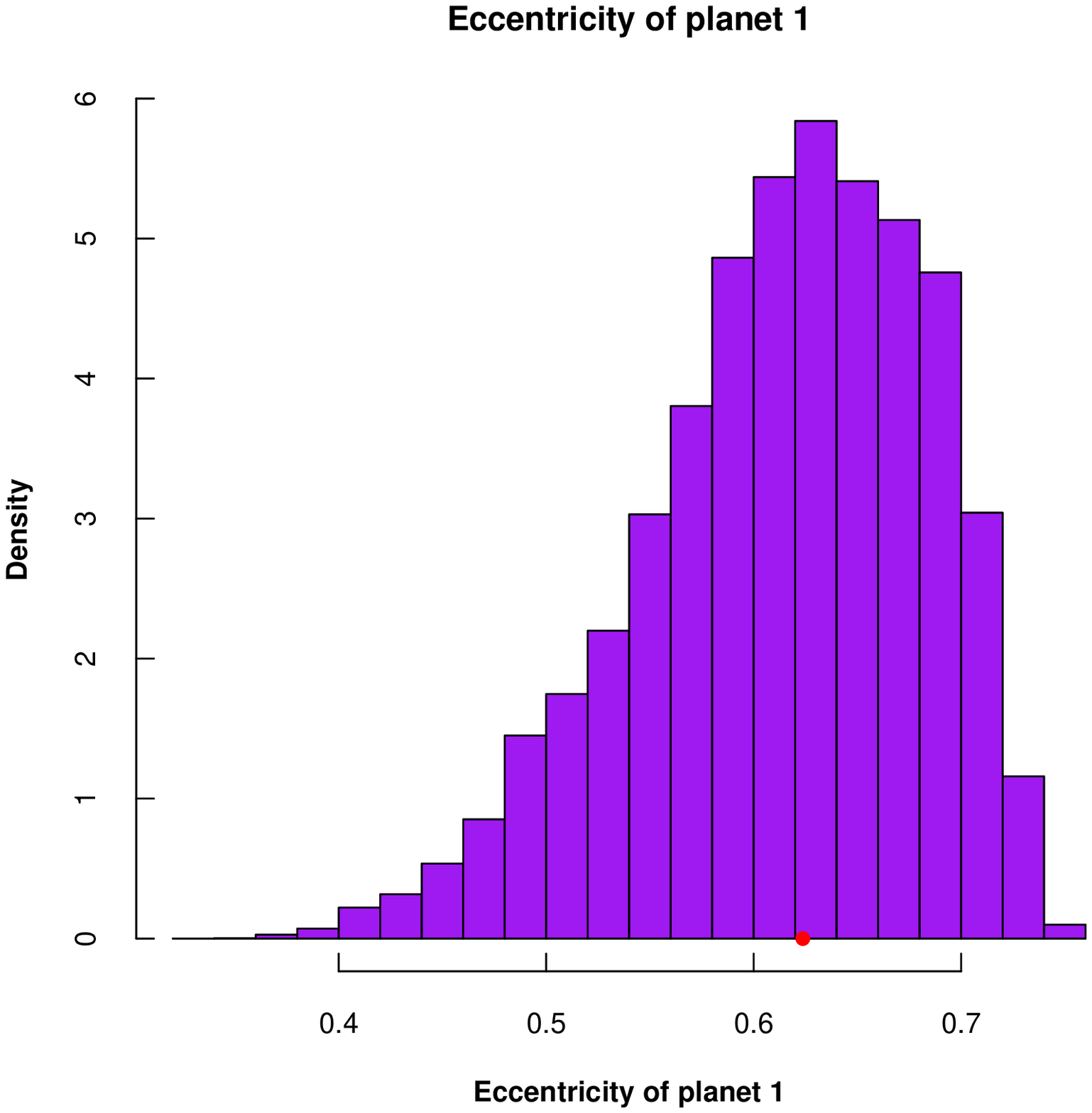}
\includegraphics[width=0.4\textwidth]{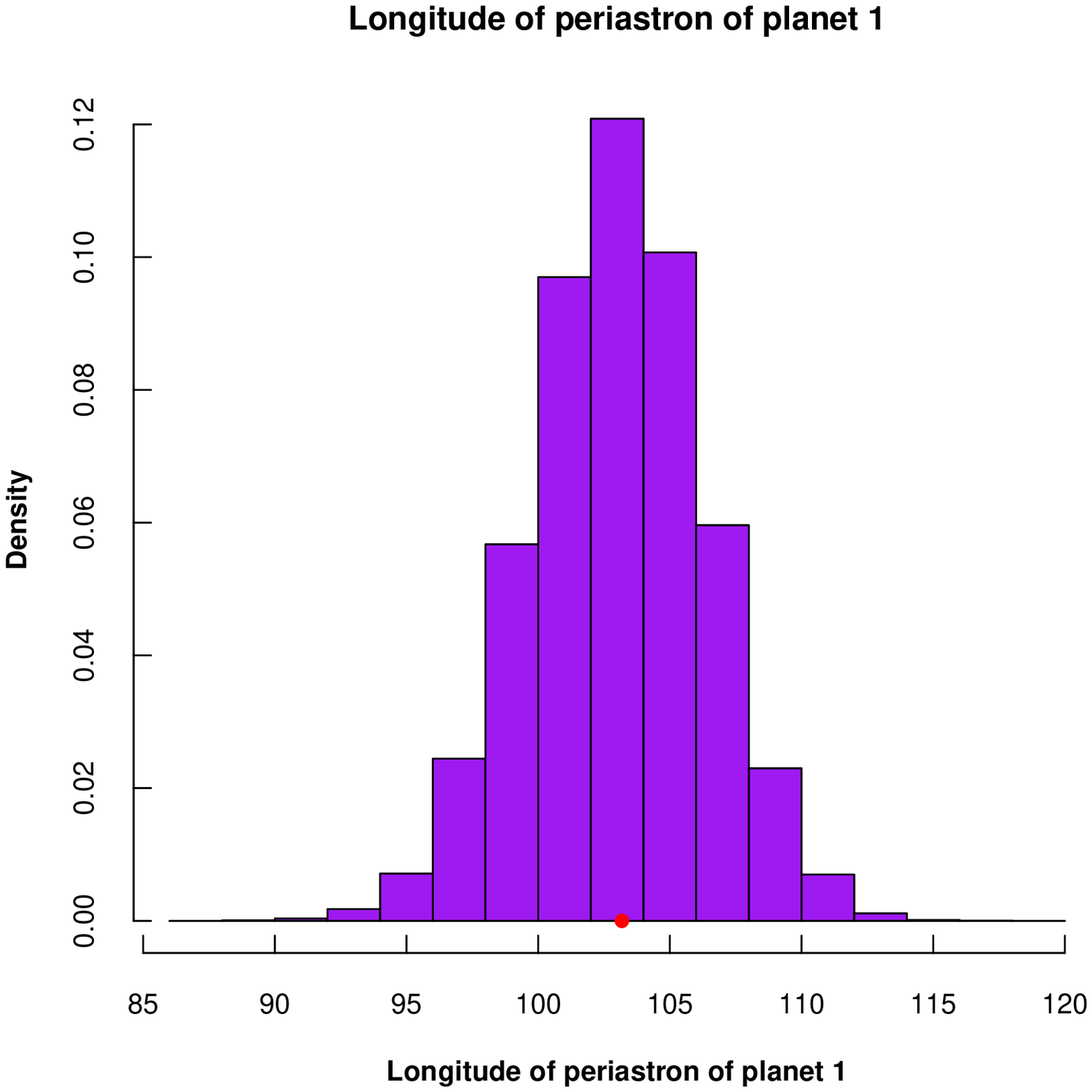}
\includegraphics[width=0.4\textwidth]{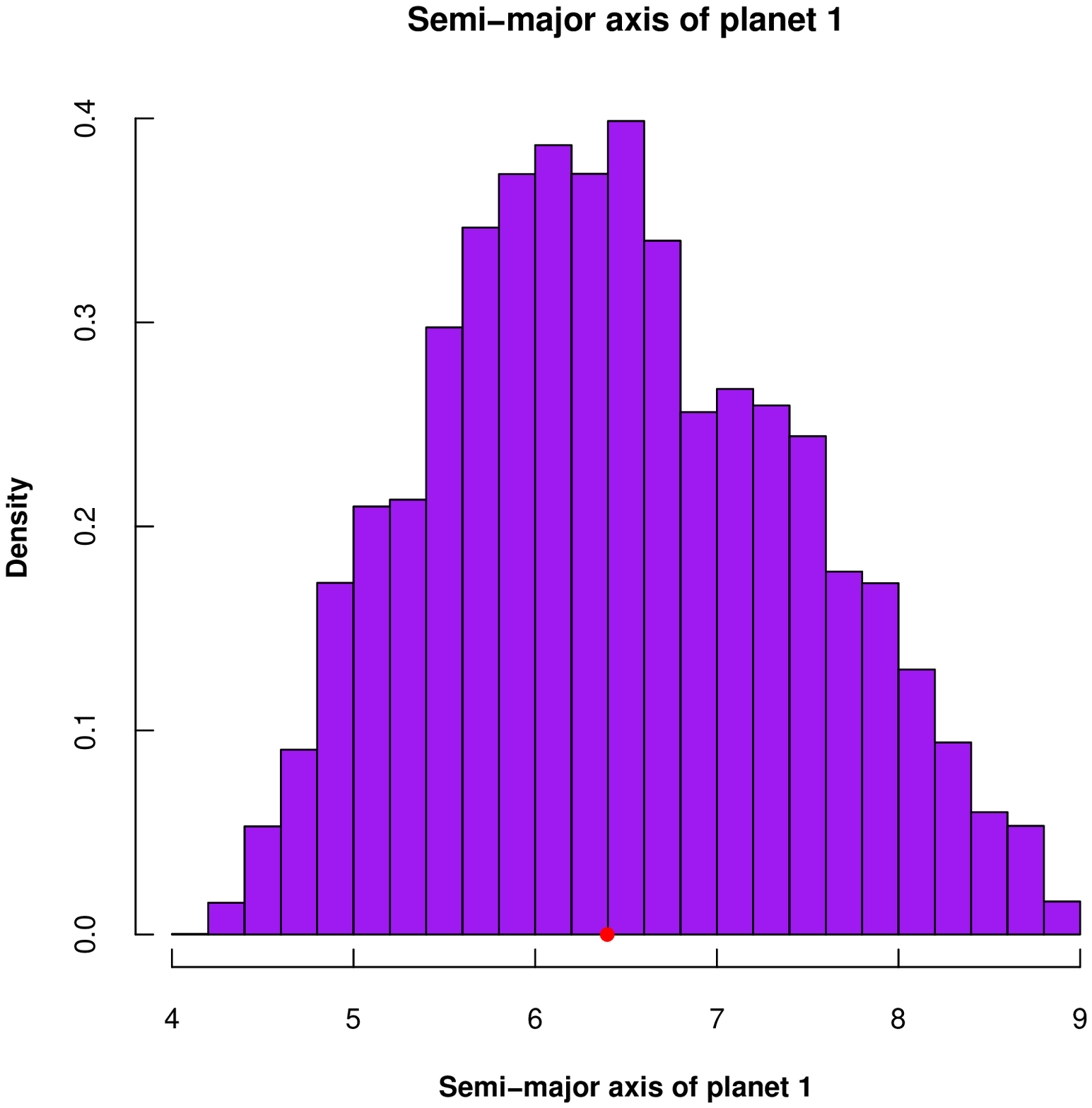}
\caption{Marginal distributions of the orbital elements for one planet
  fit resulting from
our MCMC analysis of HD 133131B RV data. The best-fit values from
Table \ref{tab:Bfit} are marked with red dots.}\label{fig:errorsB}
\end{figure*}

Armed with appropriate total errors for each RV data point, we ran the
SYSTEMIC MCMC algorithm with 2 chains, skipping the first 1000
samples (burn in), until the potential scale reduction factor $R$ was equal to
1.1 (Gelman \& Rubin 1992; Brooks \& Gelman 1997), which we took to
indicate convergence. The resulting marginal distributions of the
orbital elements are show in Figures \ref{fig:errorsA1},
\ref{fig:errorsA2_highecc} (high eccentricity fit for planet c),
\ref{fig:errorsA2_lowecc} (low eccentricity fit for planet c), and
\ref{fig:errorsB}, with the best fit values shown with a red dot. We
summarize these distributions by reporting the best fit (highest likelihood), median, and
median absolute deviation (MAD) for each parameter in Tables \ref{tab:Afit} and \ref{tab:Bfit}.

We confirmed our simple, analytic estimation of $s_j$ values for each data set by
running the same MCMC algorithm in SYSTEMIC on input data without the $s_j$ term
added in quadrature to the formal errors. The marginal posterior
distributions for the planet parameters were all consistent with those
derived above, and the distributions of the $s_j$ values were peaked
around the values we estimated. The median$\pm$MAD $s_j$ value for HD 133131A PFS is 1.66$\pm$0.63, and
for MIKE it is 13.8$\pm$3.15; for HD 133131B PFS it is 0.63$\pm$0.58. 

\subsection{Planet Detectability}

\begin{figure}[!ht]
\centering
\includegraphics[width=0.35\textwidth,angle=90]{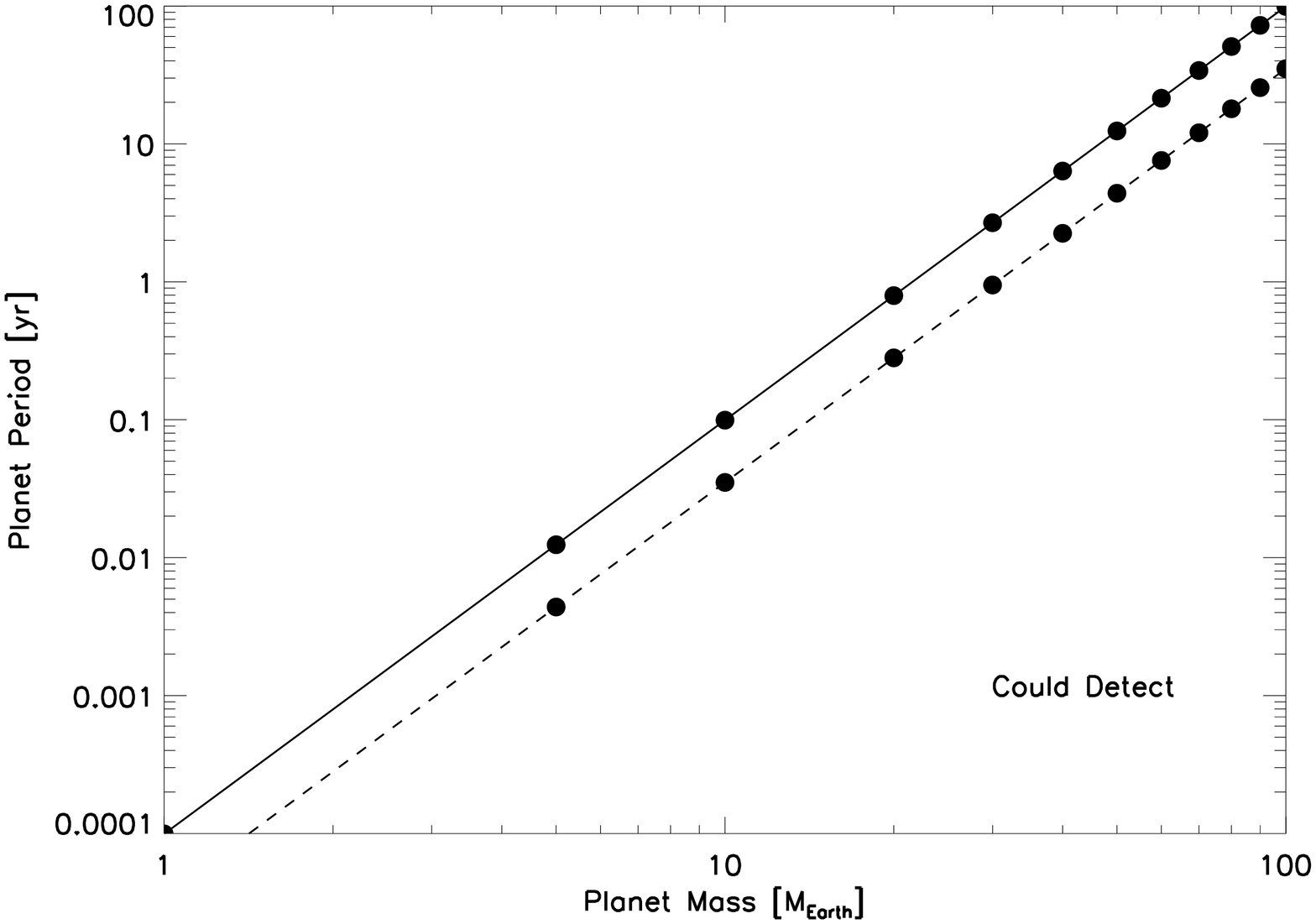}
\caption{Limits on detectable planet masses and periods around HD
  133131A or B, given our
  current time baseline of observations and RV precision. The solid
  line assumes an inclination of 90$^{\circ}$, and the dashed line
  an inclination of 45$^{\circ}$. Both lines assume zero eccentricity,
  a stellar mass of 0.95 $\mathcal{M}_{\odot}$, and an RV
  semi-amplitude $K_{\ast}$ of 2 m~s$^{-1}$.} \label{fig:detect}
\end{figure}

Given the interesting nature of the potential planet HD 133131Bc at
5.88 days, along with the stellar abundance differences described in
\S5.2, we explored what limits our current data could place on
additional planet detection around HD 133131A and B. The radial
velocity semi-amplitude, $K = (\upsilon_{r,max} - \upsilon_{r,min})/2$,
can be expressed in terms of the interesting quantities
$\mathcal{M}_{P}$ and $P$,

\begin{equation}
K_{\ast} = \frac{8.95~\rm{cm~s^{-1}}}{\sqrt{1-e^{2}}}
\frac{\mathcal{M}_{P}~\rm{sin}~i}{M_{\oplus}} (\frac{\mathcal{M_{\ast}}+\mathcal{M}_{P}}{\mathcal{M}_{\odot}})^{-2/3} (\frac{P}{\rm{yr}})^{-1/3}
\end{equation}

\noindent which we use with the following assumptions to approximate our
observational detection limit in this case: zero eccentricity, 90$^{\circ}$ or
45$^{\circ}$ inclination, 0.95 $\mathcal{M}_{\odot}$ stellar
mass, and a $K_{\ast}$ limit of 2 m~s$^{-1}$, based on our stellar
jitter and instrument RMS values. In Figure \ref{fig:detect}, we show
this estimated detection limit in planet mass-period space; the solid curve
is assuming a 90$^{\circ}$ inclination and the dashed curve a
45$^{\circ}$ inclination for the planetary orbit. As expected, a 0.018
$\mathcal{M}_{J} \sim$ 5.7 $\mathcal{M}_{\oplus}$ planet at 5.88
days $\sim$ 0.016 years is at the limit of detectability with our
current data, even assuming the maximum potential RV signal. However,
we could expect to detect a $\sim$20 $\mathcal{M}_{\oplus}$ planet in an orbit of a few hundred days, for instance,
given our current data and RV precision and the assumptions made above.

\section{New Characterization of HD 133131 A \& B}
\subsection{Stellar Parameters}
Here we derive stellar parameters -- effective temperature ($T_{eff}$),
surface gravity (log $g$), metallicity ([Fe/H]), and
microturbulent velocity ($\xi$) -- with iodine-free spectra from PFS; the instrument and data reduction
are described below. 
A traditional ionization-equilibrium balance
method was implemented, in which the correlations between [Fe I/H] and
the excitation potential ($\chi$) and between [Fe I/H] and the
reduced equivalent width [log($EW/\lambda$)] are minimized. The
difference between the mean iron abundances measured from Fe I and Fe
II lines is also minimized. This procedure is done iteratively to derive the ``best'' spectroscopic stellar
paramters. We used the publicly-available \texttt{Qoyllur-quipu}
($q^2$) Python package\footnote{https://github.com/astroChasqui/q2},
described in detail in Ram\'irez et al. (2014), a wrapper for the
spectral analysis code MOOG (Sneden 1973; Sobeck et al. 2011) that takes EWs as input to
derive stellar parameters and abundances. 

Equivalent width (EW) measurements were made by manually fitting
Gaussian functions to observed line profiles using the
\texttt{splot} task in IRAF\footnote{IRAF is distributed by the National
  Optical Astronomy Observatory, which is operated by the Association
  of Universities for Research in Astronomy, Inc., under cooperative
  agreement with the National Science Foundation.}, with the iron line list
from Ram\'irez et al. (2014) that includes Fe I lines from a wide range
of excitation potentials and 16 Fe II lines. The EWs are measured
differentially with respect to the Sun on a line-by-line basis; our
solar spectrum comes from PFS observations on 7 Jan 2016 UT of
reflected sunlight from the asteroid Vesta, and we assume $T_{eff,\odot}$=5777 K, log
$g_{\odot}$=4.44, [Fe/H]$_{\odot}$=0, and $\xi_{\odot}$=1
km\,s$^{-1}$. This differential approach reduces the impact of
uncertainties in stellar models and in atomic data, as they are
canceled out in each star$-$Sun measurement. The continuum regions for
each line were chosen to be ``clean'' and the same for the two stars
and the Sun; this is possible because HD 133131A and B are both inactive G2
stars and the data are taken with the same spectrograph set-up. EWs
were translated into abundances using the \texttt{abfind} driver in
MOOG with 1D-LTE model atmospheres linearly
interpolated from the \texttt{marcs} grid.

The derived [Fe I/H] and [Fe II/H] values for HD 133131A and B have line-to-line
scatters of 0.037/0.032 dex (71 Fe I lines) and 0.041/0.041 dex (16 Fe
II lines), respectively. $q^2$ automatically calculates errors on
$T_{eff}$, log $g$, and $\xi$ as in Epstein et al. (2010) and Bensby
et al. (2014), and [Fe/H] errors by adding the other stellar
parameter errors in quadrature (under the assumption that they are
independent) with the standard error of the mean
line-to-line scatter ($\sigma/\sqrt{N-1}$) of the [Fe/H]
values. The errors reported here are a reflection only of how well the
minimization criteria above are met, and are still subject
to systematic errors, e.g., 3D and NLTE effects (Asplund 2005) not
captured in our 1D-LTE analysis. Comparisons of abundances measured in
1D versus 3D stellar atmosphere models can differ at the 0.01-0.02 dex
level, and have been shown to increase with lower [Fe/H] (e.g.,
Ram\'irez et al. 2008; Bergemann et al. 2012; Magic et al. 2014). However,
these differences are measured in an absolute sense, so that they are
subject to systematic uncertainties in a way that our strictly
differential analysis is not, especially because HD 133131A and B are
very similar to the Sun except for their metallicities (Table \ref{params}). 

The stellar parameters derived here from a differential analysis with
the Sun are listed in the second section of Table \ref{params}, and
agree moderately well with the parameters found by Desidera et
al. (2006b), who use a similar ionization-equilibrium balance
  to derive T$_{eff}$ and [Fe/H] but derive their log $g$ values using photometric
  information. Interestingly, we find the stars to
be slighly closer to solar in $T_{eff}$ and [Fe/H], and also that the
A component is cooler and more metal rich than the B component, the
opposite of Desidera. However, given our errors, we find the A and B
stellar parameters are insignificantly different in everything except
[Fe/H] (with $\Delta$(A-B) [Fe/H]$=-0.025\pm0.021$). 

The metal-poor nature of this system puts both stars outside the
 ``solar twin'' realm, and although it is near-by ($\sim$50 pc), the system
may have experienced slight differences in Galactic chemical evolution
(especially if its age is much older; see above). As in previous
stellar ``twin'' papers (e.g., Ram\'irez et al. 2015; Saffe et al. 2015;
Teske et al. 2016), here we also derive more precise stellar
parameters by comparing the two stars strictly against each
other. In previous works where such a strict differential
method is used to compute stellar parameters, only the
``cool''-``hot'' (here, A-B) case is derived. In this paper we compare
both (A-B) and (B-A) stellar parameters, since both stars are so
similar to the Sun. In each case (A-B, B-A), we assume the (B, A) solar reference parameters and
measure only the (A, B) $T_{eff}$, log $g$, $\Delta$[Fe/H],
and $\xi$. These strictly A vs. B differential parameters
are listed in the third and fourth sections of Table \ref{params}. In
both cases, the alternative parameters are very similar to the solar reference parameters,
but the errors are reduced, particularly in the (A-B) case (last
section of table). 

\begin{deluxetable}{lcccc}
\tabletypesize{\scriptsize}
\tablecolumns{5}
\tablewidth{0pc}
\tablecaption{Stellar Parameters \label{params}}
\tablehead{ 
\colhead{Star} & \colhead{ T$_{\rm{eff}}$ } & \colhead{ log $g$}  &
\colhead{$\rm{[Fe/H]}$} & \colhead{$\xi$ }  \\
\colhead{} & \colhead{(K)} & \colhead{ [cgs] }  &\colhead{(dex)} &\colhead{(km s$^{-1}$)}}
\startdata
Desidera et al. (2006) & & & &  \\
\hline
HD 133131A & 5745&4.46&-0.33 & 1.05\\
HD 133131B  & 5739& 4.46 &-0.36& 1.05 \\
$\Delta$(A-B)  &$+$6 &0&$+$0.03& 0 \\
$\Delta$(A-B)\tablenotemark{a} &$+$5$\pm$15 &\nodata&$+$0.032$\pm$0.015& \nodata \\
\hline
\hline
This work, Solar Reference & & & &  \\
\hline
HD 133131A & 5799$\pm$19& 4.39$\pm$0.050 &-0.306$\pm$0.016 & 1.10$\pm$0.040 \\
HD 133131B  & 5805$\pm$15& 4.41$\pm$0.045& -0.281$\pm$0.013&1.12$\pm$0.030  \\
$\Delta$(A-B)  &$-$6$\pm$24 & $-$0.02$\pm$0.064& $-$0.025$\pm$0.021&$-$0.02$\pm$0.050 \\
\hline
\hline
This Work, HD 133131A Reference & & & &  \\
\hline
HD 133131A   (same as solar ref.)  & 5799$\pm$19& 4.39$\pm$0.050 &-0.306$\pm$0.016 & 1.10$\pm$0.040 \\
HD 133131B                                & 5811$\pm$12& 4.42$\pm$0.032& -0.275 $\pm$0.010&1.11$\pm$0.020  \\
$\Delta$(A-B)                           & $-$12   $\pm$22 &$-$0.03$\pm$0.059& $-$0.031$\pm$0.019& $-$0.01$\pm$0.045\\
\hline
\hline
This Work, HD 133131B Reference & & & &  \\
\hline
HD 133131A                                & 5796$\pm$13& 4.40$\pm$0.032& -0.311$\pm$0.011& 1.12$\pm$0.030\\
HD 133131B (same as solar ref.)  & 5805$\pm$15& 4.41$\pm$0.045& -0.281$\pm$0.013&1.12$\pm$0.030  \\
$\Delta$(A-B)                           & $-$9  $\pm$20 &$-$0.01$\pm$0.055& $-$0.030$\pm$0.017& 0.00$\pm$0.042\\
\enddata
\tablenotetext{a}{As reported in their Table 11, final analysis with
  propogated errors.}
\end{deluxetable}

The median $S$-index value for HD 133131A is 0.155,
while the median for HD 133131B is 0.154 (see Tables \ref{tab:rvdata_MIKE},
\ref{tab:rvdataA_PFS}, \ref{tab:rvdataB_PFS}). Taking these median
$S$-index values and converting them to $R'_{\rm{HK}}$ using the Noyes et
al. (1984) calibration results in $R'_{\rm{HK}}=$ -4.913 for A and
-4.919 for B. Several less recent chromospheric age relations (Soderblom et
al. 1991 Eqs. 1 \& 3; Donahue 1993) indicate that stars with
$R'_{\rm{HK}}\sim -4.9$ should be $\gtrsim$9.5 Gyr old, as does Figure 1
of Pace (2013, for the $T_{eff}$ of HD 133131A \& B derived
here). However, the age-metallicity relation of Rocha-Pinto et
al. (2000), for [Fe/H] $\sim -0.3$, suggests a slightly
younger age of $\sim$7 Gyr. The recent work of Bergemann et al. (2014)
measuring the age-metallicity relation of Milky Way disk stars as part
of the Gaia-ESO survey also suggests an age closer to $\sim$9.5 Gyr
for a star with the [Fe/H], [Mg/Fe] ($\sim$0.03) and $T_{eff}$ of HD 133131A \& B (see their
Figures 6 \& 7). The kinematics of the pair suggest with
$\sim$93\% probability thin disk
membership, based on UVW velocities (calculated from the coordinates, proper
motions, and parallax
of van Leeuwen 2007, and the absolute radial velocity of Gontcharvo 2006)
and the relations of Reddy et al. (2006; Eq. 1 \& 2).  

We conclude that the pair are
coeval, and likely older than the Sun, perhaps over twice as old.

\subsection{Stellar Abundances}
In the spectra of Vesta, HD 133131A, and HD 133131B  we measured absorption
lines of 19 elements in addition to Fe, and combined these EW measurements
with our derived stellar parameters in a curve-of-growth analysis
within MOOG to derive stellar abundances. The procedure was the same as
that used in Teske et al. (2016), involving an examination of the
normalized, Doppler-corrected spectra of every line with the
\texttt{splot} task in IRAF and choosing regions for the continuum
that were clean and the same in all three spectra. Carbon abundances
were derived from both C I and CH features, and lines from both
neutral and singly-ionized species were measured for Sc, Ti, and
Cr. We applied hyperfine structure corrections to the V, Mn, Co, Cu,
Rb, Y, and Ba abundances. The measured EW values for each element,
including Fe, are listed in Table \ref{lines}. The abundances and total errors for each abundance,
including line-to-line scatter and the errors propogated from each
parameter uncertainty, are listed in Table \ref{abuns}. In the case of
CH, Al I, and Zr II, where only one line is measured, we adopted the
largest line-to-line scatter value amongst the other elements with
$>$3 lines, and added this value in quadrature with the errors
propagated from the stellar parameter errors. 

The only elemental abundance not derived directly from EW measurements
was [O/H]. The oxygen triplet at 7775 {\AA} is not included in the
wavelength range of PFS spectra, so we performed a synthesis analysis
on the [O I] line at 6300 {\AA} using the MOOG \texttt{synth} driver,
as described in Teske et al. (2014). We adopt a conservative error of the
average of the other elemental abundance errors, for each set of
parameters (A-Vesta, B-Vesta, B-A, A-B; see Table \ref{abuns}). 

\begin{figure*}
\centering
\includegraphics[width=0.75\textwidth]{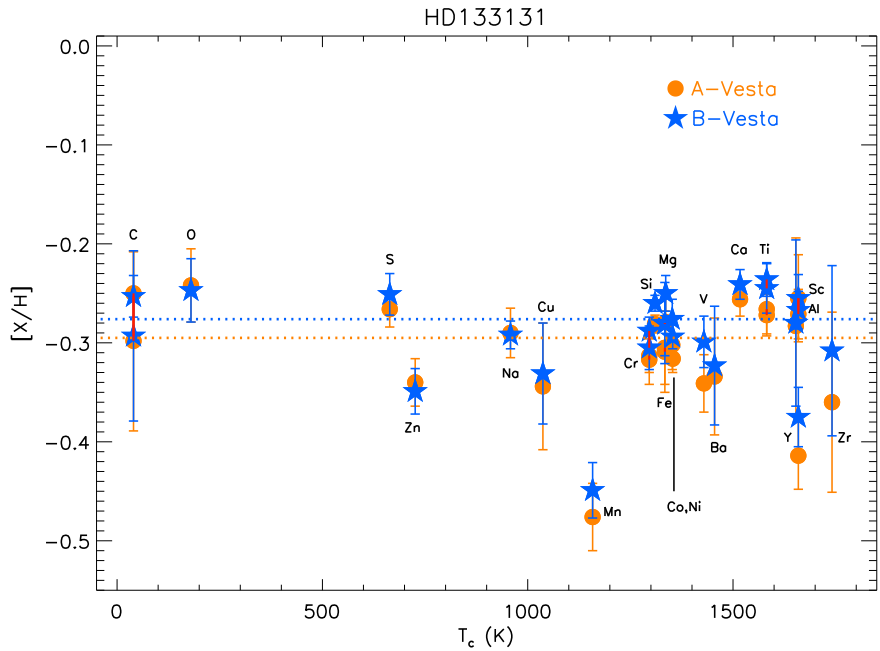}
\caption{The relative
  abundances of HD 133131A (orange circles) and B (blue stars) versus T$_c$ (Lodders 2003),
  calculated using the derived stellar parameters in Table
  \ref{params}, columns 3, 4, 5, and 6. Red lines connected multiple ionization states of the
  same species, or in the case of C, abundances derived from C I
  and CH features. Here the stars are normalized to solar
  abundances measured from Vesta; these abundances are derived
  with the parameters determined in a differential analysis with
  respect to Vesta. Dotted lines indicate the weighted mean of
  each data set (A-Vesta in orange, slightly lower, B-Vesta in
  blue, slightly higher).}
\label{fig:delta1}
\end{figure*}

\begin{figure*}
\centering
\includegraphics[width=0.75\textwidth]{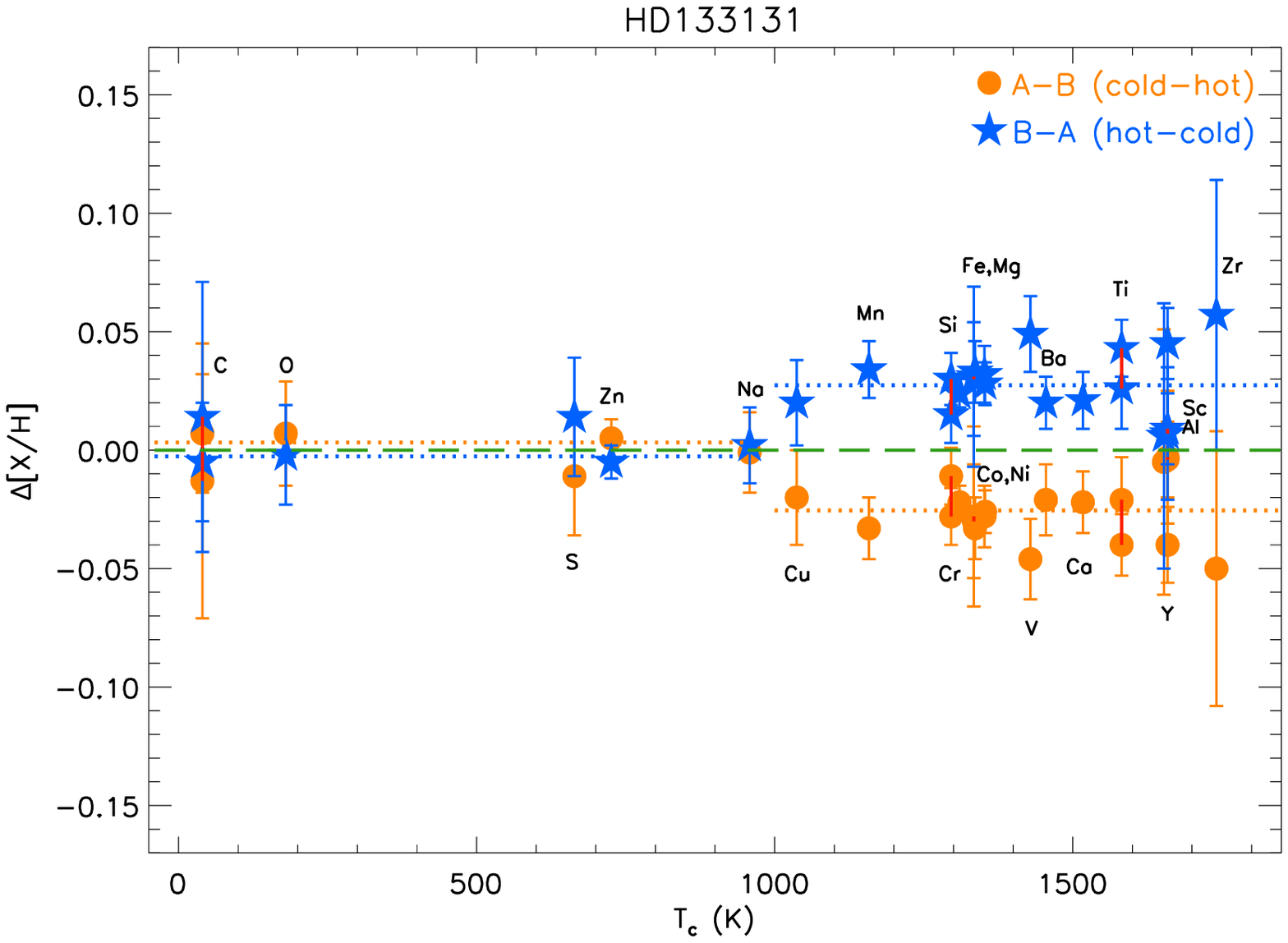}
\caption{The relative
  abundances of HD 133131A (orange circles) and B (blue
  stars) versus T$_c$ (Lodders 2003),
  calculated using the derived stellar parameters in Table
  \ref{params}, columns 7, 8, 9, and 10. Red lines connected multiple ionization states of the
  same species, or in the case of C, abundances derived from C I
  and CH features. Here the stars are normalized to each other; these
  abundances are derived with the parameters determined in a
  differential analysis with respect to B (orange circles ) or A (lighter
  blue stars). A green dashed line shows zero difference, and blue
  dotted lines show the weighted mean of elements, split at $T_c =
  1000$ K. The $\Delta$[X/H] values for elements with $T_c < 1000$ are
indistinguishable from zero in both cases.}
\label{fig:delta2}
\end{figure*}

In Figures \ref{fig:delta1} and \ref{fig:delta2}, we show the
abundance differences derived with respect to Vesta, and derived for
each star with respect to the other, plotted against the 50\%
condensation temperatures ($T_c$) from Lodders (2003) for solar
compostion gas. In each case, the corresponding
stellar parameters were used to derive the abundances -- e.g., in the
A-B abundances, the A-B stellar parameters from Table \ref{params} were
used. The dotted lines in Figure \ref{fig:delta1} show the weighted
means of the A-Vesta (orange, lower value) and B-Vesta
(blue, higher value) abundances; HD 133131A is more metal-poor overall. There are slight differences between the
weighted means of elements above and below $T_c = 1000$K, with a
$\sim$0.01 dex decrease in more refractory (higher $T_c$) element
abundances, but no obvious pattern like that seen in Mel\'endez et
al. (2009)'s analysis of the Sun versus other solar twins. In that
work, Mel\'endez et al. suggested that the deficit of refractory
elements in the Sun with respect to other solar twins was a signature
of small planet formation. More recent work (e.g., Schuler et
al. 2015; Spina et al. 2016; Nissen 2015; Adibekyan et al. 2014) has
shown that trends with $T_c$ may not be (entirely) due to planet
formation, but instead to the birthplace and age of a star. 

We can instead examine the $\Delta$[X/H] values, derived from a strict
comparison of one star in the HD 133131 system versus the other, to
avoid potential causes of abundance differences like age and birthplace; we
omit the uncertainty introduced by the Sun and comparing HD 133131 A
and B to a dissimilar star (at least in [Fe/H]). In Figure
\ref{fig:delta2}, these $\Delta$[X/H] values are plotted against $T_c$,
with a green dashed line showing zero difference and dotted lines
indicating the weighted means of elements with $T_c < 1000$ K and $T_c >
1000$ K, the approximate demarcation between volatile and refractory
elements. We show both A-B (orange circles) and B-A (blue
stars) cases to demonstrate the good
agreement between both stellar parameter derivations; the EWs are the
same in both abundance determinations. In previous stellar ``twin''
abundance studies (Ram\'irez et al.\,2011; Mack et al. 2014; Liu et
al. 2014; Tucci Maia et al. 2014; Teske et al. 2015; Ram\'irez et
al.\,2015; Saffe et al. 2015; Teske et al. 2016), the $\Delta$[X/H] abundances are always quoted from
the ``cold-hot'' star comparison. There is now an obvious decrease (A-B)/increase
(B-A) in the $\Delta$[X/H] values moving from volatile to refractory
elements. The weighted means of the $T_c < 1000$ K abundances are
0.0032/-0.0027 dex, whereas the weighted means of the $T_c > 1000$ K
abundances are -0.0255/0.0274 dex.

\begin{deluxetable}{lcccccc}
\tablecolumns{7}
\tablewidth{0pc}
\tabletypesize{\scriptsize}
\tablecaption{Measured Lines \& Equivalent Widths \label{lines}}
\tablehead{ \colhead{Ion} & \colhead{$\lambda$} & \colhead{$\chi$} &
  \colhead{log $gf$} & \colhead{EW$_{\odot}$} & \colhead{EW HD 133131A} & \colhead{EW HD 133131B}\\
 \colhead{ } & \colhead{({\AA})} & \colhead{(eV)} & \colhead{(dex)} &
 \colhead{(m{\AA})} & \colhead{(m{\AA})} & \colhead{(m{\AA})} }
\startdata
Fe I &	4389.245	&0.052&	-4.583&	71.5	&65.2&	61.6\\
Fe I&	4602.001	&1.608&	-3.154&	72.4	&59.6&	63.1\\
Fe I&	4690.140	&3.69&	-1.61&	57.7	&44.1&	45.7\\
Fe I&	4788.760	&3.24&	-1.73&	70.1	&58.2&	57.8	\\
Fe I&	4799.410	&3.64&	-2.13&	35.6&22.5&	24.5	\\
Fe I&	4808.150	&3.25&	-2.69&	28.1&17.5&	18.0	\\
Fe I&	4950.100	&3.42&	-1.56&	76.8&59.2&	60.5	\\
Fe I&	4994.129	&0.915&	-3.08&	103.6&89.7&	91.7\\
Fe I&	5141.740	&2.42&	-2.23&	90.0	&72.1&	73.4	\\
Fe I&	5198.710	&2.22&	-2.14&	97.3&85.0&	85.2\\
Fe I&	5225.525	&0.11&	-4.789&	73.2&60.0&	61.3\\
Fe I&	5242.490	&3.63&	-0.99&	88.0	&72.7&	73.8\\
Fe I&	5247.050	&0.087&	-4.961&	65.9	&53.2&	54.6\\
Fe I&	5250.208	&0.121&	-4.938&	68.0&52.8&	54.5	\\
Fe I&	5295.310	&4.42&	-1.59&	29.6&17.2&	18.5\\
Fe I&	5322.040	&2.28&	-2.89&	61.5&46.5&	47.9	\\
Fe I&	5373.710	&4.47&	-0.74&	62.4&48.1&	49.8	\\
Fe I&	5379.570	&3.69&	-1.51&	60.9&46.6&	48.3\\
Fe I&	5386.330	&4.15&	-1.67&	32.3&20.0&	21.5\\
Fe I&	5441.340	&4.31&	-1.63&	31.2	&19.5&	20.1\\
\enddata
\tablecomments{This table is available in its entirety in a machine-readable form online. A portion is shown here for guidance regarding its form and content.}
\end{deluxetable}

\begin{deluxetable}{lccccccccc}
\tabletypesize{\scriptsize}
\tablecolumns{10}
\tablewidth{0pc}
\tablecaption{Derived Stellar Abundances \label{abuns} }
\tablehead{ 
\colhead{Species} & \colhead{T$_c$} & \multicolumn{2}{c}{\underline{A-Vesta Params}} & \multicolumn{2}{c}{\underline{B-Vesta Params}} &
\multicolumn{2}{c}{\underline{B-A Params}} & \multicolumn{2}{c}{\underline{A-B Params}}  \\
\colhead{} & \colhead{} & \colhead{$\Delta$[X/H]} & \colhead{error} & \colhead{$\Delta$[X/H]} & \colhead{error} & \colhead{$\Delta$[X/H]} & \colhead{error} & \colhead{$\Delta$[X/H]} & \colhead{error} \\
\colhead{} & \colhead{(K)} & \colhead{(dex)} & \colhead{(dex)} &
\colhead{(dex)} & \colhead{(dex)} & \colhead{(dex)} & \colhead{(dex)} & \colhead{(dex)} & \colhead{(dex)}}
\startdata
C I& 40     & -0.250 &   0.042 & -0.253 & 0.021 & -0.005 & 0.025 & 0.007& 0.025\\
CH$^{a}$ & 40 & -0.298 & 0.091 & -0.293 & 0.086 & 0.014 & 0.057 & -0.013 &0.058 \\
O I$^{b}$& 180 &  -0.242 & 0.037 & -0.247 &   0.032& -0.002 & 0.021    &0.007&0.022 \\
Na I &958   & -0.290 & 0.025 & -0.292 & 0.014 & 0.002 & 0.016 & -0.001& 0.017 \\
Mg I & 1336&  -0.279 & 0.024 & -0.250 & 0.018 & 0.033 & 0.013 & -0.033
& 0.013 \\
Al I$^{a}$ & 1653 & -0.283 & 0.089 & -0.280 & 0.084 & 0.006 & 0.056 & -0.005
& 0.056 \\
Si I & 1310& -0.281 & 0.009 & -0.260 & 0.008 & 0.024 & 0.006 & -0.022
& 0.007 \\
S I &664&-0.266 & 0.018 & -0.251 & 0.021 & 0.014 & 0.025 & -0.011 & 0.025  \\
Ca I &1517& -0.256 & 0.017 & -0.241 & 0.015 & 0.021 & 0.012 & -0.022 & 0.013  \\
Sc I &1659& -0.255 & 0.044 & -0.255 & 0.024 & 0.007 & 0.028 & -0.003 &0.028\\
Sc II &1659&-0.271 & 0.025 & -0.267 & 0.019 & 0.009 & 0.015 & -0.004 &0.016\\
Ti I &1582&-0.272 & 0.021 &-0.236 & 0.017 & 0.043 & 0.012 & -0.040 & 0.013   \\
Ti II &1582&  -0.266 & 0.025 & -0.245 & 0.025 & 0.026 & 0.017 & -0.021& 0.018\\
V I &1429& -0.341 & 0.029 & -0.299 & 0.026  & 0.049 & 0.016 & -0.046
& 0.017 \\
Cr I &1296&  -0.313 & 0.017 & -0.288 & 0.014 & 0.030 & 0.011 & -0.028
& 0.012 \\
Cr II &1296&-0.317 & 0.025 & -0.305 & 0.022 & 0.015 & 0.012 & -0.011 &
0.012\\
Mn I &1158& -0.476 & 0.034 & -0.449 & 0.028 & 0.034 & 0.012 & -0.033 &
0.013\\
Fe I &1334&-0.305 & 0.037 & -0.281 & 0.032 & 0.030 & 0.024 & -0.030 & 0.024    \\
Fe II &1334& -0.309 & 0.041 & -0.280 & 0.041 & 0.031 & 0.038 & -0.028
& 0.038 \\
Co I &1352&  -0.302 & 0.025 & -0.276 & 0.02 & 0.032 & 0.012 & -0.028 & 0.013\\
Ni I &1353& -0.316 & 0.014 & -0.294 & 0.012 & 0.028 & 0.009 & -0.026 &
0.009\\
Cu I &1037& -0.344 & 0.064 & -0.331 & 0.051 & 0.020 & 0.018 & -0.020 &
0.020\\
Zn I &726&  -0.340 & 0.024 & -0.349 & 0.023 & -0.005 & 0.007 & 0.005 &
0.008\\
Y II &1659&  -0.414 & -0.034 & -0.375 & 0.030 & 0.045 & 0.015 & -0.040
& 0.016\\
Zr II$^{a}$ &1741& -0.360 & 0.091 & -0.308 & 0.087 & 0.057 & 0.057 & -0.050
& 0.058  \\
Ba II &1455& -0.334 & 0.059 & -0.323 & 0.060 & 0.020 & 0.011 & -0.021
& 0.015   \\
\enddata
\end{deluxetable}
\tablenotetext{a}{These elements have only one line measured, so the
  line-to-line to scatter value adopted in their error calculation is the greatest line-to-line
  scatter from the rest of the elements with $>$3 lines measured.}
\tablenotetext{b}{The errors on the oxygen abundances were
  conservatively estimated as the average of all the other elemental
  abundance errors, within each parameter set.}

\section{Discussion}

As demonstrated by the radial velocity observations and planet orbital
fitting in \S3 and \S4, HD 133131A hosts two planets of moderate eccentricity,
a 1.43$\pm$0.04 $\mathcal{M}_J$ minimum mass planet at 1.44$\pm$0.005 AU and a 0.63$\pm$0.15 $\mathcal{M}_J$ minimum
mass planet at 4.79$\pm$0.92 AU (assuming the low-eccentricity
configuration for planet c). HD 133131B hosts one planet of
relatively high eccentricity (0.62$\pm$0.04), with a mimimum mass of
2.50$\pm$0.05 $\mathcal{M}_J$ orbiting at 6.40$\pm$0.59 AU. The two stars are
separated by only $\sim$360 AU (Desidera et al. 2006b), making them
the most closely separated ``twin'' pair \textbf{($\Delta T_{eff} \lesssim
100$ K)} with detected planets (the
next closest are 16 Cyg A/B and HD 80606/7 at $\sim$1000
AU). This system is even more rare in that both ``twins'' host
  planets. Interestingly, although previous studies have found that close-in giant
planet host stars are more likely to have stellar companions with
separations between 50-2000 AU versus field stars (Ngo et al. 2015),
broader planet occurrence rates seem to decrease for stellar
companions as a function of separation; Wang et al. (2014) found that
stars with a companion at 100 AU are 2.6$\pm$1 times less likely to
host a planet, and stars with a companion at 1000 AU are 1.7$\pm$0.5
times less likely to host a planet. Furthermore, stellar multiplicity
seems to be negatively correlated with the presence of multiple transiting
planets, although it is unclear for separations similar to that of HD
133131A and B whether this is due to the suppression of planet formation (Roell
et al. 2012; Wang et al. 2014ab; Touma \& Sridhar 2015; Kraus et al. 2016) or the
inclination of planets being perturbed by stellar companions (Wang et al. 2014b, see Fig. 7; Wang et al. 2015). 

It is also noteworthy that both stars are relatively metal-poor (with
[Fe/H]$\sim$-0.30 dex); their metallicities fall below $\sim$90\% of
measured planet host metallicites, and they join the sample of only six other such metal-poor exoplanet
host stars in binary systems (based on \texttt{exoplanets.org}). Thus,
based only on what has been measured, the probability of HD 133131A/B
hosting planets is $<$1\%.  The most recent functional forms of the giant planet
metallicity correlation predict stars with [Fe/H]$\sim$-0.30 dex are
$\sim$3.5 times less likely to host planets than solar metallicity
stars, and $\sim$8.5 times less likely to host planets than stars with
[Fe/H]$=0.30$ dex (Mortier et al. 2013). Thus, these giant planets around metal poor stars
seem to be rare, at least based on current statistics of 0.1-25 $\mathcal{M}_J$
planets at 5-5000 day periods. The metal-poor nature of HD 133131A and
B also place them in a sparsely populated part of planet $e$ versus
host star [Fe/H] space -- only five (11) other exoplanets have both
$e>0.3$ ($e>0.2$) and host star [Fe/H]$\leq$-0.30 (out of 80 [Fe/H]$\leq$-0.30 host
stars, based on \texttt{exoplanets.org}).

As discussed in \S2, HD 133131A and B have different abundances
relative to each other, for elements with $T_c > 1000$K, by about
0.025 dex, with A more metal-poor than B (see Figure
\ref{fig:delta2}). Past studies of ``twin'' planet hosting stars have
used such measured abundance differences to try to deduce information about
planet composition and/or formation (Ram\'irez et
al. 2011; Tucci Maia et al. 2014; Teske et al. 2015; Ram\'irez et
al. 2015; Teske et al. 2016). We can explore similar arguments
here. However, these explorations depend
significantly on the assumed convection zone masses of stars, which change
over the lifetime of the star (e.g., Bahcall et al. 2001). 

For instance, we can estimate whether the difference in mass
between the two planets around HD 133131A versus around B can be
explained by the observed abundance differences using the corrected formula from Ram\'irez et
al. (2011):

\begin{centering} \label{eq5}
\begin{equation}
\Delta[M/H] = \mathrm{log}\bigg[\frac{(Z/X)_{cz}\mathcal{M}_{cz} + (Z/X)_P\mathcal{M}_p}{(Z/X)_{cz}(\mathcal{M}_{cz}+\mathcal{M}_p)}\bigg]
\end{equation}
\end{centering} 

\noindent The current convection zone mass of star A or B
$\mathcal{M}_{cz}$ is estimated to be 0.026
$\mathcal{M}_{\odot}$ from the relation in Pinsonneault
et al. (2001)\footnote{We note that these relations are based on stars
  of age 0.1-4.57 Gyr, whereas our activity-age analysis of HD 133131A
  and B indicate ages of $\sim$9.7 Gyr. It is around 10 Gyr for a 1
  M$_{\odot}$, Z=0.02 star that the surface convection zone begins to
  expand as the star leaves the main sequence and ascends the red
  giant branch.}. The planet metallicity, $(Z/X)_p$, is estimated to be 0.1, although this value can vary from
$\sim$0.03-0.5 for planets with masses $\sim$0.4-2.5 $\mathcal{M}_J$ (see
Thorngren et al. 2015, Fig. 8.). The mass difference between Ab+Ac and
Bb, $\mathcal{M}_p$ is 0.44 
$\mathcal{M}_J$
(assuming a low $e$ c planet, and ignoring
the effects of inclination in $\mathcal{M}\,sin~i$, which on average
should be the same for both stars), and we estimate
the stellar convection zone metallicity
$(Z/X)_{cz}$ by scaling Asplund et al. (2009)'s
$(Z/X)_{\odot}=0.134$ to the system [Fe/H] (-0.30 dex). Then the
required depletion/enhancement expected in HD 133131A/B is 0.09
dex, three times larger than the observed abundance
difference (although this value can range from 0.04 to 0.13 if we
include $\mathcal{M_P}$ errors). If instead we assume $\mathcal{M}_{cz}$ is 0.15 $\mathcal{M}_{\odot}$,
estimating a solar-type star's convection zone size at 15-20 Myr according to the standard model of Serenelli et al. (2011), then the
required depletion/enhancement in A/B to explain the planet mass difference is only
0.02 dex, similar to the observed $\sim$0.025 dex difference. We can instead perform
the reverse calculation, and ask what mass difference explains a 0.025
dex abundance difference, which results in 0.11 $\mathcal{M}_J$ for
$\mathcal{M}_{cz}= 0.026 \mathcal{M}_{\odot}$ and 0.65 $\mathcal{M}_J$
for $\mathcal{M}_{cz} = 0.15 \mathcal{M}_{\odot}$; the latter value is
closer to the observed mass difference between the planets around HD
133131A and HD 133131B. Referring back to Figure \ref{fig:detect}, a
0.11 $\mathcal{M}_J \sim 35~\mathcal{M}_{\oplus}$ planet with a period of up to
$\sim$3 years should be detectable in our data, and a
0.65 $\mathcal{M}_J \sim 200~\mathcal{M}_{\oplus}$ planet with a
period of up to
$\sim$tens of years should be detectable in our data, including the
assumptions of zero $e$, edge-on orientation, and a $K_{\ast} \sim$ 2
m~s$^{-1}$ limit to our RV precision. The non-detection of these
planets, given our detection limit approximation and current data, suggests that planets of these masses are on
longer orbits, if they exist.

In the estimates above, the motivation is to account for the abundance
differences between the two stars with refractory-rich mass ``missing''
from HD 133131A or ``added to'' HD 133131B at some point during the
planet formation process. The results change by an order of magnitude
depending on the assumed convection zone sizes of the stars, which is
related to the time at which refractory material is accreted/depleted. Perhaps the
good match with observations with an assumed larger $\mathcal{M}_{cz}$, and thus
earlier accretion/depletion, is a sign that this is the more likely
scenario. We note that none of the simple ``toy model'' calculations above
account for the effect of gravitational settling of heavy elements
(e.g., Bahcall 1995), enhanced mixing (e.g., Pinsonneault et al. 1989), or
radiative levitation (e.g., Pinsonneault 1997) that are known to occur in the Sun. 

Instead of assuming that the HD133131 planets have the same
heavy element mass ($(Z/X)_p = 0.1$), we can improve this estimate by considering the relation between planet mass,
and total heavy element mass derived by Thorngren et al. (2015):

\begin{centering}
\begin{equation} \label{eq6}
\mathcal{M}_z = (46 \pm 5.2)\mathcal{M_P}^{0.59 \pm 0.073}
\end{equation}
\end{centering}

\noindent where $\mathcal{M}_z$ is the heavy element mass in $\mathcal{M}_{\oplus}$ in a planet of mass $\mathcal{M_P}$ in $\mathcal{M}_J$.
Including the errors in the constants and the uncertainties in
our derived $\mathcal{M}~sini$ values, planet Ab, Ac, and Bb could have $\mathcal{M}_z$
ranging from 48-69, 30-43, and 65-95 $\mathcal{M}_{\oplus}$,
respectively. Taking the middle of each of these ranges, the heavy
element mass in Ab+Ac $\sim$95 $\mathcal{M}_{\oplus}$, and in Bb is $\sim$80
$\mathcal{M}_{\oplus}$, or $\Delta\mathcal{M} = 15\,\mathcal{M}_{\oplus} \sim$0.05 $\mathcal{M}_J$ of heavy
element material that should be in the HD 133131A planets (versus the
star) and not in the
HD 133131B planet (but in the star). If we use Equation 6 to solve for
$\mathcal{M_P}$, assuming $\mathcal{M}_{z}$=15 $\mathcal{M}_{\oplus}$, this
$=0.15 \mathcal{M}_J =\mathcal{M}_p$ in the denominator of Equation
5. Also assuming 15 $\mathcal{M}_{\oplus}$ for $(Z/X)_P\mathcal{M}_p$ in
Equation 5 , the resulting
$\Delta$[M/H] is then 0.10 dex for $\mathcal{M}_{cz}$=0.026 $\mathcal{M}_{\odot}$, and 0.02
dex for $\mathcal{M}_{cz}$=0.15 $\mathcal{M}_{\odot}$. The observed $\Delta$[M/H] (0.025)
falls in between these values, again perhaps giving a glimpse of how the
timing of planet formation and stellar convection zone size are
related. Overall, the $\Delta$[M/H] between HD 133131A and B may be an
artifact of the different interior compositions of their planets. Again referring back to Figure \ref{fig:detect}, our
  detection limit estimates suggest that a
0.04 $\mathcal{M}_J \sim 13~\mathcal{M}_{\oplus}$ planet or a 0.11
$\mathcal{M}_J \sim 35~\mathcal{M}_{\oplus}$ planet must be in orbits
with periods longer than $\sim$a few dozen days or $\sim$3 years,
respectively, since we do not detect them.

Alternatively, if $\mathcal{M}_z$ for Ab, Ac, and Bb are on the low end of our
estimates (48-69, 30-43 and 65-95 $\mathcal{M}_{\oplus}$), the two systems
could have the same amount of heavy element mass sequestered in their
respective planets, leaving the $\Delta$[M/H] unaccounted
for. There may have been small planets or rocky material that was
once in either system, but due to planet-planet scattering was ejected
(in the case of A) or accreted on to the star (in the case of B). Ford
\& Rasio (2008) discuss in detail planet-planet scattering in systems with
two unequal mass planets, and find that collisions are less frequent
versus scattering between equal-mass planets. Ejection also seems more
likely, versus planet collision, given the relatively large periods
of the detected planets (Petrovich 2015). The ``jumping Jupiter''
scenario, to account for the observed angular momentum deficit in
the terrestrial planets despite Jupiter crossing its 2:1 mean motion
resonance with Saturn (Morbidelli et al. 2009, 2010; Brasser et
al. 2009; Minton \& Malhotra 2009; Walsh \& Morbidelli 2011; Agnor \&
Lin 2012), suggests that (1) a planet like Uranus or
Neptune scattered off of Jupiter (Thommes et al. 1999; Nesvorn\'y 2011; Batygin et
al. 2012), and (2) that the angular momentum diffusion of the
terrestrial planets could have been significant (Angnor \& Lin 2012;
Brasser et al. 2013). In fact, a large suite of N-body simulations
show a $\geq$ 85\% probability that at least one terrestrial planet is lost
(accreted on to the star or ejected) in the process of the solar
system giant planet instability (Kaib \& Chambers 2016). Ford \& Rasio (2008) also point out that
an important parameter for determining whether giant planets have
circular or eccentric orbits may be the timing of the final
planet-planet scattering event, whether there is still enough material in
the disk to damp eccentricities (see also Ida \& Lin 2008; Raymond et
al. 2009). It is important to keep in mind the relatively close
separation of HD 133131A and B ($\sim$360 AU), and that this could
have affected the disk sizes, orientations, stabilities/lifetimes (see Zhou et
al. 2012 for a brief overview). 

There is always the possibility that the $\Delta$[X/H] differences
observed between the two stars are not related to planet
formation. While we avoid some complicating factors like different
degrees of dust cleansing by luminous stars (\"{O}nehag et al. 2011;
\"{O}nehag et al. 2014) and different ages (Adibekyan et al. 2014;
Nissen 2015; Spina et al. 2015) by analyzing
``twin'' stars in a binary, there is a chance that gas-dust
segregation (Gaidos 2015) may have happened differently in the two
stellar disks, and/or that the stellar composition difference is an
artifact of formation of the stars. Larger surveys of binary stars
suggest that [Fe/H] differences $\gtrsim 0.03$ are atypical (Desidera et al. 2004; 2006b), and here
$\Delta$[Fe/H]$_{A-B}$ is -0.030$\pm$0.017. Gratton et al. (2001)'s
study of a wider suite of abundances in six wide binaries found four
with abundances equal at the $\sim$0.012 dex level, but suggested the other
two stars could be planet hosts based on their larger abundance differences. Furthermore, the two stars in the
Gratton sample with the lowest $\Delta T_{eff}$ (at 11$\pm$32 and
90$\pm$13 K), have $\Delta$[Fe/H] values of 0.000$\pm$0.019 and
-0.005$\pm$0.009, respectively. Our even smaller $\Delta T_{eff}=$-9$\pm$20
K suggests the $\Delta$[Fe/H]$_{A-B}$ difference we measure is
meaningful. We also note that while Nordstr\"om et al. (2004) list the
$v~sini$ values for HD 133131A and B as 4 and 7 km~s$^{-1}$,
respectively, we see no difference in the width of absorption lines in
the two components across many unblended lines, indicating differences
in rotation are not resposible for differences in abundance measurements.

Speculation about the origin of the $\Delta$[M/H] abundance
differences aside, the context of the Thorngren et al. (2015) study again emphasizes how
unusual the HD 133131A and B planets are based on their observed
masses and host star metallicities. Though not well defined, there is
a positive relationship between planet heavy element mass and host
star [Fe/H]; a bootstrap regression gives $\mathcal{M}_z =
(31.4\pm3.4)\times10^{0.48\pm0.047 [Fe/H]}$. This relation predicts
$\sim$23 $\mathcal{M}_{\oplus}$ of heavy element mass in the HD 133131A and B
planets, much lower than most of the estimates from Equation
6 above. Understanding how giant planet heavy element mass depends on other
elements may help shed light on the disagreement between these two
predictions. Indeed, HD 133131A and B both have higher abundances of C, O, Mg, Si,
and Ti than Fe; further studies of variation in these host stars
abundances with giant planet heavy enrichment are needed.

\section{Conclusions}

In this work we present the detection of three giant planets, two
orbiting the A component of HD 133131 and one orbiting the B
component. This is the first planet detection paper based primarily on Magellan/PFS radial velocity observations, and
demonstrates the $\sim$1 m~s$^{-1}$ precision of this instrument over
its six-year baseline. The outer planets around HD 133131A and B also push
the boundary of RV planet detection -- only $\sim$a dozen other RV planets
have been detected at periods $\geq$3600 days. A full analysis of the
frequency of eccentric giant planets at long periods is beyond the
scope of this paper, but we note that of the 52 planets detected at
$\mathcal{P} > 5$ years and with $\mathcal{M} > 0.3 \mathcal{M}_J$
($\sim$Saturn mass), 18 have $e > 0.3$ and
30 have $e > 0.2$ (\texttt{exoplanets.org}); the detections presented
here add two (three in the $e>0.2$ case) planets to that group, implying an uncorrected frequency of
$\sim$40\% (or 60\% for $e>0.2$). 

Our careful differential analyses, both comparing the stars to
the Sun (A-Vesta, B-Vesta) and comparing the two stars to each other
(B-A, A-B), provides precise stellar parameters and abundances that
reveal that HD 133131A and B are solar ``twins'' in $T_{eff}$, log
$g$, and $\xi$, but are more metal-poor ([Fe/H] for both $\sim$-0.30)
and likely older than the Sun ($\sim$9.7 Gyr). Additionally, we find a
small but significant depletion of high $T_c$ ($> 1000$ K) elements in HD
133131A versus B, and explore how this could be related to differences
in planet formation, interactions between the planets and the two stars, and/or the composition of the planets orbiting the
two stars. This system is the smallest separation ``twin'' binary
system for which such a detailed abundance analysis has been
conducted. Overall, HD 133131A and B are especially noteworthy because
they are metal-poor but are orbited by multiple giant planets,
contradictory to the predicted giant planet-metallicity
correlation. The planets detected here will be important benchmarks in
studies of host star metallicity, binarity of host stars, and long
period giant planet formation.


\acknowledgments
RPB gratefully acknowledges support from NASA OSS Grant NNX07AR40G,
the NASA Keck PI program, and from the Carnegie Institution of
Washington. MD gratefully acknowledges the support of
CONICYT-PFCHA/Doctorado Nacional, Chile. The work herein is based on observations obtained at Las Campanas Observatory of the Carnegie Institution of Science. 
This research has made use of the SIMBAD database, operated at CDS,
Strasbourg, France, and the Exoplanet Orbit
Database and the Exoplanet Data Explorer at exoplanets.org. We thank
the referee for their detailed reading of our paper and helpful
comments, which improved the content and quality of the paper.

{\it Facilities:} \facility{Magellan:Clay (MIKE, PFS)}

Software: Qoyllur-quipu (Ram{\'{\i}}rez et al. 2014), IRAF, MOOG (Sneden 1973; Sobeck et al. 2011), SYSTEMIC (Meschiari et al. 2009)


\end{document}